\documentclass[aps,onecolumn,10pt,nofootinbib,longbibliography,floatfix]{revtex4-2}
\usepackage[T1]{fontenc}
\usepackage[utf8]{inputenc}
\usepackage{amsmath,amssymb,bm}
\usepackage{graphicx}
\usepackage{booktabs}
\usepackage{microtype}
\usepackage{xcolor}
\usepackage[unicode]{hyperref}
\hypersetup{
  colorlinks=true,
  allcolors=blue,
  breaklinks=true,
  bookmarksnumbered=true,
  pdfborder={0 0 0},
  pdfauthor={Slava G. Turyshev},
  pdftitle={Sustained habitation beyond Earth: health, recovery, and renewal},
  pdfsubject={Physical requirements, technology readiness, biological qualification, and productive renewal}
}
\newcommand{\gzero}{g_0}
\newcommand{\dd}{\mathrm{d}}

\makeatletter
\let\paper@nativeparboxrestore\@arrayparboxrestore
\appdef\tabular@hook{%
  \def\@arrayparboxrestore{\paper@nativeparboxrestore\raggedright\let\\\@tabularcr}%
}
\AtBeginDocument{\immediate\write\@auxout{\string\citation{apsrev42Control}}}
\makeatother

\begin{document}

\title{Sustained Habitation Beyond Earth: Health, Recovery, and Renewal}
%\title{Sustained habitation beyond Earth: health, recovery, and renewal}

\author{Slava G. Turyshev}

\affiliation{Jet Propulsion Laboratory, California Institute of Technology,\\4800 Oak Grove Drive, Pasadena, California 91109, USA}

\date{\today}

\begin{abstract}
Establishing a lasting human settlement beyond Earth requires a habitable destination at arrival and sustained renewal of its population and infrastructure. We compare Earth delivery, robotic predeployment, local resource processing, and localized manufacture under coupled protection, power, recovery, and health requirements. In a specified 1,000-resident case, class-specific retention times and 1\% annual growth require about 1.8 tonnes of equipment replacement and expansion per resident-year. Local manufacture raises mean site electricity from 11.8 to 19.5~MW; the declared retention range gives 17--26~MW at fixed processing intensity. A separate 100-tonne/year pilot reduces ten-year deliveries from 1,000 to 419 tonnes after a three-year commissioning ramp, while retaining imported inputs. Local production thus reduces freight by adding maintained capacity at the destination. Receiving capacity is bounded by replacement-adjusted supply, not passenger throughput. Published Martian transport gives approximately 63,000 tonnes of planar cover per hectare for the selected 10-mSv/year galactic-cosmic-ray allocation; Earth-like acceleration at 2~rpm requires a 224-m rotation radius. These scales favor local shielding and modular rotating habitats. Repeated ten-day wastewater outages every 35 days require capacity at least 1.4 times inflow; storage cannot correct a persistent throughput deficit. Existing evidence leaves lifelong partial-gravity and mixed-field radiation responses, gestation, development, and subsequent fertility unresolved. With sustained investment, recoverable habitats, resource pilots, and controlled-gravity investigations are credible 10--20-year objectives; broad industrial renewal requires further multidecade development. Earth-supported habitation is a credible development path. Independent population continuity requires demonstrated health, recovery, demographic resilience, and essential-product renewal within one exposure and resource history; its completion cannot yet be dated.

\end{abstract}

\maketitle

{\small\tableofcontents}

\section{Introduction}
\label{sec:scope}

The central question is whether people can establish settlements beyond Earth that support healthy lives, successive generations, and the infrastructure on which both depend. We assess how reusable transport, predeployed infrastructure, and progressively localized production can support sustained habitation, and which recovery and biological requirements remain independently limiting. Here \emph{habitation} means sustained residence, \emph{expansion} means growth in occupied capacity or reach, and \emph{relocation} means population transfer for a specified objective. Permanent habitation can retain trade with Earth; surviving interruption of essential imports is a further capability.

Continuous occupation of the International Space Station (ISS) by successive adult crews reached 25 years in November 2025 \cite{NASAISS25}. Reusable transport, robotic servicing, collected-water recovery, and assembly and fabrication demonstrations provide a development base for larger reserves, maintained installations, and repeated experiments \cite{OIGCrew2026,Jones2018,CSADextre2024,SpaceLogistics2026,Boyce2023,Gregg2024,ESAmetal2026}. The resulting opportunity extends beyond cheaper access: useful hardware can be retained, selected material flows localized, and environments tested before larger commitments. Whether those gains support a settlement depends on the complete service and health requirements.

Earth supplies pressure, gravitational loading, shielding, chemical inventories, and an interacting biosphere. Engineered habitats must supply the corresponding functions---respiration, nutrition, development, thermal regulation, infection control, and recovery---rather than reproduce the planet in miniature \cite{BVAD2022,Wheeler2017}. Imported active equipment can enable robotic placement of local cover and civil works before arrival; in-situ resource utilization (ISRU) and qualified parts production can subsequently replace recurring deliveries. Predeployment specifies when capacity is ready, whereas local sourcing specifies its material origin. Both must be evaluated through commissioning, operation, and replacement.

Integrated settlement studies couple food, atmosphere, spares, and resupply \cite{Do2016}; equivalent-system-mass methods include hardware, power, operations, and logistics \cite{Ho2022}; and supportability studies quantify failures, maintenance, and spares from ISS records \cite{Owens2016,Owens2018}. Robotic industrial models include machinery growth and import substitution \cite{Metzger2013}, while activity-based models connect human time to shared services \cite{Salotti2020}. Building on these methods, we derive and evaluate conditions under which commissioned supply, recovery capacity, and equipment renewal support the same population within one electrical and thermal boundary. Controlled comparisons isolate retirement and commissioning; joint sensitivity identifies power-incompatible combinations; and a reproduced industrial-throughput benchmark distinguishes nominal from accepted output. Biological evidence then determines what can be inferred about the health of the supported population. The contribution is the change in operating and architectural conclusions when these requirements are imposed together.

Three results organize the assessment. First, localization exchanges recurring freight for maintained site capability: the specified product line reduces ten-year delivery by 58\% after commissioning, while local manufacture of the separate 1,000-resident equipment stream raises mean electricity from 11.8 to 19.5~MW (17--26~MW over the retention sensitivity). Second, net capacity governs establishment and recovery. Retirement extends the reference build from 9 to about 13 years at 3,000 tonnes/year and prevents target attainment at 1,000 tonnes/year; ten-day outages every 35 days require operating processing capacity at least 1.4 times inflow. Third, environmental control and biological adequacy require different evidence. The lowest tested loading preserving selected mouse muscle-function and fiber-type endpoints was $0.67\gzero$ \cite{Tsuji2026}; that finding motivates controlled loading and complete-life-cycle investigation, rather than specifying a human threshold.

The numerical reference fixes a common population and stated technology, while Appendix~\ref{app:repro} separates observations, model assumptions, design sensitivities, and unresolved responses. The calculations establish conditional operating requirements; qualification concerns their simultaneous attainment together with clinically adequate function, consent, and continuity of essential services \cite{NASEMThriving2023,NASEMRadiation2021}.

The paper is organized as follows: Section~\ref{sec:framework} defines settlement stages, material balances, and advance/hold gates. Section~\ref{sec:transport} follows transit, arrival, commissioning, and declining Earth dependence. Sections~\ref{sec:radiation}--\ref{sec:reproduction} connect protection and gravity to lifelong health and population continuity. Sections~\ref{sec:lifesupport}--\ref{sec:isam} develop chemical, electrical, recovery, and industrial requirements. Sections~\ref{sec:architectures} and~\ref{sec:purpose} compare destination architectures and strategic objectives; Sec.~\ref{sec:roadmap} identifies development priorities and discriminating experiments. Section~\ref{sec:conclusions} synthesizes the feasibility judgments. Appendix~\ref{app:repro} records input provenance and reproducibility conventions.

Tables~\ref{tab:abbr} and~\ref{tab:notation} collect the recurring abbreviations and principal symbols for use throughout the analysis; specialized terms and quantities are defined at first use.

\begin{table}[!htb]
\caption{Abbreviations used repeatedly in the manuscript. Readiness terms refer to the stated function and environment, not to an entire settlement.}
\label{tab:abbr}
\begin{tabular}{p{.16\textwidth}p{.76\textwidth}}
\toprule
Abbreviation & Meaning\\\midrule
AI & Artificial intelligence.\\
ECLSS & Environmental control and life-support system.\\
EVA & Extravehicular activity.\\
FSP & Fission surface power.\\
GCR & Galactic cosmic rays.\\
ISS & International Space Station.\\
ISRU & In-situ resource utilization.\\
PRA & Probabilistic risk assessment.\\
RBE & Relative biological effectiveness.\\
SANS & Spaceflight-associated neuro-ocular syndrome.\\
SPE & Solar particle event.\\
TRL & Technology readiness level, applied only to a specified implementation and operating environment.\\
\bottomrule
\end{tabular}
\end{table}

\begin{table}[!htb]
\caption{Principal notation shared across sections. Subscripts identify materials, equipment classes, or operating cases; additional local symbols are defined at first use.}
\label{tab:notation}
\begin{tabular}{p{.09\textwidth}p{.18\textwidth}p{.67\textwidth}}
\toprule
Symbol & Typical units & Meaning\\\midrule
$N(t)$ & residents & Population history used consistently across resource, health, and renewal calculations.\\
$S_i$ & kg or t & Usable inventory of essential material $i$; $S_{i,\min}$ is its required minimum.\\
$K_j$ & qualified capacity or t & Installed functional equipment stock in class $j$.\\
$P_i, I_i$ & mass/time & Local production and imported supply of material $i$.\\
$c_i$ & -- & Effective recovery fraction of material throughput $i$.\\
$\delta_j$ & yr$^{-1}$ & Net permanent retirement rate of qualified equipment class $j$ after credited repair.\\
$\gamma$ & yr$^{-1}$ & Continuous fractional population or capacity growth rate in the stated model.\\
$k_j$ & t/resident & Qualified equipment allocation per resident for class $j$.\\
$\Sigma$ & g/cm$^2$ & Shield areal density.\\
$f_o$ & -- & Fraction of residence time outside the modeled protected region.\\
$k_w$ & -- & Waste-processing capacity divided by normal inflow.\\
$\bar p_{\rm site}$ & kW/resident & Annual-mean destination electrical demand under the stated accounting boundary.\\
$Q_{\rm reject}$ & W & Heat that must be rejected at the specified thermal boundary.\\
\bottomrule
\end{tabular}
\end{table}

\section{Settlement requirements: people, stocks, and essential services}
\label{sec:framework}
Readiness is functioning destination capacity, not dispatched cargo. Before departure, the first crew needs verified protection, power, an environmental control and life-support system (ECLSS), communications, medical response or evacuation, and reserves for the declared interruption. Imported and locally built systems can coexist; after occupation, deliveries and local output must cover consumption, net retirement, and growth.

\subsection{Feasibility definitions and a common history}
\label{sec:feasibility}
We distinguish physical admissibility, demonstrated engineering capability, physiological adequacy, and persistent population viability. For a configuration and history these form the definitional hierarchy
\[
 \mathcal F_{\rm persistent}\subseteq\mathcal F_{\rm physiological}
 \subseteq\mathcal F_{\rm engineered}\subseteq\mathcal F_{\rm physical}.
\]
Mechanics and conservation constrain the outer set; manufacture, interfaces, lifetime, and recovery constrain engineering readiness. Biological qualification requires the relevant health and functional endpoints under the resulting exposure. Persistent populations additionally require healthy development, reproduction, and aging. These definitions organize the assessment; the calculated bounds below provide its quantitative tests.

Every combined calculation uses the same configuration, operating interval, and $N(t)$. Age structure changes nutrition, care, usable volume, and available work; the infrastructure must therefore follow the population history used in the genetic and demographic assessment.

\subsection{Essential material and equipment balances}
For essential material $i$, let $S_i$ denote usable inventory. Over the declared operating interval,
\begin{equation}
 \frac{\dd S_i}{\dd t}=P_i+I_i-L_i-C_i,\qquad S_i(t)\geq S_{i,\min}.
 \label{eq:stocks}
\end{equation}
$P_i$ is local production entering usable inventory, $I_i$ imports, $L_i$ unrecovered loss, and $C_i$ net commitment to growth or inaccessible reservoirs. Recycling reduces $L_i$ rather than creating an additional production credit. During an import-free interval, $I_i=0$ and every essential stock must remain above its minimum. Finite stores bridge deficits but cannot remove persistent ones. The corresponding equation for functional equipment capacity is
\begin{equation}
 \frac{\dd K_j}{\dd t}=P_j^K-\delta_j K_j.
 \label{eq:capital}
\end{equation}
Here $K_j$ is the installed stock of qualified capacity, $\delta_j$ its net permanent retirement rate under the specified maintenance policy, and $P_j^K$ newly commissioned replacement and expansion. Commissioned additions can originate in Earth-delivered equipment, accepted local manufacture, or released spare inventory; these sources are resolved in Sec.~\ref{sec:transition}. Restoration of retained equipment is credited in $\delta_j$, not added again to $P_j^K$; temporary unavailability is treated separately through service and recovery models. Equivalently, a gross-loss formulation uses $\dot K_j=P_j^K+R_j-L_{j,\mathrm{gross}}$ with $\delta_jK_j=L_{j,\mathrm{gross}}-R_j$. Repair parts, feedstock, tools, electricity, and service demand remain charged to their own material and process balances. A replaced subcomponent and its retained parent must not both be booked as new whole-unit capacity. Where $K_j$ is represented by mass, product specification is fixed: structural material cannot substitute for an unavailable sensor or catalyst.

\begin{figure}[tb]
\centering\includegraphics[width=0.82\textwidth]{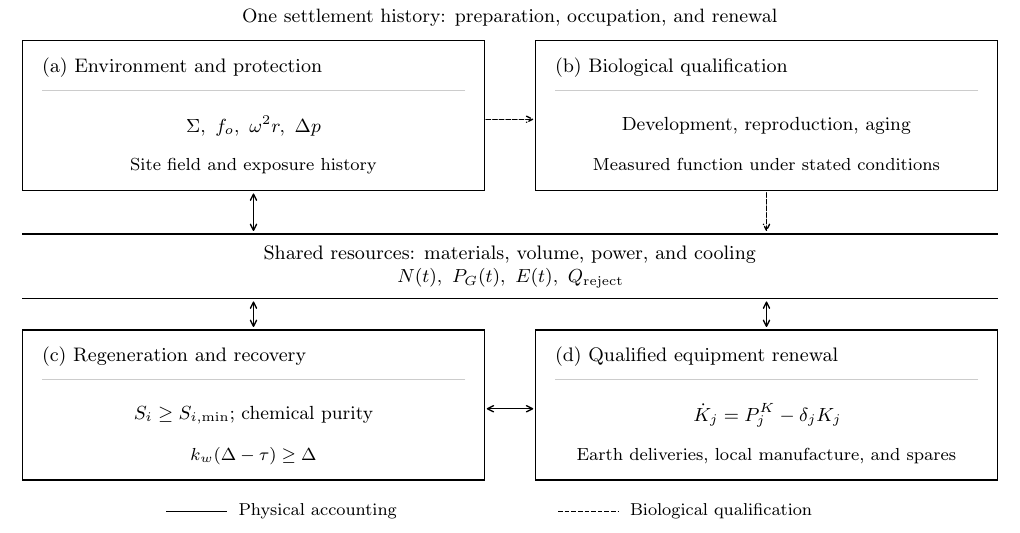}
\vskip -8pt 
\caption{Linked requirements along one settlement history. Environmental protection, biological qualification, regenerative services, and equipment supply share materials, power, volume, and cooling. Solid links indicate physical dependencies; dashed links indicate exposure-to-response evidence and biological constraints on design. Earth deliveries and local output jointly supply equipment renewal through $P_j^K$; the diagram is not a calibrated causal network.}
\label{fig:boundary}
\end{figure}

Capabilities are therefore a vector of usable stocks, product-class renewal margins, service availability, contamination, and recovery times. Dependencies cross shared power, cooling, software, feedstock, and access paths. A regional network can pool production and reserves only if its common supply boundary satisfies Eqs.~\eqref{eq:stocks} and \eqref{eq:capital} \cite{Do2016,Ho2022,Owens2018}. Figure~\ref{fig:boundary} summarizes the shared accounting and qualification links.

\subsection{Occupation, population renewal, and declining Earth dependence}
Early settlements combine Earth deliveries, local resources, and recycling. Reduced dependence means greater essential-product coverage and longer interruption endurance while maintaining health and services. Self-sustaining populations additionally complete healthy reproductive cycles and renew equipment and stocks without essential imports; a regional network may meet this requirement collectively. Table~\ref{tab:stages} gives the advance/hold gates and Sec.~\ref{sec:arrivalcargo} resolves material origins and strategic inventories.

\begin{table}[tb]
\caption{Settlement stages and advance/hold gates. Each decision uses a specified configuration, population history, and failure case; later commitments retain the preceding requirements. Detailed experiments are given in Sec.~\ref{sec:stagegates}.}
\label{tab:stages}
\begin{tabular}{p{.17\textwidth}p{.34\textwidth}p{.435\textwidth}}
\toprule
Stage & Controlling quantities & Gate for advancement\\\midrule
Design and robotic preparation & Product boundaries, exposure, resource grade, installed power, protection, and usable stocks. & Fix the configuration and failure cases; demonstrate adverse-environment output and recovery before crediting local supply.\\
Transit and arrival & Transfer, consumables, exposure, entry mass, landing reliability, and post-transit task performance. & Qualify delivery and assisted transfer/egress; verify shelter, utilities, reserves, and return or refuge before departure (Sec.~\ref{sec:arrivalfunction}).\\
Establishment & Construction throughput, commissioning time, imported support, and net retirement. & Commission adequate capacity for the arriving cohort; accepted additions cover retirement and growth [Eqs.~\eqref{eq:capital} and~\eqref{eq:productionramp}].\\
Continued occupation and expansion & Complete diet, air/water quality, shared utilities, valid clinical inputs, and reserves. & Test crop and crew branches separately; recover before depletion [Eq.~\eqref{eq:recovery}] and restore reserves before recurrence [Eq.~\eqref{eq:periodic}]; test the missed-delivery case before increasing occupancy.\\
Population regeneration & Age-specific survival/fertility, development, care coverage, and effective population. & Demonstrate life-cycle and clinical adequacy; test cohort disruption and service succession with independent ethical oversight.\\
Increasing independence & Essential product coverage, energy/feed supply, acceptance equipment, and external assistance. & Demonstrate renewal of indispensable classes and production machinery, with reserve restoration during the declared import interruption.\\
\bottomrule
\end{tabular}
\end{table}

For constant net deficit $D_i=L_i+C_i-P_i>0$, import-free endurance is $\Theta_i=(S_i-S_{i,\min})/D_i$. The earliest indispensable-material, equipment, medical, or environmental failure bounds system endurance; zero deficit removes only the corresponding depletion limit.

\subsection{Recovery must precede depletion}
If service $i$ supports a draw $D_i$ during an outage, a simple usable-buffer endurance is $T_i=(S_i-S_{i,\min})/D_i$. A necessary recovery condition is
\begin{equation}
 T_{\rm detect}+T_{\rm isolate}+T_{\rm restore}<\min_{i\in\mathcal D}T_i,
 \label{eq:recovery}
\end{equation}
where $\mathcal D$ contains the inventories actually depleted by that failure. The quantities must be evaluated for the same event; a clean-water tank cannot bridge a loss of pressure. The system must also restore reserves before the next credible interruption: post-repair throughput above normal demand is insufficient unless catch-up fits within the available operating interval [Eq.~\eqref{eq:periodic}]. Pressure breaches impose minute-scale decisions, gas processing hour-to-day scales, crop replacement months, and infrastructure renewal years. A favorable annual balance does not establish these transient margins.

At continuous population growth rate $\gamma=\dot N/N$, fixed per-resident inventories $s_i$ and machinery $k_j$ require
\begin{equation}
 \frac{P_i+I_i}{N}\geq (1-c_i)q_i+\gamma s_i,\qquad
 \frac{P_j^K}{N}\geq(\delta_j+\gamma)k_j,
 \label{eq:growth}
\end{equation}
where $q_i$ is throughput and $c_i$ effective recovery. Growth consumes real inventory and productive capacity. Health enters through the required environment and services; it is not represented by an assumed numerical biological-viability multiplier. Later sections make these quantities explicit and distinguish their measurement from their engineering selection.

\subsection{A shared-capacity test for expansion}
\label{sec:sharedcapacity}

For a fixed per-resident specification, Eq.~\eqref{eq:growth} also bounds the instantaneous expansion rate. For positive $s_i$ and $k_j$, its scalar consequence is
\begin{equation}
 \gamma\leq\gamma_{\rm supply}\equiv
 \min\!\Big\{
 \min_i\frac{(P_i+I_i)/N-(1-c_i)q_i}{s_i},\;
 \min_j\Big(\frac{P_j^K}{Nk_j}-\delta_j\Big)
 \Big\}.
 \label{eq:growthmargin}
\end{equation}
A negative term means even stationary provision is inadequate. Positive growth must also fit food, habitable volume, dispatch, cooling, work, and clinical capacity. Equation~\eqref{eq:growthmargin} therefore bounds resource provision rather than prescribing fertility or immigration. Only commissioned output can meet current demand. A technology contributes by increasing the limiting margin after its own requirements are included.

Qualification must include shared failures: a processor supplied by an external grid has not demonstrated operation during the habitat's generator outage. Local throughput is credited only with the required utilities, recovery, and support available together. Table~\ref{tab:stages} states the corresponding gates.

\section{Transit, arrival, and establishment}
\label{sec:transport}
\subsection{Reusable access: cost reduction is not destination independence}
\label{sec:access}
Compare provider cost, customer price, cost per actual payload to a stated orbit, and cost per commissioned destination function separately. A National Aeronautics and Space Administration (NASA) historical analysis found approximately twentyfold improvement in nominal low-Earth-orbit cost per maximum payload, but approximately fourfold improvement in its capsule-inclusive ISS comparison \cite{Jones2018}. These differently bounded estimates establish substantial historical change, without isolating reuse or an inflation-adjusted trend. The published Falcon~9 plan quotes \$74~million for up to 5.5~tonnes to standard geostationary transfer orbit; its larger performance limits apply to fully expendable operation \cite{SpaceX2026Pricing}. Prices and payloads must refer to the same configuration.

For repeated launches, a transparent average-cost model is
\begin{equation}
 c_L=\frac{C_r/n+C_e+C_t+C_f/\nu_L}{m_u},\qquad
 \frac{C'_{\rm sys}}{C_{\rm sys}}=1-f_L+\frac{f_L}{R_L}.
 \label{eq:launchcost}
\end{equation}
Here $C_r$ is reusable hardware cost allocated across $n$ flights, $C_e$ expendable hardware and propellant, $C_t$ turnaround and flight operations, $C_f$ annual fixed infrastructure, $\nu_L$ flights/year, and $m_u$ actual useful payload. The first relation describes provider cost, not its market price. The second holds an architecture and all nonlaunch costs fixed while its launch-price contribution, initially fraction $f_L$ of lifecycle cost, falls by factor $R_L$. With $R_L=10$, total cost falls by 18\% for $f_L=0.2$, or 72\% for $f_L=0.8$. The limiting savings as launch becomes free are $f_L$. Reuse can also increase payload production and experiment cadence; reoptimization can replace costly mass minimization with heavier, simpler shielding, reserves, or serviceable hardware. Those changes require a new architecture calculation rather than extrapolating the fixed-design elasticity.

Interplanetary delivery additionally requires aggregation, transfer, landing, and commissioning. The March~2026 lunar-lander audit described a Starship campaign with more than ten tanker flights plus depot and lander, aggregation more than 200 days before crew departure, and an approximately six-day replenishment target \cite{OIGHLS2026}. Cryogenic transfer and turnaround remained demonstrations at that audit date, not established performance. Pad and utility capacity also constrain cadence \cite{OIGInfrastructure2026}. The relevant throughput is successful commissioned payload, including supporting flights and losses.

Cheaper imports improve residence and experimentation but can delay local fabrication's financial crossover by reducing the value displaced (Sec.~\ref{sec:servicecost}). The decision therefore separates affordability from independence: local repair can retain resilience value even when continued imports are cheaper.

\subsection{Transportation requires a trajectory and a surviving payload}
For circular, coplanar heliocentric orbits of radii $r_1$ and $r_2$, the minimum-energy two-impulse transfer has semimajor axis $a_t=(r_1+r_2)/2$ and time
\begin{equation}
 t_H=\pi\sqrt{\frac{a_t^3}{\mu_\odot}},\qquad
 v_{\infty,1}=\sqrt{\mu_\odot(2/r_1-1/a_t)}-\sqrt{\mu_\odot/r_1}.
 \label{eq:hohmann}
\end{equation}
For Earth and Mars this idealization gives 259 days, a departure hyperbolic excess of 2.94~km/s, and arrival excess of 2.65~km/s. Escape from a 400-km Earth parking orbit then requires approximately 3.57~km/s of impulsive velocity increment; arrival at 125~km above Mars occurs at approximately 5.60~km/s before atmospheric braking. These reference values use the stated circular-orbit approximation. Faster transfers generally change departure energy, arrival energy, propellant, and exposure together. The 180-day leg used in the consumables and radiation benchmarks is a faster-transfer assumption, distinct from Eq.~\eqref{eq:hohmann}; the 259-day calculation supplies the separate precursor-timing illustration. Table~\ref{tab:transitdose} compares their exposure durations. Planetary ephemerides, launch windows, capture, abort, and landing constraints must be included in a mission design \cite{Cianciolo2016,JPLPlanets}.

The propulsion trade cannot be described by travel time alone. For ideal exhaust speed $v_e=g_0I_{\rm sp}$,
\begin{equation}
 \frac{m_0}{m_f}=e^{\Delta v/v_e},\qquad
 F=\frac{2\eta_pP_p}{v_e}.
 \label{eq:propulsion}
\end{equation}
At $\Delta v=10$~km/s, the mass ratios are 9.64, 3.11, and 1.405 for selected $I_{\rm sp}=450$, 900, and 3,000~s. These represent chemical-class, nuclear-thermal-target, and electric-propulsion-class comparisons, not equally available vehicles or optimized trajectories \cite{NASEMNuclear2021}. Higher exhaust speed reduces propellant at the cost of thrust at fixed power. At 1~MW and $\eta_p=0.65$, 3,000~s gives only 44~N; propulsion power and radiator mass must accompany the thruster.

For constant electric power, constant exhaust speed, and final nonpropellant mass $m_f=m_d+\alpha_pP_p$ including the power system, the ideal powered duration is
\begin{equation}
 t_b=\frac{(m_d+\alpha_pP_p)v_e^2}{2\eta_pP_p}
       \left(e^{\Delta v/v_e}-1\right).
 \label{eq:epburn}
\end{equation}
Here $m_d$ is nonpower-system dry mass; initial propellant is separately charged through $m_0=m_f e^{\Delta v/v_e}$. At the stipulated $m_d=100$~t and $\alpha_p=10$~kg/kW, increasing power from 1 to 10~MW reduces this ideal burn time from 343 to 62 days, while increasing power-system mass from 10 to 100~t and initial propellant from 45 to 81~t. Its infinite-power limit is 31 days. These powered durations exclude coast, gravity, maneuver direction, and capture. Power density, rather than specific impulse alone, controls whether electric propulsion is useful for rapid crew delivery. It can be valuable earlier for slow cargo. No fusion propulsion, torpor, or biological redesign is required for the near-term pathway considered here.

Chemical and kilowatt-class electric propulsion have operational heritage; the nuclear-thermal target and megawatt power train require integrated qualification \cite{NASEMNuclear2021}. At 10~MW and 65\% propulsion efficiency, 3.5~MW remains outside useful directed jet power; onboard dissipation must be separated from exhaust losses before radiator sizing. Reactor shielding, propellant storage, distribution, and lifetime accompany the propulsion gain.

\subsection{Crew scale, life support, and transit volume}
A transport manifest must include its own protected volume, maintenance supplies, food, gas and water stores, and landing or rendezvous interfaces. For $N_t$ travelers, duration $t_t$, dry-food demand $q_f$, and \emph{separately established} oxygen and water recovery fractions $c_O,c_W$, a simple consumable lower-bound inventory is
\begin{equation}
 M_c=N_tt_t\Big\{q_f+(1-c_O)q_O+(1-c_W)q_W\Big\}+M_{\rm buffers}.
 \label{eq:transitconsumables}
\end{equation}
The inventory excludes packaging, collection failures, hygiene beyond the specified stream, leakage, initial working fluids, and growth of biological stores. With $q_f=0.62$, $q_O=0.84$, and $q_W=4$~kg/person/day \cite{BVAD2022}, 180 days without recovery requires approximately 983~kg/person before buffers. At independently assumed 90\% recovery of both gas oxygen and the specified water stream, the value is 199~kg/person; at 98\% it is 129~kg/person, of which 112~kg is dry food. The latter oxygen performance is a design assumption, not inferred from ISS water recovery. Thus 1,000 travelers require at least 112~t of dry food for that leg, and the electrical plant must operate the regenerators throughout transit. With the same 98\% recovery assumptions, a 259-day leg instead requires 186~kg/person before buffers, including 161~kg of dry food. The extra duration changes both exposure and delivered consumables.

Habitable volume excludes space occupied by machinery. Standards require accommodation for tasks, sleep, privacy, exercise, care, and emergencies, not a universally sufficient volume \cite{NASA3001V2F}. The declared 25--100~m$^3$/person envelope gives 2,500--10,000~m$^3$ for 100 travelers. At 100~kPa and 293~K, air is 1.19~kg/m$^3$, or 3--12~t; pressure structure and shielding are larger burdens. Compartments isolate faults but add interfaces. Cohort size must fit the demonstrated vehicle and rescue architecture, and exposure history carries into residence.

\subsection{Arrival requires both delivery and usable human function}
\label{sec:arrivalfunction}
At the reference Mars entry speed, kinetic energy is approximately 15.7~MJ/kg: a 20-t \emph{entry vehicle} must dispose of about 314~GJ, and a lander delivering 20~t of useful payload has greater entry mass. Ballistic coefficient $\beta=m/(C_DA)$ controls the atmospheric column required for deceleration; stagnation heating scales approximately as $\dot q\propto\sqrt{\rho/R_n}\,v^3$. Geometric similarity gives $A\propto m^{2/3}$ and $\beta\propto m^{1/3}$: a twentyfold mass increase raises $\beta$ by 2.71, whereas holding $\beta$ fixed requires twenty times the area, or 4.47 times the diameter. Supersonic retropropulsion, decelerators, guidance, and plume interaction therefore require representative qualification. Human-class studies investigate payloads of order 20~t; the 100-t cases below are sensitivities beyond the approximately tonne-class robotic heritage \cite{Cianciolo2016}. The Moon exchanges atmospheric braking for propulsive deceleration; free-space destinations replace landing with rendezvous and assembly.

Delivery does not establish immediate crew capability. In 15 astronauts after approximately six-month ISS missions, mobility, balance, and bimanual performance deteriorated; mobility and pegboard performance approached preflight levels around 30 days after return, while balance recovery was task-dependent \cite{Tays2021}. These were terrestrial recovery observations, not a Martian rehabilitation schedule. Protected transfer, hatch operation, leak isolation, equipment handling, and emergency egress must be tested after representative transit exposure, with suits, assistance, and the destination loading included. No survival-critical arrival task should depend on unverified post-transit performance. Predeployed robots, autonomous environmental control, accessible interfaces, and any independently qualified personnel must bridge the interval until task-specific function is demonstrated. Visual and circulatory endpoints in Sec.~\ref{sec:gravity} belong in this acceptance test alongside strength and balance.

\subsection{Initial cargo, local construction, and readiness at arrival}
\label{sec:controlledcomparison}
\label{sec:arrivalcargo}

Let $M_{E,0}$ denote cumulative Earth-origin cargo successfully delivered to the destination by first-crew arrival. Separate cargo arriving earlier, $M_{\rm pre}$, from cargo accompanying that cohort, $M_{\rm crew}$:
\begin{equation}
 M_{E,0}=M_{\rm pre}+M_{\rm crew},\qquad
 M_g=M_{g,\rm equipment}+M_{g,\rm consumables}+M_{g,\rm reserves}\quad(g=\mathrm{pre},\mathrm{crew})
 \label{eq:initialcargo}
\end{equation}
with installed equipment, working consumables, and reserved stocks evaluated as disjoint cargo categories for each delivery group; held replacement parts belong to reserves. This is delivered cargo, not launch mass: propulsion stages, transit expenditure, packaging, failed shipments, and transfer losses must be added through the transportation calculation. Predeployed equipment may consume stocks or retire before crew arrival; cumulative delivery and usable inventory at arrival are consequently different quantities. Infrastructure ready at that date includes accepted local works as well as surviving imported equipment. Table~\ref{tab:deployment} separates these categories without adding local feed and the products made from it twice.

For installed renewable hardware $k$ tonnes/resident, imported hardware fraction $\phi$, imported start-up stocks $M_s$, and useful landed payload $m_L$, the restricted delivery screen remains
\begin{equation}
 n_L=\Big\lceil\frac{\phi Nk+M_s}{m_L}\Big\rceil,\qquad
 T_{\rm windows}=\left(\left\lceil n_L/c_L\right\rceil-1\right)T_{\rm syn},
 \label{eq:deployment}
\end{equation}
Here $c_L$ is successful deliveries per opportunity. The numerator omits construction-period retirement, rejected cargo, and additional seed equipment unless explicitly included in $k$; it is therefore a lower bound. $T_{\rm windows}$ runs from first to last opportunity, not through transfer and commissioning. The Earth--Mars synodic interval is approximately 26 months. Missed opportunities and commissioning failures require separate reserves and schedule margins.

For $N=1,000$ and $k=30$~t/resident, two years of dry food (453~t) and the initial air in $100,000$~m$^3$ of connected residential volume (119~t) give $M_s=572$~t. Cultivation and industrial volumes, water, nutrients, drugs, propellant, and commissioning spares require additional entries. At $m_L=20$~t, the resulting lower bounds are 1,529 deliveries for $\phi=1$, or 179 for $\phi=0.1$. At 100 successful deliveries per opportunity, they span 16 or two opportunities, respectively. The second case requires 27,000~t of accepted local hardware and any additional machinery needed to make it. The 90\% hardware localization case is an engineering sensitivity. Unlike a pressure module, much shielding and civil material can be excavated and placed without advanced manufacture (Sec.~\ref{sec:isru}).

\begin{table}[tb]
\caption{Material and infrastructure categories in establishing the reference settlement. Predeployment is a timing category, not a material origin. Stocks, flows, raw feed, and commissioned products are accounted for separately; illustrative quantities do not constitute a complete manifest.}
\label{tab:deployment}
\begin{tabular}{p{.185\textwidth}p{.365\textwidth}p{.40\textwidth}}
\toprule
Category & Physical definition & Accounting quantity or reference scale\\\midrule
Initial Earth deliveries & Equipment, starting consumables, and reserves delivered before or with the first cohort. & $M_{E,0}=M_{\rm pre}+M_{\rm crew}$; separately include transport losses and prearrival expenditure.\\
Predeployed infrastructure & Imported and locally built capacity demonstrated operational before crew departure. & Available kW, habitable volume, accepted structures, usable stores, and recovery performance; not an extra mass added to delivery totals.\\
Local material and ISRU & Excavated cover, rock, recovered water, atmospheric gases, and refined feed. & Grade, recovery, and feed/output t/year; about 63,000~t/ha of planar cover in the specified Martian shielding calculation.\\
Construction and manufacture & Installed works and accepted equipment produced or assembled on site. & Local product t/year and commissioning yield; assembly of imported modules is distinct from manufacture from local feed.\\
Operating consumption and replacement & Unrecovered material losses, complete food deficits, worn parts, and growth. & 226~t/year dry-food demand and 1,510~t/year stationary hardware replacement per 1,000 residents; origin is specified separately.\\
Strategic inventories and spares & Usable, segregated stocks covering missed deliveries, plant failures, and recovery. & 180-day dry food: 112~t; 30-day gross respiratory O$_2$ backup: 25.2~t; a ten-day specified personal-water interruption: 40~t clean supply and 40~t wastewater capacity.\\
Reduced Earth dependence & Essential needs met without new Earth delivery, including the producing equipment's own needs. & Local coverage by product, continuing imports per resident-year, stock depletion, and verified interruption endurance.\\
\bottomrule
\end{tabular}
\end{table}

The reserve cases in Table~\ref{tab:deployment} assume loss of oxygen regeneration or continued water withdrawal and collection during processor outage. Acceptance requires the actual failure and resupply horizon, including purity, shelf life, and replenishment. Crew-critical services must operate before arrival; the later factory ramp concerns additional production.

\begin{figure}[tb]
\centering\includegraphics[width=0.80\textwidth]{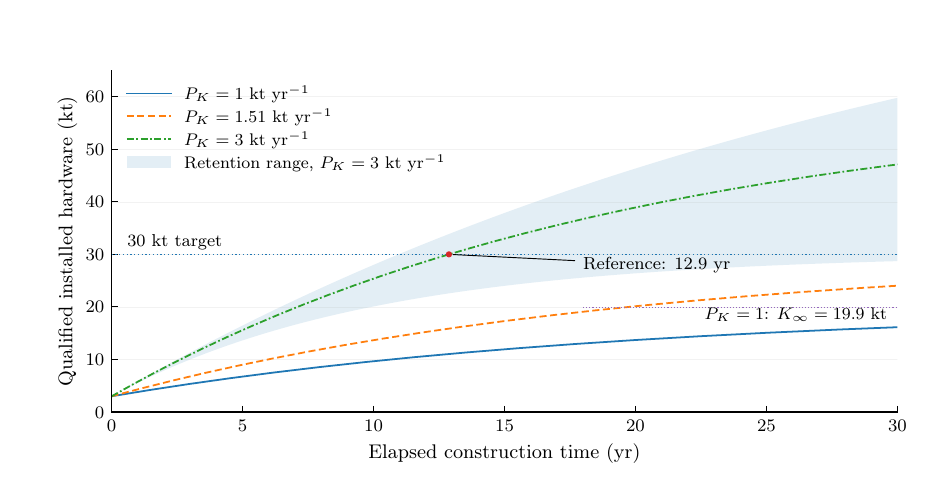}
\vskip -18pt
\caption{Establishment and equipment retention. At the reference net-retirement rate $\delta_{\rm eff}=0.0503$~yr$^{-1}$, compatible output of 1,000~t/year remains below the 30,000-t target, 1,510~t/year approaches it asymptotically, and 3,000~t/year reaches it in 12.9 years. Shading at 3,000~t/year varies the net-retention times over Table~\ref{tab:capital}'s stated endpoints: the corresponding crossings are 11.0 and 60.2 years, the latter outside the displayed 30-year interval. This is a deterministic sensitivity, not a confidence band; full trajectories are archived. Equation~\eqref{eq:bootstrap} assumes compatible production preserving the class mix.}
\label{fig:establishment}
\end{figure}

The following fixed-throughput calculation isolates the replacement burden; Sec.~\ref{sec:isrucommission} extends it to growing productive capacity. At 1,000 qualified tonnes/year, 27,000~t takes 27 years even if early equipment never retires. More decisively, for constant qualified production $P_K$ and net permanent retirement $\delta$, Eq.~\eqref{eq:capital} has solution
\begin{equation}
 K(t)=K_0e^{-\delta t}+\frac{P_K}{\delta}(1-e^{-\delta t}),\qquad
 P_K>\delta K_*\quad\hbox{to reach }K_*>K_0\hbox{ in finite time}.
 \label{eq:bootstrap}
\end{equation}
For the capital composition and net-renewal times in Table~\ref{tab:capital}, the mass-weighted rate is $\delta_{\rm eff}=0.0503$~yr$^{-1}$. When compatible production preserves that mix, the aggregate stock follows Eq.~\eqref{eq:bootstrap} with $\delta=\delta_{\rm eff}$. An initial 3,000~t and $P_K=1,000$~t/year approach only 19,900~t, below the 30,000-t target; 3,000~t/year reaches the target in approximately 13 years (Fig.~\ref{fig:establishment}). Stationary replacement at the target is 1,510~t/year, or 1,810~t/year including 1\% growth. Holding output at 3,000~t/year while removing retirement gives $(30,000-3,000)/3,000=9$ years rather than 12.9. At 1,000~t/year, the same omission gives 27 years whereas the reference retirement model never reaches the target. This paired comparison isolates retirement; it does not attribute its omission to the cited industrial models. The figure also shows the sensitivity to the retention intervals in Table~\ref{tab:capital}. These constant-throughput calculations require compatible class-level production and commissioning; a changing capital mix requires the full system of Eq.~\eqref{eq:capital}, not a fixed effective rate. Initial modules should increase net productive capacity, with explicit dependencies on electronics, catalysts, bearings, feedstock purity, and metrology \cite{Metzger2013}. Shield emplacement and fabrication have different productivity and acceptance requirements; Sec.~\ref{sec:isru} connects the local-cover campaign to complete excavation cycles.

For $n_c$ independently critical deliveries with success probability $s$, all-arrival probability is $s^{n_c}$: twenty at $s=0.99$ give 0.818. Replaceable bulk cargo instead needs loss allowance and reserves. Thus predeployment should duplicate critical functions and verify the installed shelter, power, atmosphere, water, communications, and refuge before departure. Counting all bulk shipments as irreplaceable overstates risk; ignoring the unique essential component understates it.

\subsection{From continuing deliveries to qualified local production}
\label{sec:localtransition}
\label{sec:transition}
After occupation, production serves consumption, replacement, and additional residents. For equipment class $j$, define required commissioned additions $D_j=\delta_jK_j+\dot K_j^{\rm growth}$ and qualified spare inventory $Z_j$. Earth deliveries $I_{K,j}$ and accepted local production $Q_j$ then satisfy
\begin{equation}
 \dot Z_j=I_{K,j}+Q_j-D_j,\qquad Z_j\geq Z_{j,\min},\qquad
 f_j=\min\!\Big(1,\frac{Q_j}{D_j}\Big)\quad(D_j>0).
 \label{eq:productsupply}
\end{equation}
The coverage fraction $f_j$ concerns current qualified output, not the fraction of cumulative structure made from local rock; excess production charges reserves. Manufacturing with Earth-delivered powders or components can yield $f_j=1$ while retaining essential imports. Report those feeds and production-machine replacements separately, without counting both a component and its finished parent as the same Earth cargo. Site manufacture also increases local electricity and cooling: Earth factory energy does not belong in the destination power budget (Sec.~\ref{sec:services}).

For illustration, retain the independent 100-t/year product-family pilot of Sec.~\ref{sec:isrubreak}. It imports 100~t of nonpower machinery and 22.8~t of power/cooling hardware, reaches 100~t/year accepted output, uses imported process inputs equal to 10\% of accepted output, and imports 6.14~t/year of other support including power-system replacement. Let full product demand $D=100$~t/year start at $t=0$, when the installation is delivered, and let local production rise linearly over a commissioning interval $t_r=3$~yr. Existing Earth deliveries cover the deficit; this is a noncritical product pilot, not a startup assumption for life support. With $M_F$ the initial plant mass and $s_F$ recurring non-feed support,
\begin{equation}
 Q(t)=D\min(t/t_r,1),\quad
 \mathcal Q(T)=\begin{cases}DT^2/(2t_r),&0\leq T\leq t_r,\\
 D(T-t_r/2),&T>t_r,\end{cases}\quad
 M_E(T)=M_F+DT-(1-\iota)\mathcal Q(T)+s_FT.
 \label{eq:productionramp}
\end{equation}
At ten years, 850~t is produced locally and 150~t imported as finished product; 85~t of process input, 61.4~t of support, and the 122.8-t plant bring delivery to 419~t, 58\% below continued imports. Payback is near 3.1 years. Immediate rated production requires 284~t: the 135-t difference is the commissioning penalty. Prospecting and preinstallation expenditure lie before the time origin; batching, qualification delays, and failures require additional reserves.

\begin{figure}[tb]
\centering\includegraphics[width=0.80\textwidth]{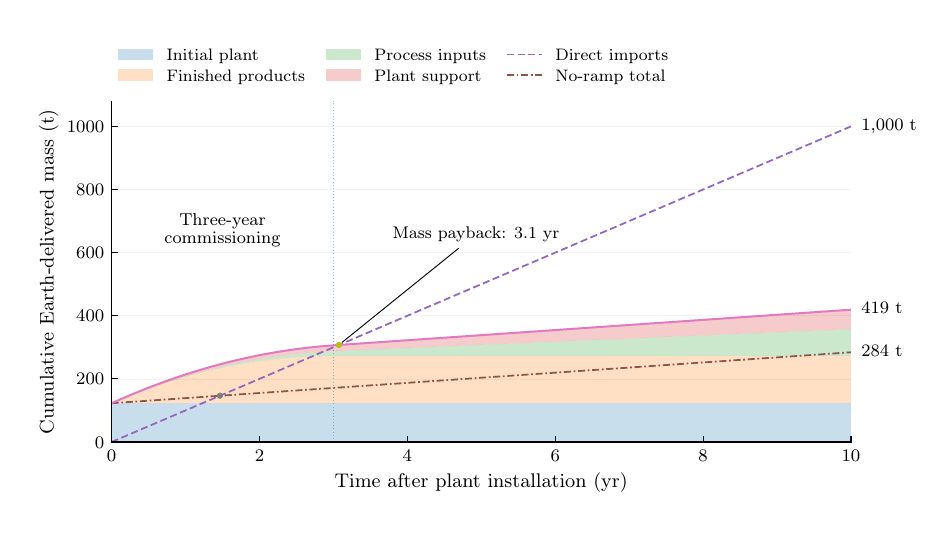}
\vskip -18pt
\caption{Commissioning and the composition of Earth deliveries. The stacked areas separate initial machinery, finished products delivered during ramp-up, imported process inputs, and continuing plant support for the three-year-ramp case. At year ten they contribute 122.8, 150, 85, and 61.4~t, respectively: 419~t in total, versus 1,000~t of direct imports or 284~t with immediate rated production. Dots mark imported-mass payback. Rated production still requires 16.1~t/year of imports. The curves follow Eq.~\eqref{eq:productionramp}; preinstallation expenditure, extra reserves, and cargo losses are excluded.}
\label{fig:transition}
\end{figure}

At rated output, 16.1~t/year of feed and support remain imported despite zero finished-product delivery: full product coverage reduces continuing imports by 84\% but does not establish independence. At 20~kWh/kg, processing draws 0.23~MW annually averaged, or 0.46~MW at 50\% duty. The useful targets are accepted output, commissioning time, support imports, useful life, and total site power, including otherwise excluded standby and recovery.

Extending the ramp from three to five or eight years at unchanged plant coefficients raises ten-year delivery from 419 to 509 or 644~t. The time origin is plant installation; development and procurement precede it. Crew-critical products require demonstrated output and stored reserves before reliance on the line.

\subsection{Population delivery and the renewal ceiling}
\label{sec:relocation}
Large-scale relocation requires receiving capacity as well as passenger delivery. Let $J$ be migrants/year, $\dot N=J+B-D$ net growth including births $B$ and deaths $D$, $\nu_cn_c$ delivered seats/year, $m_D\nu_D$ qualified cargo/year, and $P_{K,\rm loc}$ local qualified production in compatible tonnes/year. For fixed scalar capital allocation $k$,
\begin{equation}
 J\leq\nu_cn_c,\qquad k\dot N+\delta kN\leq m_D\nu_D+P_{K,\rm loc}.
 \label{eq:relocation}
\end{equation}
Births avoid passenger transport, not the receiving services and eventual productive capacity. Food, power, medical care, and product-specific commissioning constrain population in addition to this mass balance.

At constant compatible throughput $\dot M_D=m_D\nu_D$ and homogeneous net permanent retirement $\delta>0$, the replacement-limited equilibrium is
\begin{equation}
 N_\infty=\frac{\dot M_D+P_{K,\rm loc}}{q_{\rm ret}},\qquad q_{\rm ret}=\delta k.
 \label{eq:renewalceiling}
\end{equation}
Expanding qualified local production raises this constant-throughput ceiling. Here $q_{\rm ret}=\sum_j\delta_j k_j$ is stationary net replacement in tonnes per resident-year and $\delta_{\rm eff}=q_{\rm ret}/\sum_jk_j$; the scalar form assumes a fixed compatible mix. Growth adds $\gamma\sum_jk_j$. At $\dot M_D=100,000$~t/year and no local production, the independent homogeneous sensitivities $(k,\delta)=(50,0.08)$, $(30,0.08)$, and $(10,0.04)$ give $q_{\rm ret}=4.0$, 2.4, and 0.4~t/(resident\,yr), hence approximately 25,000, 42,000, and 250,000 residents. The fivefold capital and twofold rate changes jointly produce the tenfold span. In contrast, the class-resolved reference in Table~\ref{tab:capital} gives $q_{\rm ret}=1.51$ and approximately 66,000 residents on the same import-only screen. The transferable quantity is net replacement per resident, which sets these conditional stock ceilings.

For scale, the class-resolved $k=30$ reference implies 30 million tonnes of installed hardware and 1.51 million tonnes/year of stationary renewal for one million residents. Retaining the farm coefficients gives 50~km$^2$ of cultivated area. Manufacturing the full replacement-and-growth stream locally gives approximately 19.5~GW mean electricity during 1\% growth; importing it avoids that local manufacturing load but not its Earth production. In the separate all-imported hardware comparison, hypothetical qualified-destination prices of \$100 and \$1,000/kg imply \$3 and \$30 trillion for initial hardware delivery alone, excluding manufacture, passengers, stocks, financing, and operations. These test prices are not launch quotations; no scale economies are assumed.

Adding one million residents uniformly by net immigration over a century, with births balancing deaths, requires 1.81 million tonnes in the final year if renewal is imported. The century total is 105.5 million tonnes: 30 million for growth and 75.5 million replacing net-retired hardware. At 100 tonnes per cargo arrival, the final-year requirement is 18,100 arrivals; at 100 seats per passenger flight, immigration needs 100 flights/year. Industrial receiving capacity can therefore dominate passenger throughput. Local production, repair, and durability change the limiting balance directly.

An independent population preserving human continuity is a different objective from evacuating most people on Earth. Its scale must follow the reproductive, clinical, and product-class conditions below. Expansion requires net growth of commissioned receiving capacity before the cohort depending on it arrives.

\subsection{Precursor timing and the critical path to arrival}
\label{sec:precursortiming}

A stock-growth curve does not include the time required to build its factory. Solving Eq.~\eqref{eq:bootstrap} for a reachable target gives
\begin{equation}
 t_* =\frac{1}{\delta}\ln\!\Big(
 \frac{P_K/\delta-K_0}{P_K/\delta-K_*}\Big),
 \qquad P_K>\delta K_*,\quad K_* > K_0.
 \label{eq:constructiontime}
\end{equation}
For the fixed reference mix, $\delta=1.51/30$~yr$^{-1}$, $K_0=3,000$~t, $K_*=30,000$~t, and $P_K=3,000$~t/year reproduce 12.9 years. This interval starts with commissioned seed hardware and compatible production, not in 2026. Process development, delivery, prearrival wear, and later health qualification must be placed on the same dependency graph. Expanding production can shorten the constant-throughput result, but requires additional commissioned machinery rather than a change of calendar label.

Mars introduces a further discrete timing constraint. Let the first precursor launch define $t=0$, let $t_H$ and $t_c$ be precursor and crew transfer times, and require an on-site validation interval $\tau_v$ before crew launch. If useful launch opportunities are separated by $T_{\rm syn}$, a simplified lower bound is
\begin{equation}
 t_{\rm crew,arr}\geq T_{\rm syn}
 \Big\lceil\frac{t_H+\tau_v}{T_{\rm syn}}\Big\rceil+t_c.
 \label{eq:precursorwindow}
\end{equation}
Using the declared 26-month cadence and 259-day legs, validation of 0.5 year or a separately selected 687 days gives earliest arrivals 2.88 or 5.04 years after precursor launch. A missed eligible window adds 2.17 years. These logistics bounds retain 259-day legs; the 180-day crew option requires its own trajectory. The calculation assumes the precursor arrives functioning and excludes additional construction. Thus a 2036--2046 first crewed Mars mission is at most a low-confidence development opportunity with early successful precursors, whereas large-cohort relocation is not justified by that window. Imported adult outposts and the 1,000-resident industrial reference are different commitments.

\section{Environmental protection: pressure, radiation, and occupied time}
\label{sec:radiation}
\subsection{Acute hazards set the required response time}
\label{sec:acute}
Long-term reliability depends on surviving short-time failures. Pressure loss, fire, impact penetration, toxic release, and loss of cooling can defeat an otherwise favorable annual material balance. A useful containment scale follows from ideal-gas discharge to near vacuum. For a well-mixed isothermal compartment of volume $V_h$ with an opening $A_h$,
\begin{equation}
p(t)=p_0 e^{-t/\tau_p},\qquad
\tau_p=\frac{V_h}{C_d A_h\sqrt{\kappa R_sT_0}
[2/(\kappa+1)]^{(\kappa+1)/[2(\kappa-1)]}},
\label{eq:blowdown}
\end{equation}
where $C_d$ is the discharge coefficient, $\kappa$ the heat-capacity ratio, and $R_s$ the specific gas constant. This choked-flow result applies while the external-to-internal pressure ratio remains below $[2/(\kappa+1)]^{\kappa/(\kappa-1)}$. With $V_h=1,000$~m$^3$, $T_0=293$~K, $C_d=0.7$, $\kappa=1.4$, and $R_s=287$~J/(kg\,K), a 1-cm$^2$ opening gives $\tau_p=20.0$~h. A 100-cm$^2$ opening reduces it to 12.0~min. An adiabatic, well-mixed reservoir instead gives
\begin{equation}
\frac{p(t)}{p_0}=\Big[1+\frac{(\kappa-1)t}{2\tau_p}\Big]^{-2\kappa/(\kappa-1)}.
\label{eq:blowdownad}
\end{equation}
For the larger opening, a 20\% pressure loss occurs in 1.94--2.68~min between these thermal limits. The specified pressure decrement is a containment metric; physiological consequences also depend on oxygen, starting pressure, and decompression history. The response requirement follows directly: detection, isolation, and transfer to an independent refuge must operate on the breach time scale, not on the repair or resupply time scale.

At fixed dry oxygen partial pressure of 21~kPa, reducing total pressure from 101 to 56~kPa increases oxygen mole fraction from 0.21 to 0.38. Equivalent oxygen partial pressure does not imply equivalent flammability: pressure, ventilation, geometry, and materials matter. Saffire experiments measured flame and hazardous-product transport at spacecraft scale \cite{Urban2024}; inspired-oxygen and decompression constraints apply separately \cite{NASA3001V2F}. Independent compartments, fire-resistant materials, fault isolation, sensing, and atmospheric cleanup should therefore be tested as one recovery system. A refuge sharing failed ventilation, power, or cooling is not independent.

Impact protection scales with exposed area and operating time. For residual penetrating flux $\Phi_{\rm pen}$ and effective exposed area $A$, an independent-impact model gives survival probability $\exp(-\Phi_{\rm pen}AT)$. Actual flux and ballistic limits require directional particle populations, impact speed, material response, and testing \cite{Christiansen2003}. Compartmentation changes the consequence of penetration; it does not reduce its occurrence. Larger protected volumes therefore require damage isolation as well as shielding.

\subsection{Measured fields and a two-variable mission budget}
For the galactic cosmic ray (GCR) budget, the Mars Science Laboratory Radiation Assessment Detector (RAD) measured a transit dose-equivalent rate of $1.84\pm0.33$~mSv/day behind that spacecraft's shielding \cite{Zeitlin2013}. Early Curiosity surface measurements gave approximately $0.64\pm0.12$~mSv/day \cite{Hassler2014}. The published transport profile is normalized to 232~mSv/year; we retain that normalization for all annual shielding and residence calculations rather than reannualizing the rounded daily value. These are instrument-derived dose equivalents in particular particle spectra and solar-modulation conditions. Absorbed dose in gray, dose equivalent in sievert, and organ-weighted effective dose are different quantities. In particular, comparison with NASA's 600-mSv career \emph{effective-dose} limit requires the agency's dosimetric and risk methodology, not direct substitution of a detector value \cite{NASA3001,NASEMRadiation2021}.

For total transit duration $T_t$ and surface duration $T_s$, independent shielding/environment multipliers $b_t,b_s$ give
\begin{equation}
H_{\rm GCR}=b_t h_tT_t+b_s h_sT_s.
\label{eq:dose}
\end{equation}
The baseline exposure benchmark has two 180-day legs and 500 surface days, hence $T_t=360$ days and an 860-day interval. With $b_t=b_s=1$ and $h_s=232/365$~mSv/day, it gives 0.980~Sv. Table~\ref{tab:transitdose} separates this assumed crew itinerary from the ideal 259-day leg in Sec.~\ref{sec:transport}. Extending both legs while retaining 500 surface days gives 1.271~Sv over 1,018 days, a 30\% increase. Keeping 860 total days instead leaves 342 surface days and gives 1.171~Sv. These comparisons isolate duration: a realizable mission must solve return phasing, velocity increments, consumables, and shielding together. Propagation of rate errors alone uses $\sigma_H^2=(T_t\sigma_t)^2+(T_s\sigma_s)^2$, assuming one common transit-rate estimate for both legs and independent transit/surface errors. Future fields, geometry, tissue weighting, and response uncertainty require separate treatment. Solar events are evaluated by organ dose below.

\begin{table}[tb]
\caption{Explicit exposure scenarios at $b_t=b_s=1$, with $h_t=1.84$ and $h_s=232/365$~mSv/day. The last column propagates only the quoted 0.33- and 0.12-mSv/day rate uncertainties. These are duration comparisons, not ephemeris-optimized missions or effective doses.}
\label{tab:transitdose}
\begin{tabular}{p{.27\textwidth}p{.12\textwidth}p{.12\textwidth}p{.13\textwidth}p{.13\textwidth}p{.13\textwidth}}
\toprule
Case & Each leg (d) & Surface (d) & Total (d) & Transit (Sv) & $H_{\rm GCR}\pm\sigma_H$ (Sv)\\\midrule
180-day benchmark & 180 & 500 & 860 & 0.662 & $0.980\pm0.133$\\
Longer legs; surface fixed & 259 & 500 & 1,018 & 0.953 & $1.271\pm0.181$\\
Longer legs; duration fixed & 259 & 342 & 860 & 0.953 & $1.171\pm0.176$\\
\bottomrule
\end{tabular}
\end{table}

\subsection{Transport-derived shielding and occupied-time budgets}
\label{sec:shield}
Areal density $\Sigma$, particle spectrum, composition, and geometry determine protection. A published high-charge-and-energy transport (HZETRN) calculation, normalized to RAD, supplies a direct mass--dose connection: $\Sigma=280$, 560, and 840~g/cm$^2$ correspond to 81, 15, and 2.9~mSv/year of modeled GCR dose equivalent \cite{Hassler2014}. At 28~g/cm$^2$ the calculated value is 295~mSv/year, above the 232-mSv/year surface reference, illustrating secondary-particle buildup. These are radiation-transport outputs normalized to the measured field.

For the explicit 10-mSv/year GCR allocation used here, logarithmic interpolation between the two bracketing published points gives
\begin{equation}
\Sigma_{10}=560+280\frac{\ln(15/10)}{\ln(15/2.9)}
\simeq629\ \mathrm{g/cm^2},\qquad M_{\rm cover}=10\Sigma A.
\label{eq:shieldmass}
\end{equation}
In the mass expression $\Sigma$ is in g/cm$^2$ and $A$ in m$^2$. A hectare therefore requires about 63,000~tonnes at the interpolated crossing. The published 840-g/cm$^2$ point gives 84,000~tonnes and a GCR allocation margin, rather than an arbitrary three-metre prescription. The 840-g/cm$^2$ point corresponds to 3~m in 2,800-kg/m$^3$ rock or 5.6~m at 1,500~kg/m$^3$. The mass is a planar-cover inventory: side berms, penetrations, access, and the finite shelter geometry require additional transport and structural design. Regolith density and composition must be established locally \cite{Hayne2017,Rostel2020}.

\begin{figure}[tb]
\centering\includegraphics[width=0.80\textwidth]{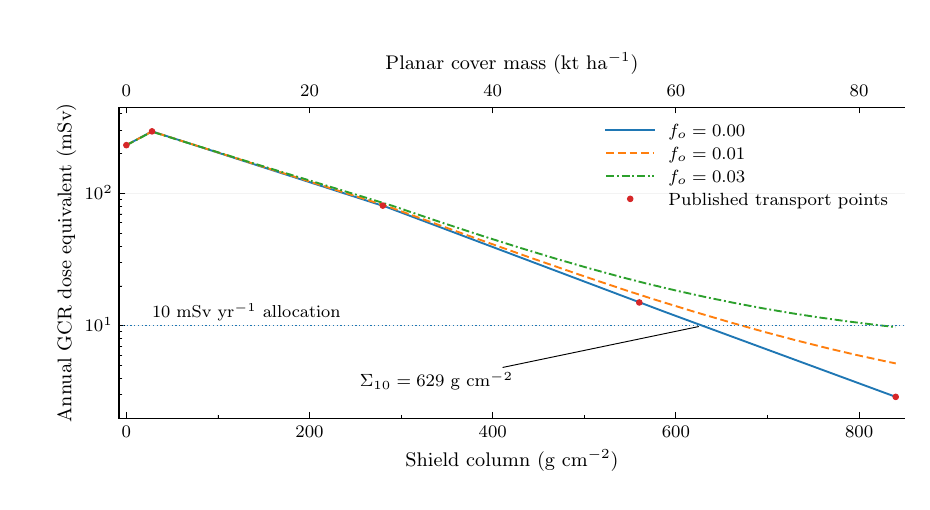}
\vskip -15pt
\caption{Shielding and occupied time in the specified Martian GCR field. Circles are published transport-profile outputs \cite{Hassler2014}; curves use piecewise logarithmic interpolation and fractions $f_o$ of annual residence outside the cover at 232~mSv/year. The dotted line is the selected 10-mSv/year design allocation, not a population safety standard. The upper axis gives planar cover mass per hectare. All source points are retained; field variability, finite geometry, and acute solar events require separate analysis.}
\label{fig:radiation}
\end{figure}

Figure~\ref{fig:radiation} couples the published shielding profile to residence outside protection. If occupants spend fraction $f_o$ outside the modeled shield, the annual GCR contribution is $(1-f_o)h_{\rm in}+f_o h_{\rm out}$. At 2.9 and 232~mSv/year, the 10-mSv/year allocation permits $f_o\leq0.031$, approximately 272~h/year. A sensitivity with both fields doubled reduces that time to about 80~h/year. Doubling is an explicit field-normalization stress test, not a confidence interval. The allocated quantity excludes local radionuclides and solar events; personnel dosimetry, solar-cycle extremes, and geometry-specific transport remain acceptance requirements. Later simulations show material and model dependence, including weak GCR attenuation over much thinner lunar regolith columns \cite{Matthia2024}. The Martian burial curve must not be exported unchanged to a lunar roof or transit vessel.

\subsection{Solar-event protection is a separate organ-dose requirement}
\label{sec:spe}
The GCR duration cases in Table~\ref{tab:transitdose} also require solar particle event (SPE) transport and organ-specific acute-dose constraints, with stated relative biological effectiveness (RBE). Table~\ref{tab:spe} gives a conditional historical-event test; event occurrence and exceedance probability are separate inputs.

\begin{table}[tb]
\caption{Conditional August 1972 solar-particle-event calculations. Deep-space values are transcribed from Tables 5 and 6 of NASA TP-3668, not read from a curve. The original cGy and cSv are converted to Gy and Sv. Absorbed dose, dose equivalent, and RBE-weighted absorbed dose are distinct quantities; the lunar statistical errors do not encompass spectrum or transport-model uncertainty.}

\label{tab:spe}
\begin{tabular}{p{.28\textwidth}p{.27\textwidth}p{.39\textwidth}}
\toprule
Event and environment & Stated shielding & Calculated event contribution\\\midrule
Deep-space astronaut model \cite{Wilson1997b} & 10~g/cm$^2$ polyethylene & Blood-forming organs: 0.10~Gy absorbed, 0.16~Sv dose equivalent. Skin: 0.40~Gy and 0.58~Sv.\\
Same source and astronaut model \cite{Wilson1997b} & 10~g/cm$^2$ aluminum & Blood-forming organs: 0.17~Gy absorbed, 0.24~Sv dose equivalent. Skin: 0.76~Gy and 1.10~Sv.\\
August 1972, lunar phantom \cite{Matthia2024} & 0, 5, 10, 20~g/cm$^2$ regolith above the phantom; lunar ground and modeled suit below/around it & Blood-forming-organ RBE-weighted absorbed dose 533$\pm$6, 88$\pm$2, 37$\pm$1, 14.3$\pm$0.8~mGy-Eq, respectively.\\
\bottomrule
\end{tabular}
\end{table}

One modeled 1972-spectrum event during either transit leg, with continuous occupancy of the 10-g/cm$^2$ polyethylene refuge, gives the tabulated organ doses. Two give 0.20~Gy blood-forming-organ absorbed dose; spacing determines consumption of acute-dose allocations. These doses cannot be added to RAD dose equivalent. Qualification requires common-field transport to lifetime effective dose and organ RBE-weighted dose over the relevant short intervals; a historical event is a stress test, not a maximum future exposure.

A four-person cylindrical refuge of radius 1.5~m and length 4~m has 51.8~m$^2$ total surface including ends. All-around polyethylene columns of 10 or 20~g/cm$^2$ require 5.2 or 10.4~t before structure; only the former is evaluated in the cited deep-space case. Harder 1989 spectra in the lunar study further distinguish acute from GCR protection \cite{Matthia2024}. The refuge needs independent air, water, cooling, and sanitation; access must beat the event rise time. Food or water shielding must be credited at minimum inventories. Spectrum, geometry, warning, and exceedance probability remain mission-specific.

\subsection{Residents, children, and longevity change the optimization}
An occupational limit for selected adults is not a validated developmental or lifelong population standard. If an engineering study allocates 10~mSv/year to the reference surface contribution, it requires $b_s\leq10/232\simeq0.0431$; over 50 years, that allocation alone integrates to 500~mSv in the stated RAD-normalized GCR design quantity, not an organ-weighted effective dose. This design allocation makes the lifetime consequence explicit and assigns substantial value to shielding that serves several generations.

Residual-risk assessment must include low-dose-rate mixed fields, heavy-ion effectiveness, cardiovascular and central-nervous-system outcomes, age, and altered gravity \cite{NASEMRadiation2021}. Material, geometry, scheduling, and transport can reduce exposure now; biological studies must bound the remaining lifetime and developmental risk. Mixed-field simulators enable mechanistic tests without reproducing lifelong exposure \cite{Simonsen2020}. Permanent residence therefore requires protected occupation and verified dosimetry rather than lifetime extrapolation of expeditionary exposure.

\subsection{Protection readiness and operational consequences}
\label{sec:protectionreadiness}
At $\Sigma=629$~g/cm$^2$, planar cover has surface mass $6.29\times10^3$~kg/m$^2$ and weight near 23.3~kPa on Mars. Arches, sidewalls, soil, and pressure containment must carry or redistribute this load. Orbit instead requires delivery or off-Earth sourcing, with centrifugal loading for a co-rotating shield [Eq.~\eqref{eq:combinedstress}]. Equal exposure allocations can therefore impose different construction requirements.

The 25~m$^2$/resident protected footprint comprises residential, clinical, circulation, and work space; the 50~m$^2$/resident crop area in Sec.~\ref{sec:food} is separate. Crops, seeds, electronics, and food quality require their own protection specification. Applying the human column to the occupied footprint gives 157~t/resident, or 157,000~t for 1,000 residents; sustained accepted emplacement at 100~t/day takes 4.3 years. Extending that column over both occupied and crop areas without footprint sharing gives 75~m$^2$/resident, 472,000~t, and about 13 years. The comparison isolates the importance of the protected footprint. Side support, access, finite-geometry transport, and actual fleet duty remain additional design inputs.

The 272-h/year exterior allowance is also operational: a stipulated 2,000-h annual outdoor workload exceeds it by a factor 7.4. Sheltered access, robotic work, redistributed duties, or a separately justified exposure allocation can resolve the conflict. Acceptance must combine installed-geometry dosimetry, minimum consumable inventories, breach/fire isolation, and a clinically justified exposure history.

\section{Gravity and physiological adequacy}
\label{sec:gravity}
\subsection{Mechanical loading, fluid regulation, and experimental evidence}
Gravity affects several physiological systems through different mechanisms. Musculoskeletal loading changes strain-dependent tissue remodeling; the hydrostatic gradient affects blood distribution and venous return; vestibular signals contribute to orientation and sensorimotor calibration. For a fluid column,
\begin{equation}
\Delta p_b=\rho_b g h.
\label{eq:bodyhydro}
\end{equation}
Using $\rho_b=1,060$~kg/m$^3$ and $h=0.35$~m, the hydrostatic contribution is 3.64~kPa (27.3~mmHg) at $1\,g_0$, 1.38~kPa at $0.38\,g_0$, and zero in sustained microgravity. This is a gravitational pressure difference, not a model of regulated intracranial pressure. In an 11-astronaut cohort, six developed stagnant or retrograde internal-jugular flow around flight day 50 and one had an occlusive thrombus \cite{MarshallGoebel2019}. The observation connects altered physical loading to a clinically relevant outcome beyond muscle loss.

Spaceflight-associated neuro-ocular syndrome (SANS) adds an ocular endpoint. In a prospective 11-astronaut cohort, mean eye axial length decreased by 0.08~mm after roughly six months in orbit; a 0.05-mm decrease persisted at one-year follow-up, alongside in-flight optic-nerve-head and choroidal changes \cite{Macias2020}. These structural observations require visual-function monitoring, not an inference that every astronaut loses useful vision. Their persistence also prevents muscle recovery from serving as a proxy for whole-organism recovery. Visual acuity, ocular imaging, operational vision, and fluid regulation must accompany gravity-countermeasure and arrival trials.

In a 17-astronaut longitudinal study, nine had not fully recovered tibial total volumetric bone mineral density one year after return \cite{Gabel2022}. A 60-day bed-rest study with 30~min/day of artificial gravity also left assessed capacities incompletely protected \cite{Kramer2021}. Adult recovery and intermittent centrifugation therefore constrain different questions from lifelong native-gravity residence.

Controlled flight studies now materially constrain the partial-gravity discussion. Lunar-level centrifugation prevented soleus mass loss under the tested conditions but did not prevent the slow-to-fast fiber-type shift \cite{Hayashi2023}. A separate mouse flight experiment compared microgravity with 0.33, 0.67, and 1~$g_0$. The 0.33-$g_0$ group showed partial protection, whereas 0.67~$g_0$ preserved the assessed muscle function and fiber-type composition \cite{Tsuji2026}. This is the most directly architecture-relevant recent observation: the lowest tested successful level for those endpoints exceeds Mars' 0.38~$g_0$.

Lunar loading is not equivalent to $1\gzero$ for skeletal endpoints. Distal-femoral trabecular bone volume fractions were approximately 30\% and 28\% below respective ground controls after 25 and 26 days at $1/6\gzero$ in Mouse Habitat Unit (MHU) missions 4 and 5, with partial protection relative to the separate microgravity experiment \cite{Okamura2024}. Differing missions, centrifuges, and postflight handling preclude a continuous human dose--response curve. The design must address the limiting function rather than select gravity from preservation of muscle mass alone.

Even with monotonic response, the discrete mouse data bracket an endpoint-specific transition between 0.33 and 0.67~$g_0$. They leave Martian loading to be tested directly. Muscle preservation alone cannot qualify skeletal, vestibular, cardiovascular, or developmental function, and nondetection of a difference is distinct from equivalence. Table~\ref{tab:bio} maps the evidence across these stages; Sec.~\ref{sec:reproduction} specifies the required life-cycle inference.

\begin{table}[tb]
\caption{Biological evidence map and architectural interpretation. The rows span adult physiology and major reproductive stages rather than claiming an exhaustive meta-analysis. Systematic reviews show that the human reproductive literature remains sparse and is dominated by analog or partial-function evidence \cite{Gimunova2024,Jain2023}. A successful endpoint does not establish organism-wide or multigenerational equivalence.}
\label{tab:bio}
\begin{tabular}{p{.19\textwidth}p{.36\textwidth}p{.385\textwidth}}
\toprule
Evidence domain & Result within the tested envelope & Decision-relevant limit or next test\\\midrule
Adult loading and recovery \cite{Gabel2022,Hayashi2023,Okamura2024,Tsuji2026} & Microgravity deconditions bone; lunar gravity provides endpoint-dependent partial protection; 0.67~$g_0$ was the lowest tested level preserving selected mouse muscle-function and fiber-type endpoints. & Adult musculoskeletal results do not identify a universal gravity threshold. Test 0.38~$g_0$ directly across organ systems, sex, duration, and clinically meaningful function.\\
Arrival function and ocular response \cite{Tays2021,Macias2020} & Postflight mobility/bimanual deficits and persistent ocular structural changes occur on different recovery time scales. & Qualify transfer, egress, visual tasks, and fault isolation after transit; neither elapsed time nor muscle mass establishes readiness.\\
Human reproductive-health evidence, systematic review \cite{Gimunova2024} & Sixteen eligible human/analog studies were identified; most addressed endocrine, menstrual, sperm, or related reproductive-health endpoints rather than pregnancy or development in flight. & Human reproductive physiology is incompletely sampled; no flight pregnancy, birth, childhood, or multigenerational equivalence dataset exists.\\
Gamete preservation \cite{Wakayama2021} & Freeze-dried mouse sperm stored on the ISS for up to five years and ten months produced offspring after terrestrial assisted fertilization. & Supports biobank concepts and storage resilience; does not qualify fertilization, gestation, or development in the space environment.\\
Preimplantation development \cite{Wakayama2023} & Two-cell mouse embryos developed to blastocysts during four-day ISS culture. & Implantation, placentation, gestation, birth, postnatal maturation, and subsequent fertility remain separate endpoints.\\
Gestation and postnatal development \cite{Wong1997,Jain2023} & Rat studies have flown portions of gestation and examined offspring after return; reviews identify pregnancy, birth, and developmental function as major unresolved domains. & Repeat under controlled gravity and measured radiation, with maternal--offspring interaction, birth, growth, and later function followed in the exposure environment.\\
Simulated microgravity and early reproductive processes \cite{Lyons2026} & Ground-based simulated microgravity changes selected sperm-navigation, fertilization, and embryo-development endpoints. & Clinostat exposure is not a uniform reduced-gravity field or space radiation environment; use it for mechanism and screening, not destination qualification.\\
Maternal flight and descendants \cite{Hong2026,Gifani2026} & Exposed F0 females retained postflight pregnancy capability, while F1/F2 reproductive and metabolic effects were reported. & Separate maternal physiology, radiation, unloading, launch/return stress, uterine environment, and inheritance; test the appropriate first unexposed generation.\\
\bottomrule
\end{tabular}
\end{table}

The architectural conclusion is to retain controllable loading rather than commit a permanent population to unqualified native gravity. Surface shielding can solve part of the radiation problem without correcting unloading; rotation can supply loading without proving developmental adequacy. Near-Earth loading is therefore an experimental and design option, not a certified biological countermeasure.

\subsection{Rotation geometry does not determine pressure-vessel feasibility}
For a person stationary in a rotating frame,
\begin{equation}
a=\omega^2r,\qquad a_C=2\omega v,\qquad \frac{\Delta a}{a}\simeq\frac{h}{r}.
\label{eq:rotation}
\end{equation}
Earth-like acceleration requires radii of 894, 224, and 55.9~m at 1, 2, and 4~rpm, respectively. At 2~rpm a 1-m/s transverse movement has a Coriolis acceleration magnitude up to $0.0427\,g_0$; a 1.8-m body experiences an approximately 0.8\% radial gravity gradient at 224~m. Tolerance is trainable but task-specific. A 12-day rotating-room study exposed four aviators to 10~rpm, establishing experimental experience beyond a universal 2-rpm rule \cite{Graybiel1965}. An incremental-acclimation experiment enrolled 11 adults in 25-min sessions; seven of the eight completing the protocol reached the imposed 25-rpm ceiling for the tested cross-coupled illusion \cite{Bretl2020}. Neither experiment establishes comfortable lifetime residence, transfer across all head-movement planes, or developmental safety. They support 2--4~rpm as a credible range for occupied-habitat experiments, not as an already qualified population envelope. The needed tests combine sustained rotation with walking, visual tasks, emergency work, sleep, docking transitions, and representative resident ages.

For a thin, rotating cylindrical pressure shell, the hoop stress contains pressure, shell self-rotation, and attached co-rotating payload terms:
\begin{equation}
\sigma_\theta=\frac{\Delta p\,r}{t}+\rho a r+\frac{\mu a r}{t},
\qquad
 t\geq\frac{r(\Delta p+\mu a)}{S_{\rm allow}-\rho a r}.
\label{eq:combinedstress}
\end{equation}
Here $t$ is shell thickness, $\rho$ material density, and $\mu$ attached payload mass per shell area; $S_{\rm allow}$ is an assumed allowable membrane stress incorporating whatever material safety allowance the designer has selected. End closures, joints, damage tolerance, fatigue, local bending, internal floors, and external-pressure cases require separate design. The expression assumes a uniform shell and distributed load, and is a lower-bound structural model rather than a fabrication prescription.

\begin{figure}[tb]
\centering\includegraphics[width=0.80\textwidth]{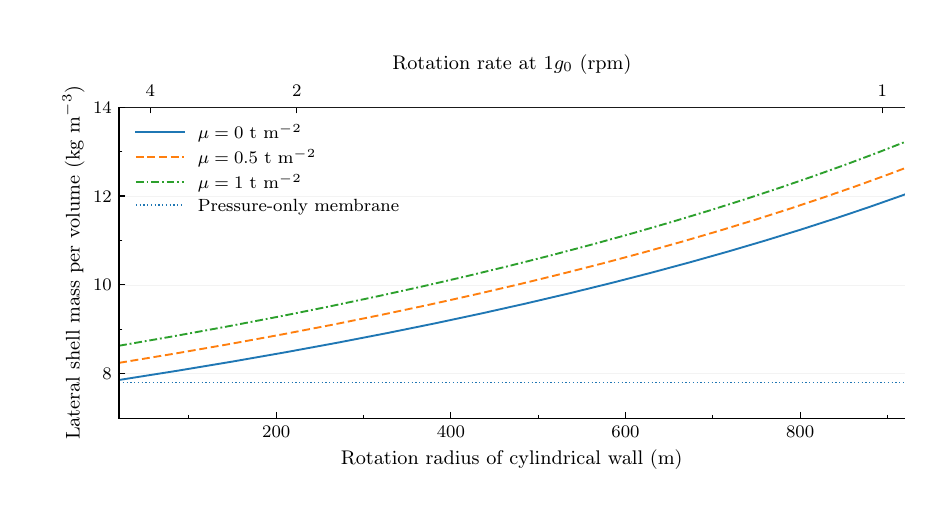}
\vskip -15pt
\caption{Pressure and rotation in the cylindrical membrane model of Eq.~\eqref{eq:combinedstress}. Shell mass per enclosed volume is $2\rho t/r$, for $1g_0$ at the wall, 100-kPa pressure, 7,800-kg/m$^3$ density, and 200-MPa allowable stress. Curves vary the attached co-rotating column $\mu$; the dotted line is the pressure-only bound. The radius range retains the 4-, 2-, and 1-rpm cases (55.9, 224, and 894~m). Ends, floors, joints, fatigue, and damage tolerance are excluded; the curves do not size a modular habitat with a separate radial load path.}
\label{fig:structure}
\end{figure}

For $r=224$~m, length $L=500$~m, $\Delta p=100$~kPa, $\rho=7,800$~kg/m$^3$, $a=g_0$, and $S_{\rm allow}=200$~MPa, shell self-rotation contributes only 17.1~MPa. Pressure alone requires $t=0.112$~m and $6.15\times10^5$~tonnes of cylindrical shell. Including self-rotation increases thickness to 0.1225~m and shell mass to $6.72\times10^5$~tonnes. Adding a co-rotating 200-kg/m$^2$ payload gives approximately $8.26\times10^5$~tonnes of shell plus that payload. Pressure, rather than shell self-rotation, controls this reference mass. Figure~\ref{fig:structure} shows the pressure--rotation coupling across rotation rates.

The mass-efficient route is to separate rotation geometry from pressure geometry. For the cylindrical pressure-only bound,
\begin{equation}
\frac{M_{\rm shell}}{V}=\frac{2\rho\Delta p}{S_{\rm allow}},
\label{eq:pressurevolume}
\end{equation}
which equals 7.8~kg/m$^3$ for the stated steel parameters, independent of radius. Merely subdividing a fixed enclosed volume does not remove pressure-vessel mass, and additional ends and joints can increase it. The advantage of a tethered or framed modular habitat is instead to \emph{decouple the rotation radius from the amount of pressurized volume}. The large cylinder above encloses $7.88\times10^7$~m$^3$; its volume divided by one circumferential floor area is $r/2=112$~m. Pressurizing only occupied modules avoids paying for a large unused central volume. Nonrotating external shielding also avoids the attached-shield rotational load, although it introduces relative-motion clearances, access, and attitude-control requirements.

For scale, two balanced 100-tonne end modules at a 224-m radius each impose approximately 0.98~MN on their radial support. At an assumed 500-MPa allowable tensile stress and 2,000-kg/m$^3$ support density, the two ideal 224-m radial members total about 1.76~tonnes before joints, redundancy, dynamics, and their own distributed loading. The support-only bound excludes complete vehicle hardware, but it establishes that the tensile members themselves do not force a city-scale experiment. Small occupied volumes can be placed at large rotation radii. Continuous rotation, docking, emergency despin, shield interfaces, and repair under rotation are the appropriate next demonstrations.

\subsection{Surface centrifuges are a third architectural option}
Artificial gravity is not restricted to free space. In a horizontally rotating surface facility, neglecting local gravity gradients, the magnitude of the effective downward acceleration is
\begin{equation}
 g_{\rm eff}=\sqrt{g_p^2+(\omega^2r)^2},\qquad
 \tan\beta=\frac{\omega^2r}{g_p},
 \label{eq:surfacegravity}
\end{equation}
where $\beta$ is the tilt of the effective vertical relative to planetary vertical. At two revolutions per minute, reaching $1\,g_0$ on Mars requires a radius of approximately 207~m and an effective-vertical tilt of approximately $68^\circ$; the corresponding lunar values are 221~m and $81^\circ$. These kinematic values require banked floors, maintained bearings and seals, stationary access, and support for pressure and protective cover. Surface rotation therefore adds a testable alternative to native-gravity residence and free-space rotation.

\subsection{Rotation energy and operational requirements}
\label{sec:rotationreadiness}
The energy needed to establish rotation is not generally the dominant power burden. Neglecting support inertia, two equal modules with total mass $M$ at radius $r$ have
\begin{equation}
 E_{\rm spin}=\tfrac12 M a r,\qquad
 L_{\rm spin}=M\sqrt{a r^3}.
 \label{eq:spinbudget}
\end{equation}
For the two 100-t modules at 224~m and $a=\gzero$, these are approximately 220~MJ (61~kWh) and $2.10\times10^9$~kg\,m$^2$/s. Ideal steady free-space rotation consumes no continuous mechanical work; bearings, flexure damping, control, drag, and docking disturbances introduce real losses. Conversely, angular momentum must be accommodated during docking, reorientation, or emergency despin. A small spin-energy inventory therefore neither proves inexpensive controls nor supplies the habitat's life-support electricity.

Modular prototypes can test deployment, spin-up, docking, fault isolation, and occupied operation while the developmental studies proceed. Surface centrifuges additionally need gravity-supported bearings or tracks, dust control, and stationary access. Section~\ref{sec:roadmap} consolidates the development priorities.

\section{Development, population continuity, and lifelong health}
\label{sec:reproduction}
\subsection{Population renewal requires completion of the life cycle}
Systematic reviews identify the gap as continuity across conception, gestation, birth, and subsequent development in a controlled environment \cite{Gimunova2024,Jain2023}. Table~\ref{tab:bio} spans those stages and distinguishes demonstrated endpoints from qualification requirements.

Gamete survival, fertilization, preimplantation development, implantation, placentation, gestation, delivery, childhood development, and adult fertility have different failure mechanisms. Mouse embryos cultured on the ISS from the two-cell stage reached blastocysts \cite{Wakayama2023}; freeze-dried mouse sperm stored on the ISS for up to five years and ten months produced offspring after terrestrial assisted fertilization \cite{Wakayama2021}. These results establish useful capabilities, including a potential role for reproductive biobanks. They do not demonstrate pregnancy or development under partial gravity. Simulated-microgravity experiments in 2026 identified effects on sperm navigation and early reproductive processes, but clinostat exposure is not identical to flight or to a spatially uniform reduced gravitational field \cite{Lyons2026}. Earlier rat spaceflight experiments also establish that portions of pregnancy can be investigated without establishing a complete extraterrestrial mammalian life cycle \cite{Wong1997}.

The 42-day maternal mouse flight study reported reproductive dysfunction in descendant generations: F1 females had lower anti-M\"ullerian hormone at postnatal day 21 and fewer litters and pups over nine months, while reduced ovarian reserve and other metabolic and reproductive disturbances persisted in F2 \cite{Hong2026}. The exposed females nevertheless became pregnant when bred five days after return, with a slight reported fecundity decline. Retained immediate pregnancy capability therefore does not establish preserved ovarian reserve or reproductive lifespan across generations. The commentary emphasizes this distinction \cite{Gifani2026}. Causal interpretation requires lineage-resolved replication separating maternal physiology, radiation, unloading, launch/return stress, and postflight breeding.

The distinction between intergenerational exposure and inheritance must be explicit. When a pregnant F0 female is exposed, the F1 fetus and its developing germ cells, which can produce F2, are also directly exposed. F3 is then the first generation without direct exposure under the usual maternal-line accounting. When a nonpregnant F0 adult and its gametes are exposed, F2 is generally the first generation beyond those directly exposed gametes. Descendant effects alone do not distinguish deoxyribonucleic acid (DNA) mutation, epigenetic transmission, altered uterine environment, or parental behavior. Embryo transfer, cross-fostering, reciprocal crosses, and matched environmental controls are needed to separate these mechanisms. A continuously exposed lineage and a returned, unexposed recovery lineage answer different questions and should both be retained.

\subsection{A quantitative design for informative experiments}
\label{sec:biodesign}
An efficient program is factorial rather than destination-specific. Initial gravity levels should include microgravity, 0.16, 0.33, 0.38, 0.67, and 1~$g_0$, with flight 1-$g_0$ and terrestrial controls; 0.33 connects to existing flight data, whereas 0.38 tests Mars directly. Exposure dosimetry must be measured, not inferred from orbital altitude. Both sexes, maternal and paternal lineages, prenatal and postnatal intervals, and complete reproductive cycles are required. Endpoints should include implantation loss, fetal growth, placental function, congenital abnormalities, bone architecture, sensorimotor development, ovarian reserve, sperm function, and offspring fertility. Human-cell and organoid experiments can clarify mechanisms, but cannot substitute for whole-organism development.

Sample size must be tied to the intended inference. For a two-group difference test with equal variance, two-sided $\alpha=0.05$, 80\% power, and a difference $\Delta=0.5\sigma$, the normal approximation gives
\begin{equation}
n_{\rm diff}\simeq\frac{2(z_{0.975}+z_{0.8})^2\sigma^2}{\Delta^2}\simeq63
\label{eq:biopower}
\end{equation}
independent observations per group. This is a \emph{difference-detection} design, not an equivalence criterion. Small flight cohorts can identify large endpoint-specific effects without qualifying small differences across the organism. Additional fibers measured within one mouse do not constitute independent animals.

For equivalence, specify the biologically acceptable two-sided margin $\Delta_E$ before data collection. Under independent normal means with \emph{known} common variance, equal groups, and true mean difference $\theta=0$, the two one-sided tests (TOST), each at $\alpha=0.05$, have power \cite{Schuirmann1987,Phillips1990}
\begin{equation}
\pi_E=2\Phi\!\Big(\frac{\Delta_E\sqrt{n}}{\sqrt2\sigma}-z_{0.95}\Big)-1,\qquad
n_E\geq\frac{2(z_{0.95}+z_{0.9})^2\sigma^2}{\Delta_E^2}
\label{eq:equivalence}
\end{equation}
when the acceptance interval is nonempty. A stipulated $\Delta_E=0.5\sigma$ and 80\% power require at least 69 independent units per group, not 63; the latter has approximately 75\% equivalence power in this idealized design. With true offset $\theta=0.2\sigma$, the corresponding normal calculation requires 138 per group. Estimated variance, multiplicity, attrition, and clustering require design-specific recalculation. The half-standard-deviation margin is an illustration, not a proposed developmental safety standard; a clinically important smaller margin raises the requirement approximately as $\Delta_E^{-2}$.

Litter clustering reduces independent information. For mean size $m$ and intralitter correlation $\rho_L$, the design effect is $1+(m-1)\rho_L$: at $m=6$, $\rho_L=0.3$, 150 pups provide about 60 independent-observation equivalents. Zero adverse outcomes in 300 independent births gives a one-sided 95\% upper bound $1-0.05^{1/300}\simeq1\%$ on incidence, not the difference from Earth. Experimental unit, lineage, margin, clustering, and attrition must therefore precede sample-size selection.

Repeatable orbital centrifugation should use prespecified endpoint and lineage protocols. Applying the illustrative 69-unit equivalence design independently to six gravity levels and two sexes gives 828 independent units before attrition or developmental clustering. Pilot studies should first estimate relevant effects and variances; confirmatory cohorts should then target the endpoints and loading ranges that discriminate architectures. Independent ethical review must address developmental risk as well as adult participation.

Facilities and independent cohorts can develop in parallel, but lineage observations require actual gestation, maturation, and follow-up. The cited nine-month reproductive follow-up illustrates that serial requirement \cite{Hong2026}. Rodent studies can inform the proposed 10--20-year campaign; human life-cycle qualification depends on its own evidence. Section~\ref{sec:roadmap} places those biological intervals within the development sequence.

\subsection{Reproductive replacement and demographic resilience}
\label{sec:demographicresilience}
With age-specific survival $l_x$ and female offspring production $m_x$,
\begin{equation}
 R_0=\sum_x l_xm_x,\qquad \bm n_{t+1}=\bm L_t\bm n_t.
 \label{eq:reproductive}
\end{equation}
For stationary vital rates, $R_0=1$ is mean reproductive replacement and the dominant eigenvalue of the Leslie operator is one \cite{Leslie1945}. It does not bound the chance of demographic failure in a finite population. Survival to maturity, voluntary fertility, age structure, and care must be assessed together. With half of births female, 95\% survival to reproductive age, and 90\% of surviving women reproducing, replacement requires 2.34 births per reproducing woman. These illustrative fractions specify a calculation, not a reproductive prescription. At 1,000 residents and the separate bookkeeping death rate $N/75$, stationary replacement is about 13 births/year; 1\% growth requires about 23. Obstetric, neonatal, and pediatric services are consequently recurring needs.

A bounded cohort perturbation shows what the mean criterion misses. Take a deterministic comparator with 1,000 people uniformly distributed across 75 annual age classes, survival to age 75, half female, and 0.1 births per woman-year at ages 20--39. This deliberately simplified population is initially stationary, with 13.33 births/year and $R_0=1$. A five-year 25\% birth reduction, followed by restoration of the original age-specific fertility, leaves five cohorts of 10 rather than 13.33. Total population initially falls by only 1.7\%, but entry to age 20 falls by 25\% twenty years later. In the annual cohort projection, the age-based ratio $(N_{<20}+N_{\geq65})/N_{20:64}$ changes from 0.667 initially to 0.683 in year 25, and the smaller parental cohorts reduce births to 12.5/year despite restored fertility. These are cohort arithmetic and an age-composition indicator, not estimated care hours or a population forecast. The calculation exposes delayed effects even without random mortality, disease, migration, or founder-age concentration.

Readiness therefore requires trajectories of reproductive-age cohorts and the services they support, not only a positive birth balance. For declared age-specific care demands $c_x$ and available qualified service hours $h_x$, a necessary aggregate check is $\sum_x h_x n_x\geq\sum_x c_x n_x$; actual simultaneous clinical and technical roles must also be covered (Sec.~\ref{sec:social}). Altered cohort size can reduce care demand now while reducing replacement of caregivers later. Nutrition, stocks, and infrastructure must use that same $N(t)$. The useful interventions are overlapping skills, protected care capacity, flexible automation, and retained exchange options; choosing a larger census or imposing births does not establish resilience. Probabilities of demographic loss require supported vital-rate and correlation models, while the deterministic perturbation already identifies margins to test.

\subsection{Genetics must follow the same population history as infrastructure}
\label{sec:population}
Genetic projections must use the same age structure, fertility, and survival history as Eq.~\eqref{eq:reproductive}; terrestrial vital rates are comparisons, not extraterrestrial developmental measurements \cite{Leslie1945}.

A neutral genetic screen uses effective, not census, population size. With unrelated founders in the idealized diploid model,
\begin{equation}
 F_G=1-\prod_{j=0}^{G-1}\Big(1-\frac{1}{2N_{e,j}}\Big),
 \qquad \frac{W_G}{W_0}=e^{-B F_G}.
 \label{eq:genetics}
\end{equation}
The first relation describes accumulation of inbreeding under the specified drift model \cite{Wright1931}; the second is a parameterized inbreeding-depression relation, not a universal human fitness law \cite{Morton1956,Kalinowski1998}. At constant $N_e=400$, 40 generations give $F_G=4.88\%$. Historical mortality estimates motivate an endpoint-specific slope sensitivity near $B=1.5$--2.5 with $B$ defined as the slope of log relative survival per unit inbreeding, rather than the diploid total $2B$; assigning $B=3$ is a stronger stress case, not a measured modern-population coefficient. At the stated $F_{40}=0.0488$, $W_{40}/W_0=0.885$--0.929 for $B=1.5$--2.5 (7.1--11.5\% lower modeled endpoint survival); $B=3$ gives 0.864 (13.6\% lower). These are relative changes in the selected endpoint, not all-cause mortality predictions. Founder relatedness, selection, migration, reproductive variance, and cryopreserved contributions require explicit treatment.

Growth changes Eq.~\eqref{eq:genetics} but also demands stocks, food, and equipment through Eq.~\eqref{eq:growth}. Published voyage estimates differ because their pedigrees, mating assumptions, and population histories differ \cite{Smith2014,Marin2018}; no universal founding number follows. Cryopreserved contributions can diversify ancestry, but require storage, clinical capability, gestation, and healthy development.

For a selected maximum inbreeding increment $F_{\max}$ over $G$ generations at constant effective size, Eq.~\eqref{eq:genetics} gives
\begin{equation}
 N_e\geq\frac{1}{2[1-(1-F_{\max})^{1/G}]}
 \simeq\frac{G}{-2\ln(1-F_{\max})}.
 \label{eq:geneticthreshold}
\end{equation}
At $G=40$, the illustrative allocations $F_{\max}=0.05$ and 0.01 imply approximately 390 and 1,990 effective breeders, respectively. Neither allocation is a human safety threshold. In the ideal unequal-sex model $N_e=4N_mN_f/(N_m+N_f)$; unequal family size, kinship and generational bottlenecks reduce the relation to census population further. Thus census headcounts of hundreds, thousands, or more cannot be defended independently of pedigree and reproductive history. A biobank can increase the number of genetically contributing founders without transporting each donor, but cannot replace gestation, healthy development, care, or the equipment that preserves and uses the bank. Maintaining reciprocal transfer capability between settlements can reduce genetic isolation while retaining operational independence from emergency resupply.

Natural evolution also cannot be treated as a planned countermeasure. Selection can change a population only through heritable differences in reproductive success; it does not remove the need for pressure, energy, molecular repair, and a compatible developmental environment. Somatic therapies and assistive technologies may reduce particular risks. Heritable redesign for low-gravity development or broad radiation resistance remains a different, unqualified intervention. Near-term feasibility should therefore be assessed for protected humans, with prospective therapies evaluated as specific additions rather than prerequisites.
\subsection{Clinical medicine, healthspan, and pharmaceutical renewal}
\label{sec:medicine}
A resident population needs obstetrics, pediatrics, infection control, dentistry, surgery, chronic care, and rehabilitation. For condition incidence $q$ per person-year, a homogeneous Poisson model gives
\begin{equation}
 P(\text{at least one event in }T)=1-e^{-qNT}.
 \label{eq:medical}
\end{equation}
The illustrative $q=10^{-3}$ and $N=1,000$ imply one expected event/year and probability 0.99995 of at least one in a decade. For condition-specific incidence $q_i$ and fatality probability $r_i$ after treatment, the residual fatal-event rate is $\lambda_{\rm med}=\sum_iq_ir_i$, retaining age and environment dependence. Allocating $10^{-4}$ fatal events/person-year to a condition with incidence $10^{-3}$/person-year requires $r_i<0.1$; other conditions consume part of the same allocation. Disability and healthy-life loss require analogous criteria. Age-specific incidence, diagnostic reliability, treatment capacity, supplies, and care teams must determine the local scope rather than extrapolation from selected astronaut cohorts \cite{NASA3001V2F,NASEMThriving2023}.

A functioning instrument need not provide a persistent medical service. A NASA-authored 2024 technology report describes the ISS portable clinical blood analyzer implementation as dependent on refrigerated cartridges with a six-month shelf life, regular software updates, and a validation access code \cite{Lewandowski2024}. This is a documented implementation constraint, not an immutable limit on diagnostic technology. The 52-month missed-delivery case therefore needs a validated extension or an alternative replenishment route, not merely more cartridges from the same expiring batch. With last-usable date $t_b^{\rm exp}$ and unused quantity $n_b(t)$ for qualified batch $b$, usable assay stock is
\[
 A_{\rm valid}(t)=\sum_b n_b(t)\,\mathbf{1}_{\{t\leq t_b^{\rm exp}\}},
\]
provided the specified storage history, calibration, and configuration remain valid. Assay demand must be met from $A_{\rm valid}$ rather than total physical stock. A six-month validity interval covers only 12\% of a 52-month interruption; sequential local batches, stability-qualified stocks, or a clinically adequate alternative must cover the remainder. Storage before arrival further consumes validity.

The complete clinical pathway includes sampling, accepted measurement, interpretation, and an available treatment; it must survive the same power and supply interruption. Its endurance is limited by the earliest exhausted or invalidated essential reagent, calibration capability, cold chain, device, trained service, or treatment input. A useful acceptance demonstration exercises this pathway with representative samples and blinded controls, interrupted external updates, a cold-chain fault, a replacement sensor or cartridge batch, and a validated fallback. Passing a device self-test or manufacturing a cartridge body is insufficient. This bounded example operationalizes the product-class autonomy test in Sec.~\ref{sec:independence}; it does not require a complete autonomous hospital before every adult expedition.

The 2023 ISS formulary had shelf-life records for 91 of 106 medicines; even the maximum labeled shelf life was at most 36 months for 54 of those 91 \cite{Diaz2024}. Expiry is a quality boundary rather than instantaneous loss of potency. Four oral formulations showed no radiation-accelerated degradation following nominal 0.5- and 1.0-Gy mixed-field exposures over follow-up to 34 months, with two replicates and incomplete final assays \cite{Plante2025}. Packaging, moisture, temperature, assays, replacement, and clinical substitution are thus actionable requirements; whole-formulary stability remains to be established.

After a three-day flight in four adults, 18 circulating cytokines/chemokines and cell-specific signatures changed \cite{Kim2024}; the Space Omics and Medical Atlas (SOMA) supports longer-term, cross-mission comparisons \cite{Overbey2024}. These identify mechanisms, not a single space-aging rate. The structured microbial and chemical distributions found in 803 samples across nine ISS modules likewise require joint surveillance of residents and process reservoirs \cite{Salido2025}. Fermentation and nutrient recovery must coexist with containment of infection, crop pathogens, biofilms, and resistance; clinical outcomes and restoration time matter more than diversity alone.

Longer life increases accumulated exposure and chronic-care requirements: at unchanged dose rate, 80 rather than 50 years adds 60\% to integrated exposure. Human aging biomarkers and rodent hypometabolism experiments do not establish rejuvenation or transport hibernation \cite{Waziry2023,Yang2023}. Present feasibility should use contemporary lifespans, preventive care, and rehabilitation; new therapies count only after their efficacy, harms, manufacture, and surveillance are established.

\subsection{Psychology, governance, and the continuity of essential services}
\label{sec:social}
Psychological health contributes directly to operational capacity. Mars-500 confined six selected men for 520 days in approximately 550~m$^3$, with heterogeneous sleep and behavioral responses \cite{Basner2013,Basner2014}. A randomized terrestrial study found cumulative cognitive impairment after 14 days at four or six hours of sleep opportunity, with inadequate self-assessment \cite{VanDongen2003}. Thus circadian lighting, protected sleep, acoustic control, privacy, and time outside assigned duties are safety provisions, not discretionary amenities. The analogs do not estimate lifelong psychiatric incidence in a diverse settlement. Relevant outcomes are clinical symptoms, sleep, errors, conflict, retention, and recovery, measured without making disclosure threaten access to care or essential supplies.

Distance changes authority as well as communications. The approximately 3--22-minute one-way Earth--Mars light time rules out Earth approval for minute-scale emergencies \cite{MarsRelay}. Local authority must be able to isolate faults while preserving independent technical review, records, and appeal for nonurgent decisions. Institutional arrangements affect hazard reporting, deferred maintenance, allocation, and whether a disagreement interrupts life support. Social--ecological research provides an empirical basis for monitoring and collective-choice arrangements \cite{Ostrom2009}; applying it here is a design inference, not a measured extraterrestrial effect size. Organizational delay enters Eq.~\eqref{eq:recovery} directly: a five-day avoidable delay in restarting the failed 4-t/day water processor consumes 20~t of clean reserve and adds 20~t of retained wastewater. Governance can therefore become limiting even when nominal hardware capacity is adequate.

A settlement also changes the relationship between residents and infrastructure owners. When one organization controls air, water, employment, transport, and health records, technical dependence can create coercive power. We identify protected minimum services, independent clinical and safety oversight, resident participation, and access to dispute resolution as ethical operating conditions; the analysis does not select a constitution. Voluntary adult risk acceptance does not establish consent for children or justify compulsory reproduction. Demographic models must accommodate voluntary fertility, disability, aging, family formation, education, and care rather than treat reproductive targets as instructions to residents.

Permanent departure adds separation from family, familiar environments, and feasible exit. Emergency return is practical only where the transport and clinical timelines permit it; elsewhere, informed consent requires a credible account of its absence. More residents can enlarge social choice but simultaneously increase housing, clinical, food, and renewal demands. No measured population size guarantees cohesion. Financing and governance must preserve critical service and replacement capacity through operator failure or funding interruption, with liabilities and succession arrangements specified. These are potentially binding conditions of long residence, not evidence that social failure is inevitable. Their tests are service continuity, protected agency, health outcomes, and response to disruption; economic and resilience benefits are compared separately in Sec.~\ref{sec:purpose}.

One continuously staffed critical position requires $8,760/h_{\rm eff}$ people when $h_{\rm eff}$ includes leave, training, and absences. At the selected 1,800 effective hours/year, this is 4.87 full-time equivalents. Simultaneous roles require separate capacity; on-call coverage requires its own response test. Automation counts after supervision and recovery demands are included. These task-level requirements belong to the failure cases in Sec.~\ref{sec:reliability}, rather than defining a universal founding population.

\section{Chemistry and life support: material cycles and service recovery}
\label{sec:lifesupport}
\subsection{Reference accounting and the role of buffers}
The material accounting uses the common $N=1,000$ reference. Adult reference flows of approximately 0.84~kg O$_2$, 1~kg CO$_2$, and 0.62~kg dry food per person per day give 307~tonnes of oxygen consumed, approximately 365~tonnes of CO$_2$ to process, and 226~tonnes of dry food per year \cite{Jones2017,BVAD2022}. Body size, activity, diet, age structure, and pregnancy change these values. Regeneration determines the fraction of these throughputs requiring new supply.

For a fully collected personal-water stream of 4~kg/person/day, recovery fraction $c$, and annual processing duration $T_y$,
\begin{equation}
M_{\rm make}=Nq_w(1-c)T_y.
\label{eq:water}
\end{equation}
At $c=0.98$, this is 29.2~tonnes/year. The ISS has demonstrated approximately 98\% recovery of collected water using its brine-processing architecture \cite{Boyce2023,NASAWater2023}; that result concerns a specified water stream and operational configuration, not whole-habitat material or industrial closure. If a fraction $1-A$ of processing time instead bypasses recovery and is discarded, $c$ is replaced by $Ac$. At $A=0.99$, the make-up requirement rises to 43.5~tonnes/year.

An isolated outage of duration $\tau$ in a constant stream $Q$ requires a wastewater buffer $B_w\geq Q\tau$ and an independently adequate clean-water reserve. With operating capacity $k_wQ$, accumulated backlog clears in
\begin{equation}
t_{\rm catch}=\frac{\tau}{k_w-1},\qquad k_w>1.
\label{eq:catchup}
\end{equation}
For $Q=4$~t/day and $\tau=10$ days, the backlog is 40~tonnes; $k_w=1.25$ needs 40 additional operating days. Restart therefore restores processing, not necessarily the pre-outage buffer margin.

\begin{figure}[tb]
\centering\includegraphics[width=0.80\textwidth]{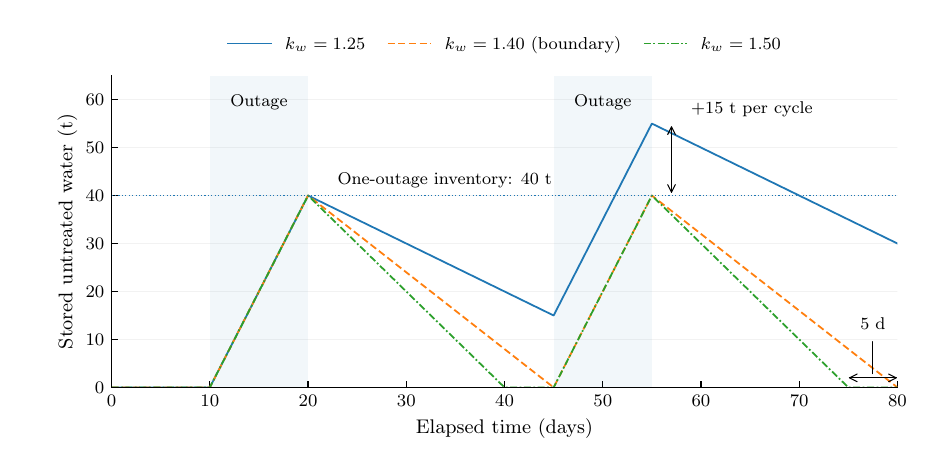}
\vskip -15pt
\caption{Recovery capacity determines whether a finite buffer remains adequate. The $Q=4$-t/day stream produces 40~t during one ten-day outage (horizontal reference). Repetition every 35 days adds 15~t per cycle at $k_w=1.25$, giving a 55-t second peak. At $k_w=1.40$ the buffer clears exactly at recurrence; $k_w=1.50$ leaves five days of recovery margin. Shading marks outages; the next starts at day 80. Equation~\eqref{eq:periodic} governs backlog, not clean-water purity, reserve availability, or recovery power.}
\label{fig:water}
\end{figure}

For identical outages with start-to-start interval $\Delta>\tau$, let $B_n$ be stored untreated water immediately before outage $n$. Assuming constant inflow, retained wastewater, instantaneous restart, and work-conserving processing at $k_wQ$, the cycle balance gives
\begin{equation}
 \begin{gathered}
 B_{n+1}=\left[B_n+Q\{\Delta-k_w(\Delta-\tau)\}\right]_+,\\
 k_w(\Delta-\tau)\geq\Delta
 \quad\Longleftrightarrow\quad
 k_w\geq\frac{\Delta}{\Delta-\tau}.
 \end{gathered}
 \label{eq:periodic}
\end{equation}
where $[x]_+=\max(0,x)$ and the inequality is the condition for bounded periodic backlog. Equality prevents drift but cannot remove inherited backlog; strict inequality provides catch-up margin. The peak inventory in cycle $n$ is $B_n+Q\tau$. Below the bound, no finite tank prevents eventual overflow under indefinite recurrence. For $\Delta=35$ days the boundary is $k_w=1.4$: the 1.25-capacity case gains 15~tonnes per cycle, while $k_w=1.5$ clears an initially empty-buffer cycle five days before the next outage (Fig.~\ref{fig:water}). These deterministic bounds apply to the stated recurrent schedule. Variable events require event-resolved inventory accounting; clean-water quality and reserve restoration remain separate requirements. Catch-up power and cooling must also remain available within Sec.~\ref{sec:powertransport}; the crop-loop example in Sec.~\ref{sec:food} shows how the failed function changes both the reserve and response time.

Gas storage does not remove the need for continuous removal of exhaled CO$_2$. In a sealed 100,000-m$^3$ atmosphere at 293~K, neglecting biological uptake and other processing, the reference 1,000 residents add approximately 0.554~kPa of CO$_2$ partial pressure per day and consume approximately 0.640~kPa/day of O$_2$. These follow directly from $\dot p_i=RT\dot n_i/V$. An illustrative CO$_2$ control band from 0.04 to 0.40~kPa is exhausted in about 16~h, whereas a 3-kPa oxygen partial-pressure decrement takes about 4.7 days. These are inventory horizons, not emergency medical thresholds; local mixing and the actual atmospheric standard must be included. CO$_2$ removal, oxygen storage, and crop uptake therefore require separate failure responses.

Oxygen and carbon loops also cannot be reduced to the presence of an electrolyzer. Sabatier reduction, CO$_2+4$H$_2\rightarrow$CH$_4+2$H$_2$O, coupled to water electrolysis recovers oxygen but commits hydrogen to methane if methane is discarded. Pyrolysis or other carbon-processing steps improve closure only if their own power, consumables, fouling, and maintenance are included \cite{Greenwood2015}. Plants couple oxygen production to carbon assimilation and food; mismatched crop and respiratory loads can require storage or separate gas processing. Integrated modeling has demonstrated the significance of precisely these couplings \cite{Do2016}.

A regenerative habitat is a controlled reactor network: elements must return in usable forms, and inaccessible precipitates or toxic metabolites are losses in Eq.~\eqref{eq:stocks}. A tractable engineering route is hybrid physicochemical and biological processing, with backup gas control independent of crop success. Its limiting metrics are usable yield, purity, catalyst life, and recovery rate, not the largest attainable recycling percentage \cite{BVAD2022,Do2016,Greenwood2015}.

The periodic bound is deterministic. For variable interruptions, the required untreated-water capacity follows the largest cumulative net inflow over a relevant interval,
\begin{equation}
 B_{\rm req}\geq\sup_{t_1<t_2}
 \Big[\int_{t_1}^{t_2}\{Q(t)-q_{\rm proc}(t)\}\,\dd t\Big]_+,
 \label{eq:bufferenvelope}
\end{equation}
where $q_{\rm proc}$ is available work-conserving treatment capacity and initial inherited backlog is added separately. Long-run average capacity greater than inflow is necessary but does not specify the storage needed for clustered faults. The event ensemble, bounded interruption set, or explicit exceedance allocation must therefore be declared. Chemical recovery is a second condition: at 98\% recovery, processing the reference 40-t backlog returns 39.2~t of water and leaves 0.8~t to be replaced or recovered through another path. An empty wastewater tank alone does not demonstrate restoration of the clean-water reserve. Power, thermal rejection, and purification must remain available during catch-up.

\subsection{Elemental recovery and chemical purity are independent constraints}
\label{sec:purity}
A high material-recovery fraction can coexist with contaminant accumulation. For a well-mixed water inventory $V_w$, contaminant concentration $C_X$, source rate $\dot m_X$, selective treatment flow $Q_X$, single-pass removal efficiency $\eta_X$, and first-order destruction rate $k_X$,
\begin{equation}
V_w\frac{\dd C_X}{\dd t}=\dot m_X-(\eta_XQ_X+k_XV_w)C_X,
\quad
C_{X,\infty}=\frac{\dot m_X}{\eta_XQ_X+k_XV_w}.
\label{eq:purity}
\end{equation}
Without removal or destruction, concentration grows linearly even when water recovery is perfect. In a specified 10-m$^3$ inventory, a source of 1~mg/day reaches a design allocation of 10~$\mu$g/L in 100 days from an initially clean state. Holding that allocation by separation requires $\eta_XQ_X\geq0.10$~m$^3$/day. The separated material must then be regenerated, destroyed, or stored outside the useful loop; moving it onto a finite sorbent does not eliminate its mass. This requirement applies separately to trace organics, metals, salts, and process-specific reaction products \cite{BVAD2022}.

The same distinction applies to nutrients. A specified 1-g/person/day phosphorus stream circulates 365~kg/year per 1,000 residents; 99\% recovery still requires 3.65~kg/year of replacement before other reservoirs. Availability in usable chemical form and contaminant assays, rather than recovery percentage alone, determine nutritional closure. Hydroponics, segregated industrial water, selective ion recovery, and polishing stages provide routes that must survive repeated cycles and contamination.

\subsection{Local propellant couples transport to the chemical plant}
Local oxygen and fuel production can reduce transport mass but cannot be treated as free use of an atmospheric resource. A combined water-electrolysis/Sabatier scheme has net reaction
\begin{equation}
\mathrm{CO_2(g)+2H_2O(l)\longrightarrow CH_4(g)+2O_2(g)}.
\label{eq:propellantchem}
\end{equation}
At 298.15~K and standard pressure, formation enthalpies and absolute entropies from the Joint Army--Navy--Air Force thermochemical tables of the National Institute of Standards and Technology (NIST--JANAF) give $\Delta H^\circ\simeq890$~kJ/mol and $\Delta G^\circ\simeq818$~kJ/mol \cite{Chase1998}. For the stoichiometric methane/oxygen products this is approximately 2.84~kWh/kg of reversible non-expansion work, before capture, purification, compression, liquefaction, heat-transfer losses, and conversion inefficiency. Fifty tonnes of that mixture require approximately 22.5~t of net water, 27.5~t of CO$_2$, and at least 142~MWh of reversible work under those boundary conditions. The oxidizer/fuel mass ratio of four is stoichiometric, not an optimized engine mixture. Excess oxygen or a different engine mixture changes the product accounting.

These are thermodynamic lower bounds, not demonstrated plant performance. Catalysts, electrolyzer membranes, dust exclusion, water purity, cryogenic storage, and restart determine actual installed capacity. Oxygen made for ascent is exported from the biological cycle; breathing demand, storage growth, and propellant production therefore require separate terms in Eq.~\eqref{eq:stocks}. A reserve-certified return-propellant plant and tank inventory must be verified before crew commitment. The Mars Oxygen In-Situ Resource Utilization Experiment (MOXIE) establishes atmospheric oxygen processing on Mars, not this complete fuel-production and storage chain \cite{Hoffman2022}.

\subsection{Nitrogen inventory, leakage, and nutritional closure}
\label{sec:nitrogen}
Atmospheric nitrogen is not consumed by breathing, but it is lost through leakage, airlock operations, venting, and maintenance. Nitrogen in food must be returned in biologically usable form; recovering N$_2$ is not equivalent to recovering nitrate or ammonium. For residential volume $V$, nitrogen partial pressure $p_N$, and temperature $T$,
\begin{align}
M_N&=\frac{p_N V\mathcal M_N}{RT},\label{eq:ninventory}\\
\dot M_N&=\lambda_\ell M_N+(1-\eta_a)\rho_NV_an_a
+(1-c_N)\dot M_{N,\rm bio},
\label{eq:nloss}
\end{align}
where $\lambda_\ell$ is the fractional inventory leak rate, $V_a$ the airlock volume, $n_a$ cycles per unit time, $\eta_a$ gas-recovery efficiency, and $c_N$ recovery of the biological nitrogen stream. All terms refer to nitrogen mass. Gas fractions lost during decompression must be measured; leakage per person is not an intrinsic physical constant.

For $V=10^5$~m$^3$, $p_N=79$~kPa, and $T=293$~K, the residential N$_2$ inventory is 90.8~tonnes. This volume excludes separately pressurized cultivation and industrial space. A sensitivity range $\lambda_\ell=10^{-5}$--$10^{-3}$~day$^{-1}$ produces 0.332--33.2~tonnes/year of leakage. It is a design range, not an extrapolation of ISS performance. Historical ISS measurements reported approximately 0.064--0.227~kg/day of total atmospheric loss in particular operating intervals \cite{Schaezler2011}; scaling those numbers to a settlement without specifying pressure, seal design, temperature cycles, and total penetrations is unjustified.

At the reference $\lambda_\ell=10^{-4}$~day$^{-1}$, leakage is 3.32~tonnes/year. Fifty daily cycles of a 10-m$^3$ airlock with 99\% recovery add 1.66~tonnes/year. A deliberately specified 100-g/person/day protein stream, with nitrogen estimated as protein/6.25, circulates 5.84~tonnes N/year; 99\% recovery leaves a further 0.058~tonnes/year. Total reference make-up is therefore 5.03~tonnes/year. Reducing airlock recovery to 95\% increases that contribution to 8.29~tonnes/year. Gas handling, not human metabolism, dominates this particular nitrogen budget.

Sample Analysis at Mars (SAM) measured $0.0259\pm0.0006$ annual-mean N$_2$ volume fraction at Gale crater \cite{Trainer2019}, equivalent to 0.249~g/m$^3$ at 600~Pa and 210~K. At 80\% capture, the 5.03-t/year make-up case needs about 0.80~m$^3$/s intake: an ordinary gas-processing rate, with dust control, purity, compression, and lifetime still to qualify. Containment sets plant scale. The Moon lacks that atmospheric route. Nitrogen-bearing volatiles including ammonia were detected in polar ejecta \cite{Colaprete2010}, but occurrence is not a production-certified reserve; imported buffer gas remains a credible initial choice.

Lower habitat pressure reduces pressure-hull and inert-gas inventories, approximately linearly at fixed volume, but requires explicit oxygen-partial-pressure, fire, decompression, and human-performance qualification. Argon or another diluent changes gas sourcing but does not substitute for biological fixed nitrogen. These coupled choices should be made at the atmosphere--structure--resource-system level rather than by optimizing gas mass alone \cite{NASA3001,BVAD2022}.

\subsection{Food, crop-water recovery, and shared-utility failures}
\label{sec:food}
Intensive cultivation gives a scale near 50~m$^2$/person, dependent on crops, light, and diet \cite{Wheeler2017,Poulet2022}. The 1,000-resident reference has 5~ha; its 30--100-m$^2$/person sensitivity spans 3--10~ha. Multiple tiers reduce footprint but not cultivated area. For lighting density $p_L$ and illuminated fraction $d_L$,
\begin{equation}
 \overline P_L=A_c p_Ld_L.
 \label{eq:lighting}
\end{equation}
The selected 200~W/m$^2$ for 16~h/day gives 6.67~MW mean lighting; the stated area and 150--300-W/m$^2$ sensitivities give 3--20~MW. Supplying 2,500~kcal/person/day corresponds to 121~W of edible energy, or 1.82\% of reference electrical lighting. Edible yield, plant respiration, inedible biomass, losses, and all essential nutrients must be demonstrated at that operating point.

An 86-day wheat-chamber study measured mean evapotranspiration near 4.5~L/m$^2$/day \cite{WheelerSager1990}. Applying that crop-specific flux to 5~ha gives 225~t/day of water circulation and 6.35~MW of latent transfer at 2.44~MJ/kg. This is internal redistribution of heat and water, not net consumption or an extra source of heat to add to lighting. A sustained 1\% loss from this selected stream would nevertheless require 821~t/year of make-up, demonstrating the leverage of effective return. A mixed-diet farm needs its own measured fluxes.

\emph{A shared-utility disturbance changes the required reserve.} Divide the reference cultivated area into ten independently isolatable sectors. One 5,000-m$^2$ sector has an illustrative constant transpiration rate $J=22.5$~t/day. Suppose its condensate-return pump fails for $\tau=1$ day while dehumidification, cooling, irrigation, and lighting remain powered. With no overflow or evaporation loss, the clean irrigation reserve falls and retained condensate rises by the same 22.5~t; the water is transferred, not consumed twice. Restart at return capacity $k_rJ$ gives
\begin{equation}
 \Delta S_{\rm irr}=J\tau,\qquad B_{\rm cond}\geq J\tau,
 \qquad t_{\rm restore}=\frac{\tau}{k_r-1}.
 \label{eq:farmreturn}
\end{equation}
Thus $k_r=1.5$ requires two further days to restore the initial reserve. The operating condenser still transfers 0.635~MW of latent heat and needs 0.212~MW of compressor work at the selected coefficient of performance of 3; lighting is 1.0~MW when on. These loads are allocated from, not added to, the main reference. Return pumping, acceptance of stored condensate, and any extra restart load must fit their actual hardware and utility allocations. Even this single-sector one-day fault requires more reserve than a five-day interruption of the 4-t/day personal-water stream.

Loss of condensation is a different fault. With constant air temperature, no other vapor sink, and initially unchanged transpiration, the time to a selected relative-humidity limit is
\begin{equation}
 t_{\rm RH}=\frac{V_f\rho_{v,\rm sat}(T)(\varphi_{\max}-\varphi_0)}{J}.
 \label{eq:humiditybuffer}
\end{equation}
For a stipulated 25,000-m$^3$ sector atmosphere at 25~$^\circ$C, saturation density is approximately 0.023~kg/m$^3$ from the ideal-gas relation and the tabulated vapor pressure \cite{NISTWaterVapor}. Raising relative humidity from 60\% to the selected 80\% control boundary stores only 115~kg, reached in approximately 7.4~min at the stated initial flux. This is an early-time control calculation, not a crop-damage deadline; surfaces, temperature, and stomatal response alter subsequent evolution. Actual trials must identify those responses and automatic standby or load-shedding performance. Closing a valve or switching off lamps cannot be credited with instantaneous cessation of transpiration.

If containment instead requires curtailment and loss of the sector, assume for reserve accounting that it supplies 10\% of the reference dry-food output. A 120-day recovery then requires 7.4~t of replacement dry food, with nutrient, seed, culture, and growth-medium reserves separately satisfied. Crew gas control must remain available independently of photosynthesis: the gross metabolic reference is 35~kg/h O$_2$ and about 42~kg/h CO$_2$ processing. These are full-backup throughput targets, not additional consumption while plants are functioning. Independence requires segregated air/water branches and verified surviving power and cooling, not merely separate crop rooms sharing one indispensable plant. The 40-t personal-water buffer in Fig.~\ref{fig:water} therefore cannot be interpreted as a whole-farm outage reserve.

Regenerative integration has experimental precedent: Lunar Palace 365 reported 98.2\% material closure and complete recovery of oxygen and human-use water over 370 days under its terrestrial accounting \cite{Fu2025}. Biosphere~2 exposed a previously missed oxygen sink involving microbes and building chemistry \cite{Severinghaus1994}. Direct sunlight and microbial protein can reduce electrical or area demands, but require full nutrient, thermal, and contamination balances \cite{Leger2021}. Repeated mixed-diet operation through defined faults and restart, rather than a harvest or closure percentage alone, should qualify additional occupancy.

\subsection{Perchlorates and the qualification of local agricultural inputs}
Phoenix measurements found approximately 0.4--0.6~wt\% perchlorate in sampled Martian soil \cite{Hecht2009}. Processing 1,000~tonnes of comparable material therefore involves 4--6~tonnes of perchlorate, although this composition cannot be assigned uniformly to all Martian regolith or airborne dust. Regolith cultivation is not equivalent to terrestrial soil agriculture. Experiments demonstrate crop-specific inhibition and incomplete restoration by simple washing under tested conditions \cite{Oze2021}. Hydroponic or other controlled substrates can avoid much of the initial regolith-to-food pathway. Where local material is used, extraction or chemical/biological destruction must be demonstrated with a complete water, salt, energy, and waste balance; clean irrigation water alone does not establish safe edible tissue.

A recent simulant experiment provides quantitative remediation alternatives: in magnesium-perchlorate-spiked Johnson Space Center (JSC) Mars-1 simulant, treatment at 470~$^\circ$C or three 1:5 solid--liquid leaching steps followed by distillation nearly eliminated the tested perchlorate inventory \cite{Coker2026}. The result is process- and salt-specific, not certification of all Martian substrates. For an assumed heat capacity of 0.8~kJ/(kg\,K), heating material from 20 to 470~$^\circ$C requires 100~kWh/t of sensible heat before reaction enthalpy, losses, or heat recovery. Three washes correspond to 15~tonnes of circulating wash water per tonne of material, not necessarily that much net water consumption. These balances favor reuse of process heat and water, or initially avoiding untreated regolith in the food loop. In the same experiment, microbial perchlorate reduction improved through three selection cycles but became negligible in the fourth and fifth \cite{Coker2026}. Biological remediation must therefore be qualified across serial regeneration, rather than by the performance of its most effective cycle.

\subsection{Dust microphysics links exposure, power, and equipment renewal}
\label{sec:dust}
Dust links local construction to indoor exposure and equipment retirement through grain formation, contact mechanics, charging, and transport \cite{Turyshev2026}. Grain size, roughness, temperature, and disturbance determine different loads on seals, optics, radiators, and filters. Consequently, Eq.~\eqref{eq:capital} requires measured component degradation rather than a universal dust penalty.

The disparity between adhesion and weight explains why low surface gravity does not make attached fines easy to remove. For a sphere--surface contact, the Johnson--Kendall--Roberts and Derjaguin--Muller--Toporov limits give pull-off force $F_a=\xi\pi W_sR_*$, with $\xi=3/2$ or 2, work of adhesion $W_s$, and effective asperity curvature $R_*$ \cite{Johnson1971,Derjaguin1975}. Relative to a grain's weight,
\begin{equation}
\frac{F_a}{m_d g}=\frac{3\xi W_sR_*}{4\rho_d a^3g}.
\label{eq:dustadh}
\end{equation}
For the explicitly specified contact $a=1~\mu$m, $R_*=10$~nm, $W_s=0.05$~J/m$^2$, $\rho_d=3,100$~kg/m$^3$, and lunar $g=1.62$~m/s$^2$, pull-off force is 2.4--3.1~nN whereas weight is $2.1\times10^{-14}$~N, a ratio of $(1.1$--$1.5)\times10^5$. Contact geometry and surface state must be measured to apply this scaling to actual grains. Mechanical separation, traveling electric fields, coatings, and dust-tolerant mechanisms address different terms in the detachment problem; NASA's electrodynamic dust shield has demonstrated removal on lunar hardware \cite{NASAEDS2025}. Flight demonstration establishes capability under those conditions, while long-service qualification requires cleaning efficiency, wear, power consumption, and repeat-cycle performance.

NASA-STD-3001 Volume~2 Revision~F specifies 0.3~mg/m$^3$ for lunar particles smaller than 10~$\mu$m during intermittent daily exposure up to six months, and 0.1~mg/m$^3$ for corresponding Martian dust exposure up to 30 days; the Martian rationale uses a 24-h time-weighted average \cite{NASA3001V2F}. These adopted exposure limits have defined durations rather than lifetime or developmental validation. The Martian assessment uses a working transport example of 100~g per crewmember per extravehicular activity (EVA), approximately 1\% respirable \cite{MarsDust2026}. In a perfectly mixed 100-m$^3$ volume, a 1-g respirable bolus produces 10~mg/m$^3$. With no new input and effective removal $k=6$~h$^{-1}$,
\begin{equation}
C(t)=C_0e^{-kt},\qquad t_{100:1}=\frac{\ln100}{6}=0.768\ \mathrm{h}.
\label{eq:dust}
\end{equation}
For sustained operations, the more relevant requirement includes the continuing source:
\begin{equation}
\frac{\dd C}{\dd t}=\frac{\dot M_{\rm in}}{V_h}-kC,
\qquad \dot M_{\rm in}\leq kV_h C_*.
\label{eq:dustsource}
\end{equation}
At the selected $C_*=0.1$~mg/m$^3$, $V_h=1,000$~m$^3$, and $k=6$~h$^{-1}$, steady ingress must remain below 0.60~g/h. Fifty entries per day would consume that entire allocation at 0.288~g of respirable dust per entry. Spatial plumes, resuspension, and time-weighted exposure must be assessed separately. Suit isolation, source capture, and segregated clean/dirty interfaces reduce this source before it reaches the shared air loop.

A specified 20\% loss of collector throughput requires 25\% more area at fixed output; reducing radiator emissivity from 0.9 to 0.7 requires 29\% more emitting area at fixed temperature and negligible incoming radiation. Cleaning and degradation therefore enter installed power and cooling. Representative dust tests must combine temperature cycling, gas or vacuum, contact loads, and repeat operations, measuring ingress, filter pressure drop, leakage, optical/thermal performance, and replacement rates.

Chemical qualification is product-specific: resource grade, retained contamination, selectivity, consumables, and the shared utilities must remain acceptable through duty and recovery. Table~\ref{tab:isru} identifies the resource routes; Secs.~\ref{sec:food} and~\ref{sec:medicine} show why nominal output alone cannot establish food or clinical service.

\section{Energy, infrastructure, and scalable renewal}
\label{sec:powertransport}
\subsection{One electrical budget for services and manufacturing}
\label{sec:services}
The principal engineering case uses 1,000 residents, 50~m$^2$ of crop area per resident, and the illumination and wheat-flux reference in Sec.~\ref{sec:lifesupport}. These quantities establish a transparent accounting point, not a complete nutritionally qualified farm. Its 6.67-MW average lighting load is supplemented by 1.5~MW allocated to gas and water processing, 0.5~MW to crop and habitat circulation, 1.0~MW to food processing, domestic, clinical, and information services, and 2.12~MW of dehumidifier work. The last quantity follows from 6.35~MW of latent transfer and a selected coefficient of performance of 3. The nonmanufacturing total is therefore $p_h=11.8$~kW per resident, obtained from these declared components. The gas-processing and auxiliary allocations are engineering assumptions informed by system-accounting practice, not demonstrated large-habitat measurements \cite{BVAD2022,Hanford2006}. Holding these 3.0-kW/resident auxiliary services, the transpiration per cultivated area, and the coefficient of performance fixed, the stated 30--100~m$^2$/resident and 150--300-W/m$^2$ lighting sensitivities give 7.3--27.2~kW per resident. Appendix~\ref{app:repro} gives the endpoint calculation; this area-based flux scaling is a scenario, not a crop-response model. Nutritional yield, purity, and installed hardware must be satisfied jointly.

Let $k_j$ be tonnes of qualified capital per resident and $q_{K,j}=(\delta_j+\gamma)k_j$ the required commissioned replacement and expansion in tonnes per resident-year. Write $\bar Q_j$ for annual accepted equipment output made at the destination, $e_{K,j}$ for its kWh/kg intensity, and $\bar p_{\rm other}$ for per-resident site loads outside the declared service and product-processing boundaries. Then
\begin{equation}
 \bar p_{\rm site}=p_h+\frac{10^3}{Nh_y}\sum_j e_{K,j}\bar Q_j+\bar p_{\rm other},
 \qquad N\bar p_{\rm site}\leq\bar P_{\rm avail},\quad Na_c\leq A_c.
 \label{eq:capacityscreen}
\end{equation}
Here $h_y=8,760$~h/year and powers are in kW. For steady reserve inventories and local coverage $f_j$, $\bar Q_j=f_jNq_{K,j}$; replenishing depleted spares or building production facilities adds the corresponding outputs. Yield $y_j$ relates one-pass gross intensity to accepted output through $e_{K,j}=e_{g,j}/y_j$. Include local extraction, refining, fabrication, rejected work, and acceptance in that coefficient only where those operations actually occur. Earth production of imported equipment is excluded from site electricity but included in whole-program cost and energy. Imported feed still needs the local processing it undergoes. Recovery, construction, idle plant, and storage enter $\bar p_{\rm other}$ only if not already included elsewhere; repair and material return are likewise counted once. This annual-energy condition does not establish dispatch or a complete industrial network \cite{Ho2022,Owens2018}.

Table~\ref{tab:capital} assigns the 30-t inventory to six classes rather than imposing one lifetime on all hardware. The central class times reflect the design intention that retained pressure structure outlasts active processing equipment; their intervals test that intention without claiming flight qualification. For a fixed representative age mix and maintenance policy,
\begin{equation}
 q_{\rm ret}=\sum_j\frac{k_j}{L_j},\qquad
 \delta_{\rm eff}=\frac{q_{\rm ret}}{\sum_jk_j},\qquad
 q_K=q_{\rm ret}+\gamma\sum_jk_j.
 \label{eq:renewalmix}
\end{equation}
The selected inventory gives $q_{\rm ret}=1.51$~t/(resident\,yr), $\delta_{\rm eff}=0.0503$~yr$^{-1}$, and $q_K=1.81$~t/(resident\,yr) at $\gamma=0.01$~yr$^{-1}$. Pairing the longest or shortest listed times gives $\delta_{\rm eff}=0.0310$--0.0998~yr$^{-1}$ and $q_K=1.23$--3.29; these are sensitivity endpoints, not confidence bounds. The industrial class contributes 50\% of reference mass and approximately 50\% of stationary replacement. The aggregate rate is consequently sensitive to industrial retention, not just the life of pressure walls. Net-retention times require age-resolved component records and a repair policy; component failures and retained-parent replacement cannot be counted as the same loss \cite{Owens2018}.
The industrial row deliberately aggregates heterogeneous production
equipment. Its terrestrial service-life reference anchors only the
order of magnitude of the declared retention interval; the
settlement value must ultimately be replaced by age-resolved
records for the actual machine, vehicle, tooling, control, and
metrology classes.

At $y=0.8$ and $e_g=30$~kWh/kg, local manufacture of all 1,810~t/year requires 7.75~MW, giving 19.5~MW with the service reference. Half local manufacture gives 15.7~MW and 905~t/year of finished-equipment imports; importing the full stream leaves 11.8~MW and 1,810~t/year. The comparison holds population, installed equipment, farm, and gas/water services fixed. Imported feed, idle machinery, and excluded site processes remain in their respective accounts. Full-local manufacture gives 17.1--25.9~MW over the retention endpoints and 14.4--37.6~MW over the independent $e_g=10$--100-kWh/kg sweep. Uniform $\delta=0.08$ supplies a separate stress case: 2.7~t/(resident\,yr) and 23.3~MW. As a terrestrial process reference, 848~TWh for 63.7~Mt of primary aluminium in 2019 gives 13.3~kWh/kg \cite{Mohammad2021}; smelting excludes upstream and downstream operations \cite{IAISmelting2024}. The habitat coefficient requires its own material, recycling, finishing, and acceptance mix, weighted by replacement flow. Whole-economy material stocks have different boundaries and cannot establish a lower bound for this electrical demand.

\begin{table}[tb]
\caption{Capital and net-renewal assumptions. $k_j$ is selected hardware in tonnes per resident; $L_j=1/\delta_j$ is a declared net capacity-retention time, with sensitivity interval in parentheses. These are engineering scenarios, not measured lifetimes or mean times between failures. The envelope is a planning allocation; local bulk shielding and external transport assets are separate.}
\label{tab:capital}
\begin{tabular}{p{.205\textwidth}p{.095\textwidth}p{.035\textwidth}p{.125\textwidth}p{.435\textwidth}}
\toprule
Capital class & Envelope (t/resident) & $k_j$ & $L_j$ (yr) & Boundary and interpretation\\\midrule
Cultivation equipment & 1--5 & 3 & 20 (10--30) & 50~m$^2$/resident at 20--100~kg/m$^2$; generation excluded. Fixtures outlive some lamps and pumps.\\
Pressure modules and framing & 1.5--5 & 4 & 50 (25--100) & Pressure-only lateral membrane is 0.78~t at 100~m$^3$/resident; ends, joints, and services add mass.\\
Power hardware, excluding thermal systems & 0.6--5 & 3 & 20 (10--30) & 30--50~kW/resident installed at 20--100~kg/kW; 1.5--2.6 times reference mean, not proven contingency capacity.\\
Thermal systems & 1.1--3.9 & 2 & 25 (15--50) & Mars-B emitting area; 5--10~kg/m$^2$, 25-kg/kW warm loop, 0.1--0.3-t supports; upper end at 50~kW/resident.\\
Life support, chemistry, and medicine & 1--5 & 3 & 10 (5--20) & Processing, monitoring, treatment, and sterile capacity. Consumable flows are charged separately.\\
Industry, vehicles, tooling, metrology & 4--25 & 15 & 20 (10--30) & 
Process-breadth allowance. The cited {U.S.} asset assumptions give metalworking service lives of 12--28~yr by industry \cite{Giandrea2022AssetLives}; they set a terrestrial scale, not extraterrestrial net-retirement measurements. Actual classes require class-specific replacement records.
\\\midrule
Sum; mass sensitivity & 9.2--48.9; 10--50 & 30 & --- & Correlated subsystem requirements prevent independently selecting all lower endpoints.\\
\bottomrule
\end{tabular}
\end{table}

Power-plant expansion itself creates replacement demand. The following homogeneous feedback limit assumes that both the nonpower equipment and the power equipment are manufactured locally. When components remain imported, their Earth manufacturing cannot be fed back into site electricity. Write $\bar p=\bar p_{\rm site}$ in this limit. For nonpower capital $k_o$ in tonnes/resident, power hardware specific mass $\alpha_P$ in kg/kW, installed-to-mean ratio $r_P$, and common replacement-plus-growth coefficient $d=\delta+\gamma$, substituting $k=k_o+\alpha_Pr_P\bar p/10^3$ gives
\begin{equation}
 \bar p=\frac{p_h+10^3 e_K d k_o/h_y}{1-\chi_P},
 \qquad \chi_P=\frac{e_Kd\alpha_Pr_P}{h_y}<1.
 \label{eq:energyfeedback}
\end{equation}
The inequality is a necessary condition for this homogeneous feedback model. Distinct equipment classes require distinct intensities and lifetimes. Shielding and thermal hardware must be placed in the corresponding capital classes, not counted twice. Improved yield, durable equipment, and heat integration can reduce the burden: raising $y$ from 0.80 to 0.95 reduces qualified-product electricity by 15.8\% at unchanged gross intensity; reducing $\delta$ from 0.08 to 0.04 at fixed $k=30$ and $\gamma=0.01$ reduces renewal from 2.7 to 1.5~tonnes per resident-year. The technology used to obtain those improvements has its own mass and service requirements.

\subsection{Joint sensitivities and the dispatch boundary}
\label{sec:jointsensitivity}
\begin{figure}[tb]
\centering
\includegraphics[width=0.80\textwidth]{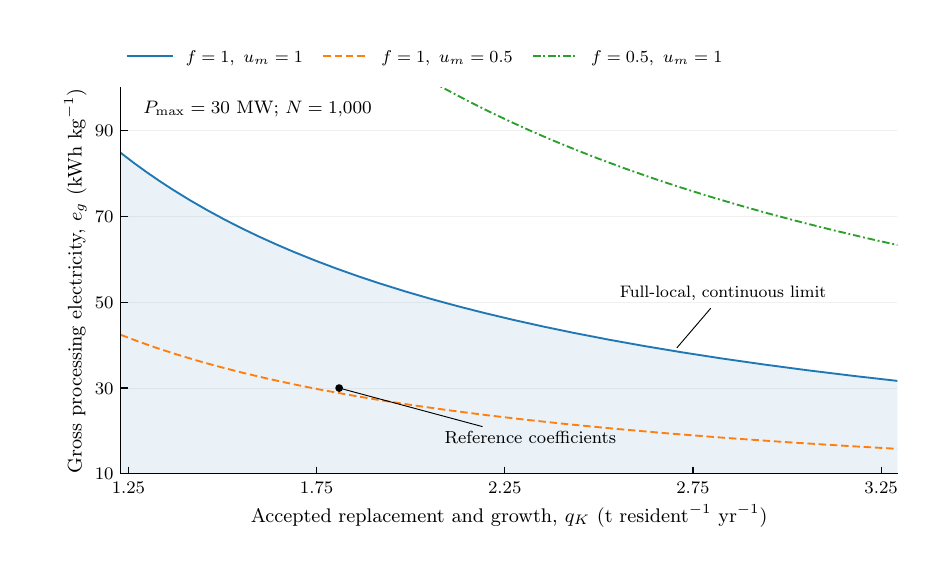}
\caption{Joint renewal, process-energy, and operating-duty screen for 1,000 residents. The horizontal domain spans the declared class-retention endpoints; the fixed service and yield assumptions are those of Sec.~\ref{sec:services}. Curves bound gross processing intensity at 30~MW available during crop illumination. The shaded region meets only this instantaneous-power condition for full local, continuous manufacture; reduced manufacturing duty contracts it, while continued imports relax it. The point marks the reference coefficients. No sampling probabilities, success fractions, reserve adequacy, or unmodeled plant resizing are assigned.}
\label{fig:jointpower}
\end{figure}

The retention and energy sensitivities must also be combined. With the same six class masses, 1\% growth, $y=0.8$, and fixed habitation services, their joint corners give 13.5--58.8~MW for full local manufacture, compared with 17--26~MW when only retention varies. This is a deterministic domain, not a probability interval. Figure~\ref{fig:jointpower} projects the class-retention choices onto their required accepted flow $q_K$; within the domain each point still requires the corresponding productive equipment and heat rejection.

Let $f$ be a uniform local-production fraction and $u_m$ the operating fraction of a manufacturing stream scheduled to overlap the illuminated period. Its machines must then process at $1/u_m$ times the calendar-average rate. For synchronized crop lighting, the fixed nonmanufacturing peak is $p_{h,\rm on}=15.12$~kW/resident. Against a specified available site power $P_{\max}$, Eq.~\eqref{eq:capacityscreen} gives the necessary boundary
\begin{equation}
 e_g\leq \frac{y h_y u_m}{10^3 f q_K}
       \left(\frac{P_{\max}}{N}-p_{h,\rm on}\right),
 \qquad f>0,
 \label{eq:jointpower}
\end{equation}
with $P_{\max}$ in kW. At $P_{\max}=30$~MW, $N=1,000$, $f=1$, and $q_K=1.81$~t/(resident\,yr), the limit is approximately 58~kWh/kg for continuous manufacture and 29~kWh/kg for half-time manufacture. Thus the 30-kWh/kg reference fits the former schedule but exceeds the latter's concurrent-power allowance. This screen holds installed stocks fixed; it rejects power-incompatible combinations without automatically enlarging generation or machinery. Additional storage, larger equipment, and their renewal require a revised stock and dispatch calculation. Catch-up after failures needs margin below the boundary. Population health, complete nutrition, and product-class availability remain separate gates.

\subsection{Generation-free intervals and dispatch}
\label{sec:dispatch}
Solar collection scales as $A_{\rm PV}=P/(S_0\eta_{\rm PV}d^{-2})$, before illumination, dust, incidence-angle, storage, and distribution losses; $S_0\simeq1,361$~W/m$^2$ \cite{KoppLean2011}. A 10-MW load at 30\% conversion requires approximately 24,500~m$^2$ at one astronomical unit (AU) under normal uninterrupted illumination. A selected 50\% net illumination/delivery factor doubles that area. Jupiter and Saturn receive approximately 1/27 and 1/90 of the 1-AU flux, so reliable energy density becomes increasingly consequential with distance \cite{JPLPlanets}.

For continuous load $P$ and generation-free interval $\tau$, delivered storage must cover $P\tau$. A 10-MW load through 336~h requires 3,360~MWh. At 0.25~kWh/kg delivered battery specific energy, the cells alone weigh 13,440~tonnes. Hydrogen--oxygen reactants at 50\% fuel-cell efficiency require approximately 1,800~tonnes of reactants for the same energy, before tanks, conversion hardware, thermal control, and reserves. Electrolysis and storage losses raise charging power. These inventories motivate location-specific illumination, load reduction, distributed generation, and fission; they do not establish a preferred installed storage system.

Fission removes dependence on local sunlight but retains conversion, cooling, fuel, controls, and supportability requirements. The Kilopower Reactor Using Stirling Technology (KRUSTY) experiment demonstrated a kilowatt-class nuclear system with load-following and transient tests, not a megawatt-electric extraterrestrial grid \cite{Poston2020}. Scaling must be assessed at the complete plant level. For either source, time-dependent power must satisfy
\begin{equation}
 P_G+P_d=P_L+P_c,\qquad
 \dot E=\eta_cP_c-P_d/\eta_d,\qquad 0\leq E\leq E_{\max}.
 \label{eq:dispatch}
\end{equation}
Charge/discharge rates, restoration, and the declared outage state constrain the solution; a repeated operating cycle also requires $E(T)=E(0)$. Average energy alone cannot establish those conditions.

The full-local-manufacture reference draws 22.9~MW while synchronized crop lighting is on, despite its 19.5-MW annual mean. Three equally sized banks with 16-hour photoperiods starting eight hours apart flatten lighting to 6.67~MW, provided the biological and auxiliary loads permit this scheduling. Otherwise a lossless constant-generation scheme needs a 53.3-MWh delivered energy swing. At 90\% round-trip efficiency, constant generation rises to approximately 19.8~MW and the optimized delivered swing is 49.7~MWh. The smaller swing results from increased generation. These nominal schedules do not include emergency reserves. Confining the full manufacturing stream to eight dark hours raises its active demand to about 23.2~MW and total site demand to about 28.4~MW, assuming the 5.12-MW nonlighting service load continues. The machinery must also provide three times the continuous-equivalent processing rate; annual-energy savings do not follow from this rescheduling.

A solar baseline is competitive where the measured illumination and end-of-life array output meet the load with acceptable storage, maintenance, and replacement. Fission becomes a candidate where the corresponding darkness or attenuation burden dominates, provided its full plant is qualified. A hybrid assigns only demonstrably curtailable demand to solar-dependent production and maintains essential services through the specified interruption; its reactor, storage, and restart capacity must satisfy Eq.~\eqref{eq:dispatch} together. The preferred mix minimizes maintained system burden at matched service and outage tolerance, not nominal generator mass.

\subsection{Close the heat balance at the same operating point}
\label{sec:thermal}
We parameterize thermal conversion as $\eta_e=\phi(1-T_c/T_h)$, with $T_h=1,100$~K, $T_c=650$~K, and $\phi=0.50$. This half-Carnot assumption gives $\eta_e=0.205$; it is not demonstrated space-plant performance. Brayton or Stirling conversion requires component efficiencies, temperature profiles, pressure losses, materials, and mass to be specified together \cite{Barrett2005,NASEMNuclear2021}.

For the full-local-manufacture reference at 19.5~MW delivered, reactor input is approximately 95.5~MW thermal and conversion heat approximately 76.0~MW. Ultimately dissipated electrical work adds approximately 19.5~MW at the lower-temperature loop. Crop latent heat is internal transfer and must not be added again; compressor work is already counted. The balance assumes no net chemical-energy export or sustained accumulation. Exported enthalpy, sunlight admitted directly, and stored chemical energy would enter explicitly.

For a gray radiator,
\begin{equation}
 \frac{Q}{A_{\rm emit}}=\epsilon\Big[\sigma T_r^4-F_g\sigma T_g^4-F_sL_s\Big]-\alpha_sG+h(T_r-T_a).
 \label{eq:radiator}
\end{equation}
Ground/sky view factors sum to one, $L_s$ is incident sky infrared, $G$ solar input, and $h$ the convective coefficient. Cold space sets incoming terms to zero; a Martian surface does not \cite{Martinez2021}. Table~\ref{tab:heat} uses shaded radiators, $F_g=F_s=0.5$, $\epsilon=0.9$, and no convective credit. The Mars cases are explicit boundary scenarios, not climatological quantiles. Emitting surface is not panel planform: two sides help only if both see the assumed environment.

\begin{table}[tb]
\caption{Full-local-manufacture energy reference for 1,000 residents. All columns use the class composition and retention times in Table~\ref{tab:capital}, 1\% growth, 80\% qualified yield, and 30-kWh/kg gross processing. The equipment replacement-and-growth stream is processed at the destination. All dissipated electrical work is assigned to 300~K in this bounding temperature allocation. Values are rounded from unrounded calculations. Mars boundary temperatures and sky fluxes are assumptions, not site measurements. The areas are emitting surfaces.}
\label{tab:heat}
\begin{tabular}{p{.39\textwidth}rrr}
\toprule
Quantity & Cold space & Mars A & Mars B\\\midrule
Ground temperature (K) & --- & 210 & 270\\
Sky infrared (W/m$^2$) & 0 & 40 & 100\\
Nonmanufacturing electricity (MW) & 11.8 & 11.8 & 11.8 \\
Renewal electricity (MW)          &  7.75 &  7.75 &  7.75 \\
Total electricity / warm-loop heat (MW)
                                  & 19.5 & 19.5 & 19.5 \\               
Reactor thermal input (MW) & 95.5 & 95.5 & 95.5\\
Conversion heat (MW) & 76.0 & 76.0 & 76.0\\
300-K emitting area ($10^3$ m$^2$) & 47.3 & 56.5 & 83.9\\
650-K emitting area ($10^3$ m$^2$) & 8.34 & 8.40 & 8.51\\
\bottomrule
\end{tabular}
\end{table}

At the principal Mars-B operating point, total emitting area is approximately 92.4~m$^2$/resident. Panels at 5--10~kg/m$^2$ weigh 0.46--0.92~t/resident; a 25-kg/kW internal warm-loop allowance contributes approximately 0.49~t/resident \cite{Hanford2006}. Hot-loop supports, degradation, and installed reserve raise the allocation. A 30--50-kW/resident nameplate range is not automatically an independent-failure reserve: the cooling, distribution, controls, and surviving generator train must provide that capacity in the stated contingency. Higher throughput consequently increases generation, heat rejection, and equipment renewal together.

Assigning every electrical dissipation to 300~K maximizes area only within the specified fixed-load comparison with rejection temperatures at or above 300~K. It is not a universal radiator bound. If fraction $\chi_T$ of manufacturing dissipation can be intercepted at process temperature $T_m$, let $f(T)$ be the net radiative flux in Eq.~\eqref{eq:radiator}. Then
\[
 A=\frac{P_h+(1-\chi_T)P_{\rm mfg}}{f(300)}
   +\frac{\chi_TP_{\rm mfg}}{f(T_m)}
   +\frac{Q_{\rm conversion}}{f(650)},
\]
where $P_h$ is nonmanufacturing service power, $P_{\rm mfg}$ is manufacturing power, and all heat rates are in watts. At $\chi_T=0.75$, $T_m=450$~K, and the same Mars-B boundary, the three emitting areas are approximately 59,000, 3,040, and 8,510~m$^2$: 70,500~m$^2$ in total rather than 92,400~m$^2$, a 24\% reduction at unchanged total rejection. The fraction, heat-exchanger approach temperatures, material compatibility, and pumps are additional design requirements. No cooling work is omitted or manufacturing heat assumed available above its actual temperature.

\subsection{Contingency generation and fuel renewal}

Mean energy is not a contingency standard. 

For the full-local-manufacture 22.9-MW synchronized peak, four independently available 10-MW generators would leave 30~MW after one generator loss, 7.1~MW above that load before degradation and catch-up work. With two unavailable, 20~MW could supply the reference nonmanufacturing peak of 15.1~MW by suspending manufacturing, but lost renewal would accumulate and later require catch-up capacity. Independence of cooling, switchgear, controls, and fuel delivery is necessary for this redundancy argument. The example is a capacity test, not a demonstrated four-reactor architecture; installed thermal hardware must support its actual dispatch and reserves.

At approximately 200~MeV per fission, fully fissioned heavy nuclei release $8.2\times10^{13}$~J/kg; the 95.5-MW thermal reference fissions about 37~kg/year. Actual loaded fuel is larger and depends on criticality, enrichment, and burnup \cite{Poston2020}; this is not a core inventory. Long-lived imported cores can support decades of residence but not indefinite fuel independence. Solar supply instead incurs collection, storage, degradation, and module renewal. The comparison is the maintained site system, not fuel versus panel mass alone.

\subsection{The scale gap to a maintained settlement grid}
\label{sec:powerreadiness}
NASA's earlier fission surface power (FSP) concept specified 40~kW, less than 6~t, and ten-year operation; its later solicitation specified at least 100~kW with closed-Brayton conversion. The January~2026 NASA--Department of Energy (DOE) agreement targets lunar-reactor development by 2030 \cite{NASAFSP2024,NASAFSP2025,NASAFSP2030}. These program requirements precede demonstrated surface operation. The 19.5-MW reference is approximately 490 or 195 times those outputs. Its qualification requires system-scale protection, distribution, cooling, fuel, maintenance, and reserve capacity, rather than simple replication of prototypes.

Higher rejection temperature reduces area only where the available process temperature, materials, and conversion or heat-pump work permit it. The low-temperature biological load still requires its own sink. Surface-power and thermal-loop demonstrations precede the maintained industrial grid; Sec.~\ref{sec:roadmap} gives the corresponding program horizons.

Ten days without reference replacement-and-growth production defer 49.6~t of accepted output. Clearing the deficit in the next ten days requires twice mean throughput, with simultaneous machine, feed, work, power, and cooling capacity. The grid gate is joint load following and reserve restoration after a specified generator, distribution, or cooling failure, including fuel and hardware renewal.

\section{Reliability, operational autonomy, and maintenance}
\label{sec:reliability}
Suppose a stationary catastrophic-event process has hazard $\lambda_c$. A century-survival target $R_{100}=0.99$, introduced here as a design example rather than a societal risk standard, requires
\begin{equation}
R(T)=e^{-\lambda_cT},\qquad
\lambda_c\leq-\frac{\ln(0.99)}{100}\simeq1.0\times10^{-4}\ \mathrm{yr^{-1}}.
\label{eq:hazard}
\end{equation}
If no catastrophic events occur over $E$ independent, exchangeable habitat-years, the one-sided 95\% Poisson upper bound is $-\ln(0.05)/E$, the exposure-time analogue of the rule of three \cite{Hanley1983}. Consequently,
\begin{equation}
E\geq\frac{-\ln(0.05)}{-\ln(0.99)/100}
\simeq3.0\times10^4\ \text{habitat-years}.
\label{eq:rulethree}
\end{equation}
A successful decade-long prototype cannot by itself statistically establish a century-scale safety claim. Pooling operating time from nonexchangeable designs or correlated facilities does not provide the stated bound. Nor does the target encompass every individual health outcome: it concerns catastrophic system failure only.

Nuclear probabilistic risk assessment (PRA), including NUREG-1150 and risk-informed licensing, combines event/fault trees, failure mechanisms, common causes, uncertainty, and defense in depth \cite{NUREG1150,NRC174}. Specified U.S. Nuclear Regulatory Commission (NRC) decisions use core-damage and large-early-release guidelines of order $10^{-4}$ and $10^{-5}$ per reactor-year; these are not observed frequencies or habitat standards. Aviation guidance similarly combines qualitative assurance with catastrophic-condition probabilities of order $10^{-9}$ per flight-hour \cite{FAA1309}, rather than demonstrating them by a billion failure-free hours.

For a habitat, the transferable evidence structure is an initiating-event model linked to tested physical barriers, subsystem failure and repair data, human-response times, and conditional consequences after buffer depletion. Each top-level hazard needs an auditable decomposition, including common power, cooling, software, contamination, and staffing dependencies. Accelerated-life tests constrain mechanisms only where acceleration preserves them; fault injection tests detection and recovery; full-system operation updates rather than replaces these models. Independent review, configuration control, sensitivity to poorly known parameters, and defense in depth prevent a precisely calculated fault tree from being mistaken for demonstrated safety. For two repairable, independently failing units, each with failure rate $\lambda$ and mean repair time $\tau$, the rare-overlap approximation gives a loss-of-function rate
\begin{equation}
\lambda_{\rm overlap}\simeq2\lambda^2\tau,
\label{eq:redundancy}
\end{equation}
before common-cause failures. At $\lambda=0.1$~yr$^{-1}$, reducing repair time from seven days to one day reduces this rate from $3.84\times10^{-4}$ to $5.48\times10^{-5}$~yr$^{-1}$. For the specified component parameters, seven-day repair misses the illustrative hazard target even before common-cause failures, whereas one-day repair leaves margin for other contributors. Whether a service interruption becomes catastrophic depends on survival-buffer endurance. Maintainability and physical containment therefore have direct quantitative value. Identical redundant units exposed to the same contamination or software defect cannot be assigned the independent-failure benefit.

The event model should distinguish loss of a redundant unit, loss of service, and harm after reserves are exhausted. If the conditional restoration time is $T_R$ and usable buffer endurance is $T_B$, a simple initiating-event approximation is
\begin{equation}
 \lambda_{\rm harm}\simeq\sum_a\lambda_a\,
 \Pr\Big\{T_{R,a}>T_{B,a}\mid a\Big\},
 \label{eq:bufferhazard}
\end{equation}
for disjoint initiating-event classes $a$, with overlaps and common causes modeled separately. It is the tail of restoration time, not merely its mean, that matters. Common contamination can simultaneously shorten buffer endurance and lengthen repair, so their joint distribution must not be factorized without evidence. For an initiating rate of 0.1/year, a selected $10^{-4}$/year harm allocation would allow at most $10^{-3}$ conditional failure-to-restore probability for that initiating class alone. This is an inverse requirement, not a measured failure rate. Near-term fault injection and component qualification can test the mechanisms and update a risk model; no 10--20-year prototype history by itself certifies an autonomous population's century-scale safety.

\subsection{Quantifying increasing self-sufficiency}
\label{sec:independence}
Three quantities distinguish self-sufficiency: product coverage $f_j$ [Eq.~\eqref{eq:productsupply}], complete Earth deliveries per resident-year, and verified interruption endurance. Imports include feed, reagents, fuel, and production machinery without double-counting parents and components. Local construction mass measures none of these. Full product coverage can coexist with imported critical inputs, as the commissioning pilot demonstrates; independence must be resolved by essential class.

With essential imports set to zero, throughput $q_i$ per resident, local make-up $P_i$, and per-resident inventory $s_i$ give the recovery requirement from Eq.~\eqref{eq:growth}
\begin{equation}
 c_i\geq1-\frac{P_i/N-\gamma s_i}{q_i}.
 \label{eq:closurethreshold}
\end{equation}
A right-hand side above one signals inadequate local production even at perfect recycling. No universal 100\%-closure requirement follows: local feedstock can replace losses. For the 4-t/day water stream, 98\% recovery loses 29.2~t/year; 99.9\% loses 1.46~t/year. A plant supplying 20~t/year at zero inventory growth requires at least 98.63\% effective recovery, including collection. These efficiencies do not establish purity or recovery after faults. With no make-up, a 200-t usable reserve lasts 6.85 years at 98\%, or 137 years at 99.9\%; even the latter is finite-stock endurance, not an indefinitely closed cycle.

For food output $Y_j$ with nutrient concentration $c_{nj}$, complete nutrition requires $\sum_j c_{nj}Y_j\geq Nd_n$ for every nutrient at the specified intake and bioavailability. The same mix must satisfy light, water, element, and heat balances. Microbial protein need not replace carbohydrates, lipids, vitamins, or dietary diversity \cite{Leger2021,Wheeler2017}. A 1\% deficit in the dry-food reference is 2.26~t/year per 1,000 residents; a complete 120-day interruption requires 74~t stored before losses. Nominal full production therefore does not remove food, seed, culture, and nutrient reserves.

Manufacturing independence is a product-class condition. Let $\bm d$ denote annual essential outputs, including equipment replacement, and let $A_{ij}$ be qualified output of class $i$ consumed in producing one unit of class $j$. A nonnegative production vector must satisfy
\begin{equation}
 (\bm I-\bm A)\bm x\geq\bm d,\qquad
 \rho(\bm A)<1,\qquad
 \bm e^T\bm x\leq E_{\rm available},\qquad
 \bm x\leq\bm x_{\max}.
 \label{eq:industrialclosure}
\end{equation}
The spectral condition characterizes productive networks under the stated linear, fixed-coefficient model; it is not evidence that the necessary processes exist. Raw feedstocks, human or robotic service, cooling, metrology, quality control, and growth impose additional constraints. Imported items set components of $\bm d$ that local production does not yet cover. A 99\% mass-localized factory can remain completely dependent on its final 1\% of electronics, seals, catalysts, or pharmaceutical inputs \cite{Metzger2013,Owens2018}. Selective standardization, modular controllers, local repair, and long-lived imported precision components are credible intermediate steps, but an explicit finite endurance replaces the term ``self-sustaining'' until those dependencies can be renewed.

Spares can be sized rather than assumed. For $n$ independent installed units, failure rate $\lambda_u$, and replacement lead time $T_L$, Poisson demand has mean $\mu=n\lambda_uT_L$. A selected stockout probability $\epsilon$ requires
\begin{equation}
 \Pr\{\operatorname{Poisson}(\mu)\leq s\}\geq1-\epsilon.
 \label{eq:sparestock}
\end{equation}
For 20 units with a stated 30-year mean time to failure and a five-year unsupported horizon, $\mu=3.33$ and eight spares exceed 99\% coverage for that failure class. If qualified local repair reduces the replenishment lead time to 0.1 year, the same model needs one spare. The comparison excludes common-cause defects, aging, repair failure, and demand from other systems. Its engineering lesson is robust: shortening verified repair lead time can reduce inventory much more than modestly improving average reliability. Repair processes and their own spares must appear in Eq.~\eqref{eq:industrialclosure}, not outside the boundary \cite{Owens2016,Owens2018}.

The result is a graded, testable independence claim: report essential product coverage, usable-stock horizons, clinical and repair scope, and outside assistance. Time without deliveries establishes only that interval unless replacement and reserve restoration are also demonstrated.

The product-network calculation is a screening model for a specified essential network. Its spectral-radius condition is useful only with a nonnegative, consistently defined input matrix, a declared product specification, and separate utility and service limits. It does not establish the availability of a semiconductor process or pharmaceutical line. Near-term independence should therefore be reported as increasing, audited interruption endurance and coverage of bounded functions. Several decades of development may expand that coverage, but the final low-mass dependencies can govern the entire system. An acceptance test must count access to Earth expertise, software validation, calibration artifacts, and hidden consumable stores as support where they are indispensable. The stronger autonomy claim is warranted only for the functions and interruption conditions actually sustained, not extrapolated from total material closure.

\section{Resource-based and serviceable infrastructure: robotics, production, and renewal}
\label{sec:isam}
Serviceable installations preserve structural and thermal investment through diagnosis, modular exchange, and requalification. The relevant outputs are restored service-hours, accepted products, avoided imports, and conditional loss reduction. Robotic industrial bootstrapping has established precedents \cite{Metzger2013}; here its support machinery remains inside Eqs.~\eqref{eq:capital} and \eqref{eq:industrialclosure}. Tables~\ref{tab:serviceability} and \ref{tab:manufacturability} distinguish demonstrated functions from larger capabilities, rather than assigning one maturity level to an industry.

\subsection{Local resource conversion changes the source of installed mass}
\label{sec:isru}
ISRU changes the supply architecture. Imported robots, power, and processors can convert much larger local stocks before occupation, but products have different acceptance conditions: geometric and geotechnical stability for cover, strength and leak tightness for pressure vessels, and distinct production chains for controllers. Table~\ref{tab:isru} ranks product-specific opportunities rather than assign a settlement-wide localization percentage.

The product demand and spare balance are given by Eq.~\eqref{eq:productsupply}. Local output $Q_j$ additionally requires an assayed resource route. For raw feed $F_r$, constituent mass fraction $g_r$, usable recovery $\eta_r$, and constituent requirement $a_{rj}$ per accepted product mass,
\begin{equation}
 F_r\geq\frac{a_{rj}Q_j}{g_r\eta_r}.
 \label{eq:isruclass}
\end{equation}
The inequality is for one specified constituent and route; shared feed, coproducts, reagents, and material return require one process balance, not full independent credit for every output. Excavation, beneficiation, conversion, finishing, inspection, and storage enter the local electricity calculation in Eq.~\eqref{eq:capacityscreen}. Manufacturing with imported powder is not extraction; oxygen removed from an oxide is not automatically collected, purified, and stored oxygen. The full-local-manufacture 19.5-MW reference already charges its specified product processing. Other resource operations add power only where excluded from those intensities.

\begin{table}[tb]
\caption{Product-resolved localization priorities. ``Candidate'' denotes an engineering route, not demonstrated industrial availability at the destination. The evidence column distinguishes flight operation, terrestrial experiments, and unqualified networks. Imported subcomponents can remain decisive despite a high locally supplied mass fraction.}
\label{tab:isru}
\begin{tabular}{p{.165\textwidth}p{.25\textwidth}p{.25\textwidth}p{.255\textwidth}}
\toprule
Local product and use & Route and limiting specification & Evidence and credible progression & Imports and remaining gate\\\midrule
Loose cover, berms, graded roads, trenches & Excavate, haul, place, and survey; bearing, slope, settlement, and verified column mass. & Prototype low-reaction-force excavation; surface campaign remains to be qualified \cite{Schuler2019,Giel2025}. & Drives, seals, bearings, tools, power and recovery; measure delivered tonnes per complete duty cycle.\\
Paving, aggregate, ceramic blocks, nonpressurized shelters & Sort, compact, sinter or melt; thermal cracking and erosion control. & Laser-melted simulant elements about 0.25~m across are demonstrated on Earth \cite{Gines2023}. & Initial heaters/optics and controls; plume loading, cyclic damage, joints, and independent pressure liner.\\
Water, oxygen, nitrogen, propellant inputs & Site-specific ice/volatile extraction or atmospheric separation; purity and stored usable yield. & Martian CO$_2$-to-O$_2$ flight demonstration; lunar resource observations and ground conversion \cite{Hoffman2022,Colaprete2010,ESAoxygen2020}. & Pumps, sorbents, membranes, catalysts, electrolytes, tanks; certify grade and recovery through adverse seasons.\\
Glass, mixed metals, qualified structural stock & Melt/chemical reduction, separation, refining, forming, heat treatment, and assay. & Ground oxygen/metal co-production; alloy grade and qualified throughput remain route-specific \cite{ESAoxygen2020}. & Refractories, electrodes, additives, machine tools and standards; mixed metal is not a qualified engineering alloy.\\
Pressure, thermal, mechanical spares & Hybrid casting/forming, machining, additive processing, joining, leak and proof testing. & Flight printing is a process precedent; product acceptance remains family-specific \cite{Prater2019,NASA6030}. & Initially import critical liners, valves, seals, precision bearings, coatings, sensors, and actuators until each route qualifies.\\
Power and industrial expansion & Local supports, shielding and eventually conductors; assemble imported active conversion modules. & Modular assembly has ground precedent \cite{Gregg2024}; complete local power-plant manufacture is unproved. & Initial solar cells, reactors/fuel, converters, electronics and storage internals; long-term renewal must include these classes.\\
\bottomrule
\end{tabular}
\end{table}

The earliest large mass substitution is generally civil work, not semiconductor manufacture. At the Mars-specific $629$-g/cm$^2$ shielding crossing in Sec.~\ref{sec:radiation}, one hectare requires about 63,000~t placed locally. Importing the same mass would require about 630 successful 100-t payload deliveries before their containers or supporting flights. A two-year emplacement interval requires 3.6~t/h calendar-average throughput, or 7.2~t/h while operating at 50\% availability. These are fleet acceptance targets, not extrapolated lifetime performance of a small prototype. Overburden also imposes approximately 23~kPa of distributed loading on Mars; lateral cover, arches, slopes, penetrations, and foundations require structural design. The Martian radiation profile does not prescribe a lunar shielding depth.

Counter-rotating bucket drums reduce the reaction-force constraint of low-gravity excavation. One NASA prototype report lists 67~kg dry mass and 0.761~Wh/kg delivered regolith, before thermal-vacuum and lifetime qualification \cite{Schuler2019}. A terrestrial preprint measured short filling runs near 778~kg/h; an assumed 10-m haul and estimated vehicle power gave 172~kg/h at 0.86~Wh/kg for the complete modeled cycle \cite{Giel2025}. Neither establishes multiyear field output. Traverse, charging, wear, and recovery set fleet production.

Loose shielding does not require melting. Thermal processing should be reserved for surfaces and components that need cohesion, abrasion resistance, or shape. Laser melting of simulant produced interlocking paving elements; ten 10-mm compression specimens had mean strength about 94~MPa with 56~MPa standard deviation, while thermal-shock resistance remained unqualified \cite{Gines2023}. This does not establish a landing pad's plume resistance or a habitat's tensile pressure boundary. For scale, a stated 1,000-m$^2$, 50-mm-thick consolidated layer at 2,800~kg/m$^3$ contains 140~t. Heating it through 1,200~K with mean heat capacity 0.8~kJ/(kg\,K), plus 400~kJ/kg effective phase-change enthalpy, requires 53~MWh absorbed. At 30\% net coupling this becomes 176~MWh; 100~kW available half the time needs about 147 days before handling and acceptance. These selected bulk-heating coefficients are not a laser-process fit. Concentrated sunlight can supply heat directly, but collection, tracking, darkness, and optics renewal remain in the plant boundary.

\subsection{Feed grade, energy, and the mass advantage of a seed plant}
\label{sec:isrubreak}
Resource grade sets throughput before a robot count is chosen. For water output $Q_w$, ice fraction $g_w$, and net extraction yield $\eta_x$, feed demand is $F=Q_w/(g_w\eta_x)$. At 100~t/year and $\eta_x=0.8$, assumed 5\% and 1\% ice require 2,500 and 12,500~t/year of feed. As a transparent thermal estimate, with $c_r\Delta T=120$~kJ/kg of regolith, water sublimation enthalpy $L_s=2,830$~kJ/kg, and heating efficiency $\eta_h=0.5$,
\begin{equation}
 e_w=\frac{c_r\Delta T/(g_w\eta_x)+L_s}{3600\eta_h}
       \quad\mathrm{kWh/kg\ of\ recovered\ water}.
 \label{eq:isruwater}
\end{equation}
The two grades give 3.2 and 9.9~kWh/kg before excavation, uncollected-vapor losses, separation, compression, and storage. Ice abundance at one sampled site is not a reserve certificate \cite{Colaprete2010}. For an independently modeled lunar ilmenite route, excavation through liquid-oxygen storage at 10\% ilmenite requires 24.3~kWh/kg centrally; 100~t/year therefore requires 2.43~GWh/year, or 0.28~MW mean process power \cite{Leger2025}. Metal refining and polymer production require their own routes. Ground molten-salt experiments at 950$^\circ$C produce oxygen and mixed metals from simulant, but the agency report explicitly identifies subsequent alloy selection and possible refining as unresolved \cite{ESAoxygen2020}. High-purity reagents, process atmospheres, electrodes, and analytical standards cannot be omitted from increasing industrial closure.

A mass crossover can include the seed plant's power supply rather than crediting local feed alone. Let accepted local output be $Q$ tonnes/year, displacing $\mu Q$ of functionally equivalent imported product. Thus $\mu<1$ permits a heavier local alternative. Let $\iota Q$ be recurring imported feed and process consumables, $M_0$ the incremental nonpower seed mass, and $s_0$ its annual support imports. With accepted-product intensity $e$, operating fraction $u$, power/cooling specific mass $\alpha$ in kg/kW, and net retirement $\delta_P$ of that power installation,
\begin{equation}
 P_{\rm inst}=\frac{10^3eQ}{h_yu},\qquad
 M_P=b_PQ,\qquad b_P=\frac{\alpha e}{h_yu},\qquad
 M_E(T)=M_0+M_P+T(s_0+\iota Q+\delta_PM_P).
 \label{eq:isrumass}
\end{equation}
$M_E$ is cumulative imported mass from commissioning through $T$ years; any precommissioning wear, deployment loss, extra storage, or uncounted support must be added. $M_0$ includes extraction, handling, processing, acceptance, and safe standby, excluding only the separately sized power/cooling plant. Its rated process envelope must cover $Q$. Comparing with direct product imports $\mu QT$ gives
\begin{equation}
 Q>\frac{M_0+Ts_0}{T(\mu-\iota)-(1+\delta_PT)b_P},
 \qquad T(\mu-\iota)>(1+\delta_PT)b_P.
 \label{eq:isrubreak}
\end{equation}
At fixed duty fraction and process coefficients, a nonpositive denominator means that increasing annual output \(Q\) cannot achieve imported-mass payback within the stated horizon. Feed abundance alone cannot fix it.

For a deliberately specified pilot envelope, take $M_0=100$~t, $s_0=5$~t/year, $e=20$~kWh/kg, $u=0.5$, $\alpha=50$~kg/kW including cooling, $\delta_P=0.05$~yr$^{-1}$, $\iota=0.10$, and $\mu=1$. At $Q=100$~t/year, processing averages 0.23~MW and requires 0.46~MW operating power; $M_P=22.8$~t. Section~\ref{sec:transition} and Fig.~\ref{fig:transition} compare the resulting 284-t and 419-t ten-year import totals for immediate and ramped output. At rated output, the ten-year crossover is about 17~t/year. At $\mu=0.5$, the same operation instead saves 43\% and crosses near 41~t/year. Equation~\eqref{eq:isrubreak} varies processing intensity and imported input; the archive retains the full sensitivity. These are product-family targets, not measured refinery performance or settlement-wide localization fractions. The plant's power and cooling enter the site's dispatch balance; its output substitutes for the same manufacturing stream rather than creating a second production credit.

Monetary advantage also requires avoided destination delivery to exceed plant development, fabrication, operation, finance, and risk. For immediate rated output, the 716-t ten-year net saving permits undiscounted additional cost $\Delta C_{\rm nonlog}<716,000c_D$, with marginal destination-delivery cost $c_D$ in dollars/kg. Hardware and bulk-cargo unit costs need not match; Sec.~\ref{sec:servicecost} treats discounted service. Local processing reduces shipment exposure but can replace it with dependence on an irreplaceable plant controller.

At 100~t/year the same mass model gives a ten-year break-even duty of 0.023 for $\mu=1$, or 0.068 for $\mu=0.5$. These thresholds retain the $1/u$ power-hardware scaling; the actual process must still attain its required active throughput. Accepted output, resource grade, and fault recovery can therefore bind before imported power mass.

\subsection{Predeployment and growth of the production network}
\label{sec:isrucommission}
Operational exchange, terrestrial robotic assembly, and flight polymer/metal deposition provide distinct starting capabilities \cite{CSADextre2024,Gregg2024,Prater2019,ESAmetal2026}. Integrating these with assay, storage, and acceptance for bounded products is a credible engineering campaign over the next few decades. An unattended industrial network must additionally renew precision hardware, pharmaceuticals, and metrology. Its measurable advances are availability, accepted yield, fault recovery, and qualified output per installed plant mass.

Robotic predeployment follows a dependency order: survey and assay; communications, pilot power, tools and recovery; access, a separated landing area, berms and foundations; qualified production and storage; pressure liners and services; then leak, load, purity, and outage tests before occupancy. Robots can place arrays, towers, tanks, imported reactors, process modules, and shielding well before they can manufacture their active components. The first lander cannot rely on a pad it has not built: designate an initial expendable landing area with stand-off from the later works. Repeated heavy landings require a verified plume/erosion design, not simply a sintered surface. Factory construction, dust traffic, habitat ventilation, and radiation-cover access must be planned together.

Industrial growth can relax the fixed-throughput ceiling in Eq.~\eqref{eq:renewalceiling}. If productive plant stock $K_F$ supplies suitable new plant at rate $u_FK_F$, with share $\beta_F$ reinvested and net retirement $\delta_F$, its conditional growth rate is
\begin{equation}
 \frac{\dot K_F}{K_F}=\beta_Fu_F-\delta_F>0,
 \qquad Q_j+I_{K,j}\geq(\delta_j+\gamma)K_j\quad\hbox{for every essential class}.
 \label{eq:industrialgrowth}
\end{equation}
Here $u_F$ counts commissioned, compatible plant capacity, not tonnes of arbitrary output. The scalar condition expresses the industrial-bootstrapping opportunity; the class conditions prevent surplus aggregate or steel from standing in for unavailable actuators, cells, electronics, fuel, drugs, or metrology \cite{Metzger2013}. No general local manufacturing prohibition follows from current imports. Conversely, closing a bulk-material loop does not demonstrate that the machinery reproducing that loop can itself be renewed. Growth powered by imported active modules remains a useful intermediate architecture, with those imports explicitly budgeted.

\subsection{Controlled comparison with a published industrial baseline}
\label{sec:publishedbenchmark}
A quantitative comparison should preserve the predecessor's boundary. The robotic bootstrapping study already includes power, generational retirement, assembly, and imported electronics \cite{Metzger2013}. Its Table~2 specifies an initial set with eight printers at 0.5~kg/h each. Summing the listed equipment and electrical entries reproduces 7.675~t and 64.36~kW; at its 70\% solar duty, nominal parts output is 24.5~t/year. The broader study varies printer rate and duty and retains imported complex components until its assumed production transitions. The comparison below reproduces that initial throughput calculation, not its recursive six-generation industrial forecast.

For unchanged hardware, let $r$ be each printer's nominal rate, $u$ powered duty, and $y_A$ the fraction of nominal output accepted without additional processing. Then $Q_A=8ru h_y y_A/10^3$ tonnes/year. Table~\ref{tab:publishedbenchmark} changes these factors one at a time and jointly. The published 0.5-to-0.1-kg/h rate sensitivity reduces output fivefold; lowering duty from 0.7 to 0.4 reduces it by a further factor 1.75. An explicitly added $y_A=0.8$ sensitivity gives 2.24~t/year when combined with both changes. The active electrical loads are held fixed and assumed off outside powered duty. The displayed energy ratio assigns the whole set's electricity to accepted parts, leaving its coproducts unallocated; it therefore benchmarks the stated set, not a transferable coefficient for a complete habitat industry.

The divergence is driven here by \emph{accepted throughput}, not initial imported mass. A fivefold printer-rate reduction increases the whole-set electrical ratio fivefold at unchanged active load; a duty reduction lowers output and annual electricity together. Increasing nominal printer count also adds equipment to reproduce. The decisive validation is accepted, class-compatible output per complete duty cycle, followed by installed-asset and assembly closure. The separate 9-to-13-year and reachable-to-unreachable comparisons in Sec.~\ref{sec:controlledcomparison} isolate retirement under this paper's stock boundary; the 135-t commissioning penalty in Sec.~\ref{sec:localtransition} isolates ramp-up. They do not impute those omissions to the published model or convert robotic industrial growth into biological qualification.

\begin{table}[tb]
\caption{Initial-set throughput benchmark. The nominal rate and duty and their lower alternatives are from the published industrial study \cite{Metzger2013}; the 80\% acceptance fraction is an additional sensitivity introduced here. Hardware and the active 64.36-kW load are unchanged. The energy ratio includes all listed set loads and does not allocate coproduct energy.}
\label{tab:publishedbenchmark}
\begin{tabular}{p{.27\textwidth}p{.105\textwidth}p{.095\textwidth}p{.10\textwidth}p{.14\textwidth}p{.14\textwidth}}
\toprule
Changed assumption & $r$ (kg/h) & $u$ & $y_A$ & Accepted output (t/yr) & Set ratio (kWh/kg)\\\midrule
Published nominal basis & 0.5 & 0.7 & 1.0 & 24.5 & 16.1\\
Printer rate alone & 0.1 & 0.7 & 1.0 & 4.91 & 80.5\\
Solar duty alone & 0.5 & 0.4 & 1.0 & 14.0 & 16.1\\
Acceptance alone & 0.5 & 0.7 & 0.8 & 19.6 & 20.1\\
Three changes combined & 0.1 & 0.4 & 0.8 & 2.24 & 101\\
\bottomrule
\end{tabular}
\end{table}

\subsection{Robotic servicing, modularity, and peer recovery}
\label{sec:servicing}
Ground-operated Dextre already exchanges external ISS equipment, including approximately 100-kg batteries and electrical modules \cite{CSADextre2024}. Commercial rendezvous and life extension also have operational heritage: the Mission Extension Vehicle (MEV) spacecraft provide client propulsion from their own thrusters, rather than refueling or rebuilding the client. The operator reports completion of MEV-1's first five-year service and transfer to a second client in 2025. Its July 2026 Mission Robotic Vehicle (MRV) launch establishes deployment, not completed autonomous repair or assembly \cite{SpaceLogistics2026}. These observations support servicing for specific access geometries and interfaces, not arbitrary spacecraft repair. Conversely, NASA's cancellation of On-orbit Servicing, Assembly, and Manufacturing~1 (OSAM-1) after cost, technical, schedule, and partnership problems illustrates the difficulty of servicing an unprepared client; it is not a physical prohibition on servicing \cite{NASAOSAM2024}.

Maintainability requires accessible replacement units, captive fasteners, grapple points, isolatable fluid/electrical interfaces, and diagnostic and calibration access. Added connectors must justify their mass and failure modes through restoration. Refurbishment restores function; remanufacturing restores a specified product through disassembly, processing, and acceptance. Refilling cannot repair a degraded pressure boundary: pumps, valves, and lines must remain replaceable. Cryogenic transfer separately requires a heat-leak and refrigeration budget; Sec.~\ref{sec:transport} supplies the campaign duration rather than that thermal sizing.

For $n$ units of mass $m$, each requiring intervention at rate $\lambda_u$, let a fraction $c$ of interventions successfully restore the same service using an exchanged submodule of mass $m_x$. Annual Earth-delivered support is
\begin{equation}
 \dot M_E=n\lambda_u\bigl[c m_x+(1-c)m\bigr]+\dot M_S,
 \label{eq:modularimports}
\end{equation}
where $\dot M_S$ includes the servicer's replacement hardware and consumables. The initial incremental installation $M_S$ includes access hardware, power, thermal control, handling, and metrology. In a declared example, $n=100$, $m=0.5$~t, $\lambda_u=0.08$~yr$^{-1}$, $c=0.8$, and $m_x=0.1$~t reduce replacement imports from 4.0 to 1.44~t/year. Adding 0.30~t/year of service-system imports gives 1.74~t/year. A 10-t incremental installation then repays its imported mass in 4.4 years; over ten years it imports 27.4 rather than 40~t. On these coefficients the ten-year mass advantage begins near 51 compatible client units. These values are an architectural sensitivity, not measured servicing economics; the assumed post-repair reliability, access success, and service-system burdens require verification. The benefit is replacement of a small failed function, not automation alone.

Teleoperation can support this architecture without making fast contact control depend on Earth. A round-trip delay $\tau_d$ consumes phase $2\pi f\tau_d$ at frequency $f$; allocating at most $45^\circ$ of phase lag to that delay gives $f<1/(8\tau_d)$. For illustrative round trips of 0.24~s and 2.56~s, this is 0.52 and 0.049~Hz. High-bandwidth force limiting, collision avoidance, and retreat therefore belong on the robot; a remote operator can specify goals and authorize uncertain actions. Increasing autonomy is required as communication delay, outages, task count, and fleet size make continuous supervision impractical. It need not imply an unconstrained general-purpose controller.

A service robot is itself a dependency. Peer recovery needs passive capture, externally releasable joints, interchangeable actuators/controllers, independent power/data, and accessible calibration. A decisive trial is bidirectional restoration after specified faults, including communication loss, followed by acceptance tests. This fleet-level capability is not established by the cited servicing demonstrations.

Equation~\eqref{eq:redundancy} quantifies its potential. For two independent robots with $\lambda_u=0.2$~yr$^{-1}$, reducing restoration time from 48 to 6 hours reduces overlap loss frequency from $4.4\times10^{-4}$ to $5.5\times10^{-5}$~yr$^{-1}$. Adding an illustrative common-cause contribution of $10^{-3}$~yr$^{-1}$ reduces the overall improvement from eightfold to only 1.36-fold. Shared software, inaccessible common tools, and a single charging or calibration station can erase most redundancy benefit. Recovery architecture must therefore include diversity, isolation, and replacement of shared equipment, not just additional robots.

\begin{table}[tb]
\caption{Robotic servicing and construction: demonstrated scope, quantitative leverage, and next qualification. Numerical examples refer to the explicitly assumed architectures in Sec.~\ref{sec:isam}, not reported performance of the cited missions.}
\label{tab:serviceability}
\begin{tabular}{p{.175\textwidth}p{.28\textwidth}p{.24\textwidth}p{.235\textwidth}}
\toprule
Capability & Evidence and scope & Leverage and relevant scale & Remaining gate\\\midrule
Teleoperated exchange and autonomous rendezvous & Dextre: operational external maintenance. MEV: operational docking and propulsion life extension, not general repair \cite{CSADextre2024,SpaceLogistics2026}. & Shorter verified repair lead time reduces outage overlap and spare inventories; useful even for one high-value client. & Repeated contact safety, access, isolation, propellant, and recovery after failed docking.\\
Modular replacement; refurbishment & Selected ISS exchange operations are routine. Complete unattended disassembly and remanufacturing remain unproved. & In the 100-unit case, 4.0 becomes 1.74~t/year including service support; 51 compatible units recover the 10-t installation within ten years. & Connector lifetime, repair coverage, restored performance, and disassembly/requalification yield.\\
Robot-to-robot recovery & Enabling manipulation exists; no cited demonstration qualifies a self-maintaining orbital fleet. & Two-day to six-hour repair reduces the independent overlap term eightfold; shared failures can dominate. & Reciprocal fault-injection trials, dead-robot capture, independent tools, and controller exchange.\\
Assembly and construction & Automated Reconfigurable Mission Adaptive Digital Assembly Systems (ARMADAS): three-robot terrestrial assembly; not orbital construction \cite{Gregg2024,NASAARMADAS2024}. & 4,000 panels at two accepted hours each require about 139 days with four robots at 60\% duty. & Flight metrology, load paths, joints, delivery integration, and isolated fluid/power circuits.\\
Artificial intelligence (AI)-enabled diagnosis, control, and logistics & Task-specific learned flight control has ISS experimental evidence; broader maintenance decisions remain application-dependent \cite{Stewart2025}. & Reduces detection, diagnosis, and queue time only when false actions and support demand remain bounded. & Fault-specific detection and false-alarm rates, safe fallback, configuration control, and independent acceptance.\\
\bottomrule
\end{tabular}
\end{table}

Remote assistance can limit fleet scale. For $n_R$ robots with intervention rate $\nu_R$ and mean operator time $h_R$, $n_R\nu_Rh_R<n_oh_o$ bounds workload for $n_o$ operators supplying $h_o$ hours each, before queueing margin. The specified 100 robots require 50 operator-hours/day at one 30-minute intervention per robot-day, or 8.3 at five minutes. Supervision must meet the task's latency: Earth operators can support lunar work but cannot close a rapid Mars contact loop. Intervention tails, peer recovery, and compatible spares determine sustained net service.

\subsection{Construction beyond launch-sized elements}
\label{sec:assembly}
In-space assembly relaxes fairing-envelope and deployment constraints and allows gradual expansion and module replacement. It does not remove pressure stress, centrifugal loading, dimensional tolerances, fatigue, or heat rejection. The strongest initial targets are repeated truss units, radiator panels, power elements, shielding containers, and replaceable equipment decks; a pressure boundary additionally needs leak-tight joints and verified fracture resistance. Specialized manipulators, crawlers, gantries, and fixtures can exploit repeated geometry more efficiently than a humanoid required to emulate every terrestrial task.

The ARMADAS experiments demonstrate this design philosophy on Earth: three robots assembled a meter-scale structure from repeated units during approximately 100 hours of operation spread over several weeks \cite{Gregg2024,NASAARMADAS2024}. Extrapolation to orbit requires vacuum-compatible mechanisms, reliable reaction-force paths, thermal metrology, and verification of every load-bearing connection. For scale, a $10^5$-m$^2$ radiator divided into 25-m$^2$ panels has 4,000 units. Four robots at 60\% productive duty, each requiring two hours per installed and accepted panel, complete the assembly workload in approximately 139 days. Panel delivery, common tooling, checkout queues, and robot servicing must also fit that schedule; the example is not an ARMADAS performance prediction.

If 4,000 independent interfaces were all essential, an escaped critical-defect probability of $10^{-4}$ per interface would give only $(1-10^{-4})^{4000}=0.67$ probability of no escaped defect. Achieving 0.99 on that deliberately stringent series criterion requires about $2.5\times10^{-6}$ per interface. Real arrays should instead tolerate isolated defective panels and include bypassable electrical branches and independently isolatable thermal loops. Modularity thus has two competing effects: more interfaces, but the possibility of containing failures and replacing limited sections. Automated alignment and fastening must be accompanied by load-path, electrical-continuity, and leak testing. Assembly count alone is not accepted capacity.

\subsection{Fabrication, remanufacturing, and material return}
\label{sec:fabrication}
Orbital fabrication should be a hybrid workshop, not a universal printer. Machining standard blanks provides precision interfaces; forming and joining produce large structures; additive processes provide otherwise inaccessible geometries and pool spare inventories across a qualified part family. Thermal components require controlled channels, surface properties, cleanliness, and fluid proof tests. Bearings, seals, motors, power electronics, sensors, and catalysts require additional processes and imports until their own production is qualified. On-orbit inspection and refurbishment can preserve the expensive installed structure while replacing these shorter-lived functions.

Polymer printing has flight heritage: the phase-I/II analysis attributed initially differing specimen properties to process settings rather than an engineering-significant microgravity effect within that experiment \cite{Prater2019}. Metal deposition has reached ISS process demonstrations. The European Space Agency (ESA) reports retrieval of its fourth metal sample in July 2026 from a 180-kg printer; this is not evidence of a complete 180-kg flight-qualified factory \cite{ESAmetal2026}. NASA reports early printed products and Refabricator bonding difficulties, illustrating that deposition and repeated material reuse are separate capabilities \cite{NASAAM2025}. In reduced-gravity vacuum testing, 69 of 70 attempted laser welds were executed successfully; the tests establish joining feasibility under those conditions, not long-duration orbital weld qualification \cite{NASAWelding2024}. None supplies settlement-scale accepted throughput, duty, or equipment life.

Process scrap and retired-product return are distinct streams. For accepted output $Q$, mass yield $y$, scrap-return fraction $r_s$, qualified end-of-life recovery $\zeta$, and same-material retired flow $Q_{\rm ret}$,
\begin{equation}
 I_{\rm virgin}=\frac{Q}{y}-r_s\Big(\frac{Q}{y}-Q\Big)-\zeta Q_{\rm ret}.
 \label{eq:materialreturn}
\end{equation}
The relation assumes steady flow, controlled composition, and adequate processing capacity; excess recovered feed is stored rather than represented by negative imports. At stationary replacement, $Q_{\rm ret}=Q$, $y=0.8$, $r_s=0.9$, and $\zeta=0.81$ give $I_{\rm virgin}/Q=0.215$. With no retired-product return the ratio is 1.025: printing alone then consumes slightly more delivered feed than importing the accepted product, before considering inventory pooling, packaging, or geometry. During growth, return cannot be credited for products not yet retired; at the reference fixed-mix net replacement of 1.51 tonnes per resident-year and growth demand of 0.30, the same material coefficients give 0.349 rather than 0.215. A changing product mix requires class-specific return streams.

Segregated alloys, documented polymers, and clean disassembly preserve reusable feed. Oxygen pickup, volatile loss, chain degradation, inclusions, and microstructural change require assay and controlled reprocessing \cite{NASA6030,NASAAM2025}. Ten cycles retain $0.95^{10}=0.60$ of an initial parcel at 95\% recovery, or 0.90 at 99\%; new feed must cover the remainder. Chemical rejection can dominate collection losses.

\begin{figure}[tb]
\centering
\includegraphics[width=0.80\textwidth]{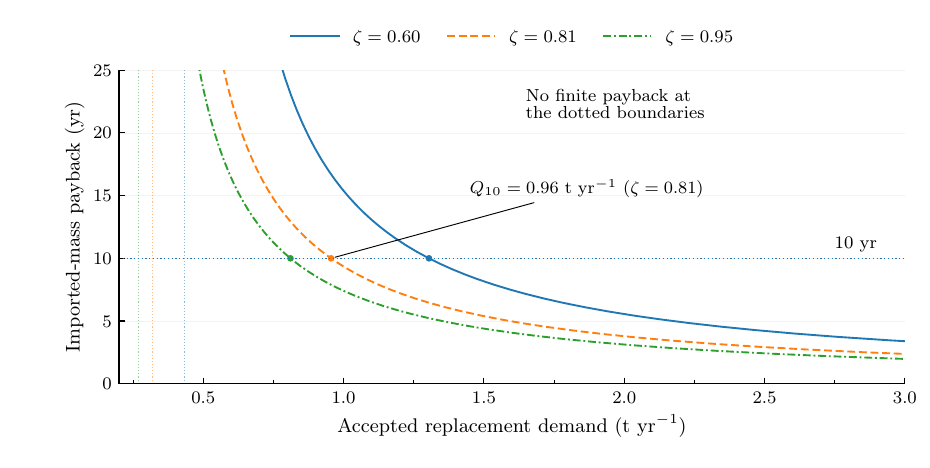}
\vskip -15pt
\caption{Imported-mass payback of a specified fabrication and inspection installation. Curves assume stationary replacement, 80\% accepted yield, 90\% process-scrap return, a 5-t installation, and 0.25-t/year support imports; $\zeta$ is qualified end-of-life material return. Vertical dotted lines mark the lower demand boundaries for positive payback, and circles mark ten-year crossings. The focused demand range retains every boundary and crossing; near-boundary payback times above the plotted range diverge. Growth reduces retired-material availability. Processing capacity, material identity, qualification, and monetary cost are separate constraints.}
\label{fig:manufacturing}
\end{figure}

For a 5-t fabrication/inspection installation, 0.25~t/year of support imports, and stationary feed fraction 0.215, ten-year mass payback requires accepted demand above 0.96~t/year (Fig.~\ref{fig:manufacturing}). At 1~t/year, payback takes 9.3 years. Processing at the assumed 50~kWh/kg gross intensity plus 1~kW idle load uses 71~MWh/year, or 8.1~kW on average. Charge these requirements only where the main renewal budget excludes the same processes. Each product family requires its own yield, material-return, energy, and inspection specification.

\begin{table}[tb]
\caption{Production and acceptance capabilities for a serviceable habitat or orbital industrial cluster. A process demonstration is not equivalent to qualification of every product made by that process. The recommended early product families are engineering priorities, not a certified manifest.}
\label{tab:manufacturability}
\begin{tabular}{p{.14\textwidth}p{.27\textwidth}p{.265\textwidth}p{.245\textwidth}}
\toprule
Capability & Evidence and maturity & Quantitative or architectural value & Main remaining gate\\\midrule
Additive and hybrid fabrication & Polymer and metal deposition flown; reduced-gravity welding tested \cite{NASAAM2025,ESAmetal2026,NASAWelding2024}. & Common feed/blank stock substitutes for selected distinct spares; the stated installation crosses ten-year mass break-even near 1~t/year accepted output. & Process window, powder/fume confinement, finishing, material properties, and calibrated metrology.\\
Refurbishment and remanufacturing & Exchange is operational for selected equipment; closed-loop orbital rebuilding is not established. & Preserve tanks, frames, decks, and serviceable heat exchangers while replacing worn functions; reduce net permanent retirement $\delta_j$. & Damage diagnosis, disassembly, surface restoration, contamination removal, and requalification.\\
Spare parts and consumables & Tool printing is demonstrated; full supply of seals, electronics, drugs, and catalysts is not. & Reducing qualified replenishment lead from five years to 0.1 year changes the stated 20-unit spare stock from eight to one. & Material identity, shelf life, substitutes, software compatibility, and clinical/chemical acceptance.\\
Recycling and reuse & Component-scale flight investigations; repeatable industrial material return remains unqualified \cite{NASAAM2025}. & The worked stationary loop needs 0.215~kg virgin feed per accepted kilogram, rather than 1.025 without end-of-life return. & Sorting, purity, alloy/polymer degradation, energy, and recovery of mixed assemblies.\\
Inspection and acceptance & Terrestrial aerospace qualification practice is established; deployment in an autonomous factory is incomplete \cite{NASA6030}. & Prevent a low defect rate from producing a large fleet-level escape probability; qualify output rather than count builds. & Demonstrated detection, reference artifacts, independent proof/leak tests, traceability, and false rejection.\\
Extraterrestrial feedstock & Resource measurements and specific process experiments; an operational export chain remains unproved \cite{Leger2025,Metzger2013}. & Bulk cover is useful at the surface immediately; orbital use requires competitive delivered feed at sufficient sustained demand. & Reserve certification, mining, assay, storage, ascent/capture, equipment renewal, and buyer specification.\\
\bottomrule
\end{tabular}
\end{table}

\subsection{Acceptance, fault diagnosis, and autonomous operations}
\label{sec:acceptance}
NASA-STD-6030 requires process/material qualification, configuration control, postprocessing, inspection, and objective acceptance; in-process monitoring does not generally replace post-build nondestructive evaluation \cite{NASA6030}. Pressure and thermal products need dimensions, proof/leak tests, volumetric or surface inspection, and fatigue/fracture allowables for the actual process history. A model prediction cannot replace that evidence.

Let $q_0$ be the preinspection fraction of a specified dangerous defect, $d$ its detection probability, and $f$ the false-rejection probability for acceptable parts. The defect probability conditional on acceptance is
\begin{equation}
 p_{\rm esc}=\frac{q_0(1-d)}{(1-q_0)(1-f)+q_0(1-d)}.
 \label{eq:acceptance}
\end{equation}
With stipulated $q_0=0.02$, $d=0.99$, and $f=0.01$, $p_{\rm esc}=2.1\times10^{-4}$, and 1,000 independent accepted parts have an 18.6\% probability of containing at least one such defect. Holding the other inputs fixed, keeping that batch probability below 1\% requires detection above approximately 99.95\%. This is a defect-class example, not a predicted habitat failure rate. Process control, multiple complementary tests, containment, and replaceability are therefore more useful than a single generic ``99\%-accurate'' inspection claim. Test dependence and correlated build defects must be measured rather than multiplied away.

Task-specific artificial intelligence (AI) in free-flyer control has ISS experimental evidence \cite{Stewart2025}; broader learned operations require validated transfer and long-horizon reliability \cite{AISafety2026}. Evaluate missed faults, false interventions, restoration quantiles, and assistance, with independent command limits and recoverable safe states. In the serial example, reducing diagnosis from eight to one hour, with four hours each for queueing and exchange and two for acceptance, changes restoration from 18 to 11 hours. A 30-day parts delay instead gives 738 versus 731 hours, less than 1\% improvement. Local spares or repair address that dominant delay; diagnosis alone does not.

\emph{Radiation-qualified control and compute refresh.} The electronic system requires a mission-specific assessment of total ionizing dose, displacement damage, and single-event effects at its parts and boards; the human dose-equivalent allocation is a different qualification quantity \cite{ECSSRHA2025}. Error correction, protected power switching, independently powered safe-state control, and verified restart are candidates for retaining service after faults. They must meet the required availability under test, not just survive a nominal exposure. Planned replacement limits cumulative exposure only if exchange occurs within the qualified interval, including stored-spare exposure and delivery delay. It does not remove destructive single-event risk before that exchange. Changed device processes, lots, bias conditions, and layouts require appropriate reassessment. New compute should therefore enter through replaceable, independently accepted modules while safety functions remain available through a qualified control path; component refresh must be charged once to the equipment and spares balances.

\subsection{Lifecycle economics and the scale of useful infrastructure}
\label{sec:servicecost}
Economic comparisons must preserve the same delivered service and risk requirements. Let $C_0$ be incremental service-system capital including launch and commissioning, $C_f$ annual fixed operating and renewal cost, $\nu$ compatible interventions per year, $s$ the probability of restoring the contracted service, $C_{\rm rep}$ the avoided cost of a delivered and commissioned replacement, and $C_v$ the variable intervention cost. Additional collateral risk is represented by $p_DV_D$, beyond ordinary replacement after an unsuccessful repair. With discount rate $r$ and horizon $T$, a stationary screening condition is
\begin{equation}
 \nu\Big[sC_{\rm rep}-C_v-p_DV_D\Big]
 > \frac{r}{1-(1+r)^{-T}}C_0+C_f.
 \label{eq:servicecost}
\end{equation}
All support, propellant, qualification, remote assistance, and end-of-life obligations belong in the costs. Failed repair still incurs replacement. Human-safety requirements are constraints, not tradable expected dollar losses. Replacement obsolescence, downtime value, client lifetimes, and changing demand require a discounted event schedule beyond this screen.

In hypothetical common-date dollars, $C_0=\$100$~million, $C_f=\$2$~million/year, $r=0.05$, and $T=10$ years give an annualized fixed burden of \$15.0~million. With $s=0.95$, $C_v=\$1$~million, and $p_DV_D=\$0.1$~million per attempt, avoiding an \$8-million replacement needs about 2.3 interventions/year to break even. Avoiding a \$2-million replacement needs about 18.7. At a selected 0.08 interventions per client-year these correspond to approximately 29 and 234 compatible clients. These are not market price estimates: they identify the demand and value data needed before claiming commercial viability. A long-lived common service hub can be valuable to expensive compatible assets before a general autonomous factory is economic; cheaper replacement and scarce compatible demand can reverse the conclusion.

Compatibility includes orbit. At 7.7~km/s, an impulsive plane change $\Delta v=2v\sin(\Delta i/2)$ costs approximately 0.13~km/s for $1^\circ$, or 1.34~km/s for $10^\circ$. A widely distributed constellation is not automatically one low-cost maintenance market. Co-located infrastructure, common spares, reachable service orbits, replaceable payloads, and scheduled delivery can improve utilization while reducing transfers. Launch-cost reductions benefit both servicing and replacement; they cannot be assumed to favor local manufacturing exclusively.

For successive hardware generations, replace the stationary screen by a discounted schedule of actual exchanges. Count a shared hub once only while its structures, tools, and interfaces remain qualified for the serviced generations; charge upgrades, mixed-generation spares, acceptance tests, software migration, and final disposal as they occur. Credit retained equipment by remaining delivered function, not nominal age. Record whether displaced modules are retired, refurbished, or reassigned in Eq.~\eqref{eq:capital}; reassignment and a replacement cannot both be credited as the same new capacity. Common interfaces can preserve service demand across generations, but qualification and economic advantage must be reevaluated when the product or mission changes.

Local lunar use avoids an export chain. Orbital supply adds processing, storage, ascent, transfer, capture, and assay: even a selected 3-km/s maneuver at 450-s specific impulse has mass ratio 1.97 before vehicle and reserve allocations. The lunar liquid-oxygen model's 24.3~kWh/kg \cite{Leger2025} cannot be assigned to metals or electronics. Accessible authorized salvage avoids extraction and ascent, but needs passivation, material identity, and contamination control. Dependable demand and matched accepted products determine whether either route undercuts Earth supply.

\subsection{Industrial maturity and the next scaling decision}
\label{sec:industrialmaturity}
Scale-up should follow demonstrated function: prepared-interface exchange before arbitrary servicing, excavation and reduction pilots before routine resource supply, and accepted part families before broad manufacturing. The next decision is whether commissioning, utility demand, and renewal of the production equipment support the intended output. Table~\ref{tab:trl} and Sec.~\ref{sec:roadmap} give the evidence and conditional horizons.

Require matched product specifications, metered utilities and human assistance, a measured commissioning trajectory, and replacement of a limiting plant component. Further growth must renew controls and metrology as well. Equation~\eqref{eq:industrialgrowth} relaxes the fixed-throughput ceiling only when output builds commissioned compatible capacity; imported active modules can support that intermediate stage.

\section{Architecture selection: match the environment to the development stage}
\label{sec:architectures}
Table~\ref{tab:architectures} compares protected, powered habitats, not bare planetary surfaces. Earth-adjacent orbit offers accessible variable-gravity and recovery experiments; the Moon offers nearby surface-production tests; Mars offers broader feedstocks with greater transport and autonomy burdens. This functional comparison does not select a unique destination. The comparison is evaluated through the feasibility definitions in Sec.~\ref{sec:feasibility}: an architecture that is physically admissible may still fail the engineered, physiological, or persistent-population tests.

\begin{table}[tb]
\caption{Architecture-level comparison, present maturity, and next discriminator. Environmental values are rounded from planetary and satellite data \cite{JPLPlanets,JPLSatellites,NASATitan}; solar flux ratios follow $d^{-2}$. Horizons are conditional judgments based on the cited evidence and declared program targets, not agency schedules or confidence intervals.}
\label{tab:architectures}
\begin{tabular}{p{.11\textwidth}p{.24\textwidth}p{.31\textwidth}p{.26\textwidth}}
\toprule
Architecture & Physical advantages and costs & Present maturity \& conditional horizon & Next discriminator and evidence\\\midrule
Earth-adjacent modular orbit & Selectable rotational loading; short logistics. Initial active hardware is imported; radiation depends strongly on orbit. & Earth-supported adult occupation is operational. Uncrewed and bounded occupied rotation experiments are credible 10--20-year objectives; a permanent rotating settlement is not dated. & Long-duration occupied rotation with measured health and maintainability; modular deployment, docking, despin, refuge, and shield-interface tests.\\
Lunar surface & $0.165\,g_0$; negligible atmosphere; local shielding; approximately 1.3-s one-way light time. Nighttime power and volatile access are site-dependent. & Short expeditions have nearer announced program targets. Durable supplied outposts, civil works, and commodity pilots are plausible 10--20-year objectives; broad local industry is multidecade. & Repeated landing and return, site-grade resources, darkness power, dust control, protected refuge, net accepted product output, and machine survival. Native-gravity reproduction remains unqualified.\\
Martian surface & $0.38\,g_0$; approximately 600-Pa atmosphere; water-bearing materials, CO$_2$, N$_2$; approximately 43\% of 1-AU sunlight before attenuation. & No integrated human chain exists. A first expedition within 10--20 years is an optimistic conditional possibility; maintained occupation and industry are later multidecade objectives. & Human-class uncrewed landing, validated receiving site, stored qualified products, return or refuge, transit health, 0.38-$g_0$ biology, and recovery through missed opportunities.\\
Outer satellites & Approximately $0.13$--$0.14\,g_0$ for Callisto/Titan; Jupiter/Saturn solar flux approximately $1/27$ and $1/90$ of Earth's. Titan: 94~K, approximately 150~kPa. & Robotic missions are feasible; crewed outposts are not a 10--20-year baseline. Several decades of prerequisite power, transport,  autonomy development would be required before a credible program. & Multi-year power, repair, propulsion, clinical autonomy, location-dependent radiation, low-gravity qualification, and resource-to-product networks.\\
Venus cloud-level aerostat & Near-terrestrial gravity and pressure can coincide at selected altitudes, but corrosive aerosols, deployment, water, solids, and ascent remain limiting. & Robotic atmosphere and materials experiments are plausible 10--20-year work. A crewed aerostat is a systems concept with multidecade integration needs; permanent settlement has no defensible date \cite{HAVOC}. & Large-volume entry/deployment, envelope life, power, water and solid-feedstock logistics, ascent/return, station keeping, and loss-of-lift recovery.\\
\bottomrule
\end{tabular}
\end{table}

Orbital biological and repair trials avoid making heavy Mars landing a prerequisite. Mature-habitat economics are different: imported bulk material, pressure structure, and exposed collector/radiator area can dominate free-space costs and reliability. Modular rotation instead reduces dependence on unqualified partial-gravity adaptation. Surface shielding can use local mass, with excavation, geometry, and maintenance charged separately.

The modeled 0.28-MW requirement for 100~t/year of lunar liquid oxygen applies to its specified grade and reduction chain (Sec.~\ref{sec:isru}); each additional material class requires its own qualified route.

Near-Earth orbital facilities isolate controllable gravity and shorten recovery paths; lunar operations test excavation, dust, resources, and power. Mars adds a wider feedstock set but delayed rescue and human-scale landing. MOXIE produced oxygen from its atmosphere \cite{Hoffman2022}; the 12-g/h reported peak \cite{MOXIE2023} is about 2,900 times below the reference 35-kg/h gross respiratory demand. That benchmark sizes full respiratory backup. At separately assumed 98\% oxygen recovery, make-up falls to 0.70~kg/h before growth, exports, leaks, and reserve restoration. Demonstrated water recovery supplies no oxygen-recovery coefficient. Product lifetime, intake, power, heat, and storage determine scale-up.

Human operations at distant moons require the same capabilities with longer unattended operation and weaker solar supply; reliable fission, robotic maintenance, and a validated life-cycle environment are enabling infrastructure. Titan's dense atmosphere changes pressure containment and aerodynamics but supplies neither breathable oxygen nor warm conditions; hydrocarbons are not free energy without an oxidant and a complete production balance. Radiation environments within giant-planet systems are strongly location-dependent, so neither the scientific interest of an ocean world nor the nominal inventory of water establishes a suitable human base.

Planetary transformation is a different scale from protected local habitation. For Mars, adding global pressure $\Delta p$ requires
\begin{equation}
\Delta M=\frac{4\pi R_M^2\Delta p}{g_M}
\simeq3.9\times10^{15}\Big(\frac{\Delta p}{1\ \mathrm{mbar}}\Big)\mathrm{kg}.
\label{eq:terraform}
\end{equation}
This mass relation follows directly from hydrostatics and planetary geometry. Published inventory, forcing, and industrial-throughput analyses distinguish atmospheric mass from warming and sustained oxygen production \cite{Jakosky2018,Terraform2026}. Particle-warming models address a distinct climate mechanism \cite{Ansari2024,Richardson2026}. Warming is not equivalent to creating oxygen, adequate pressure, a stable nutrient cycle, or sufficient gravity. No pathway in this paper depends on planetary terraforming. Robotic scientific investigation and contamination control should precede irreversible alterations of sites relevant to the search for indigenous life \cite{PlanetaryProtection}.

\subsection{Comparative viability and the scale of implementation}
\label{sec:scenarioviability}
Table~\ref{tab:architectures} separates destination advantages from demonstrated capability and the next discriminating test. Section~\ref{sec:roadmap} consolidates program horizons; the operating origin of a stock or commissioning calculation is not the start of its development program.

The near-term lunar program supplies an important scale reference. NASA currently targets an Artemis~III orbital demonstration in 2027 and an Artemis~IV landing in early 2028, with two astronauts on the surface for about a week \cite{NASAArtemisIII2026,NASAArtemisIV2026}. These are announced mission targets, not achieved capabilities or dates for a permanent outpost. Small adult expeditions may therefore precede the 10--20-year habitat-development window, while continuous occupation, local industrial supply, and reproduction remain separate gates. NASA's Moon to Mars architecture likewise describes required elements and capabilities rather than a validated settlement manifest \cite{NASAMarsArchitecture2026}.

At the same solar assumptions, a 19.5-MW load requires approximately 1.3~km$^2$ of collectors near Jupiter and 4.3~km$^2$ near Saturn before illumination, storage, incidence-angle, and atmospheric losses. Ideal circular-orbit transfers at 5.2 and 9.58~AU take about 2.7 and 6.1 years one way. Those durations make clinical autonomy and unsupported hardware life central to a crewed architecture. Dragonfly's 2028 launch and 2034 arrival objectives describe a robotic Titan campaign \cite{NASADragonfly2026}.

A Venus aerostat exchanges pressure-vessel burden for buoyancy, deployment, corrosion, and ascent requirements. In a declared equal-temperature ideal-gas example, $p=100$~kPa and $T=300$~K give gross breathable-air lift in a CO$_2$ atmosphere of
\begin{equation}
 \Delta\rho=\frac{p(\mathcal M_{\rm CO_2}-\mathcal M_{\rm air})}{RT}
 \simeq0.60\ \mathrm{kg/m^3},\qquad M_{\rm supported}\leq\Delta\rho V.
 \label{eq:venuslift}
\end{equation}
Using molar masses 44.01 and 28.97~g/mol, 100~t requires at least $1.66\times10^5$~m$^3$ of lifting volume before envelope, gondola, gas losses, or reserve lift. The selected pressure and temperature are an illustrative local state, not simultaneous global values at a fixed altitude. The High Altitude Venus Operational Concept (HAVOC) identifies a mission concept, but deployment of that volume, acid-resistant life, water and solid-feed supply, and ascent to the return vehicle are unqualified at the relevant human scale \cite{HAVOC}. Near-terrestrial gravity is an advantage, not a substitute for that transport and resource chain.

At an assumed global atmospheric addition of $10^{12}$~kg/year, Eq.~\eqref{eq:terraform} gives approximately 3,900 years per millibar before losses or warming. The interval scales inversely with net addition rate. Planetary pressure, warming, and chemistry therefore require separate throughput assessments from local protected habitation \cite{Jakosky2018,Terraform2026,Ansari2024,Richardson2026}.

The baseline uses protected habitats and contemporary human physiology. Terraforming, routine human torpor, fusion crew transport, broad heritable adaptation, and identity transfer remain speculative alternatives rather than inputs to its feasibility balances.

\section{When can autonomous habitation provide strategic resilience?}
\label{sec:purpose}
Scientific exploration, economically supported residence, and independent population continuity have different success criteria. A station receiving regular Earth deliveries can deliver science or commercial services without being a refuge from terrestrial collapse; a refuge need not accommodate mass evacuation. Strategic comparison must preserve the objective, horizon, and resource boundary.

\subsection{Conditional survival, construction time, and correlated failure}
Let $E$ be irreversible loss of the terrestrial population over horizon $T$, and $q_X=\Pr(\hbox{extraterrestrial continuity}\mid E)$. For an already operational population, the reduction in total-population loss is exactly $\Pr(E)q_X$, without assuming independent failures. For a developing program, write instead
\begin{equation}
 \Delta p_X(T)=\int_0^T f_E(t)\,q_X(t)\,\dd t,
 \label{eq:resilience}
\end{equation}
where $f_E(t)$ is the first-loss event density and $q_X(t)$ includes readiness at that date and survival through the ensuing disruption. A colony completed after the event contributes nothing to preventing that event's loss. With a stipulated constant event hazard giving 1\% cumulative probability in 100 years, zero independent capability for 30 years, and $q_X=0.9$ thereafter, the calculated reduction is 0.63 percentage points, not the 0.90 points obtained by assuming immediate readiness. These probabilities are a decision sensitivity, not an estimate of extinction risk or colony reliability.

A population depending on continuous imports of controllers, medicines, fuel, or validation services can have low $q_X$ after terrestrial supply failure. Shared pathogens, software, command systems, conflict, and astronomical events can correlate losses. Separation is protective only against specified mechanisms, with independent reserves, renewal paths, and containment over the relevant replacement times.

To compare a relocation capability $X$ with an alternative $A$, use the same horizon, population-loss endpoint, resource basis, and treatment of newly induced risks. In a simplified equal-severity model with event probability $p_E$, effective conditional reduction $e_A$ from prevention or terrestrial resilience, social value scale $V$ for avoiding that loss, nonresilience benefits $B$, and lifecycle costs $C$, $X$ is preferred only if
\begin{equation}
 Vp_E(q_X-e_A)+(B_X-B_A)>C_X-C_A.
 \label{eq:strategy}
\end{equation}
For development programs, replace $p_Eq_X$ by Eq.~\eqref{eq:resilience} and use a time-resolved alternative. $V$ is a normative valuation; safety and consent remain constraints. Any difference in induced catastrophe risk must be deducted from the claimed reduction. With no co-benefit difference, stipulated $(C_X-C_A)/V=10^{-3}$, $e_A=0.5$, and $q_X=0.9$ require $p_E>0.25\%$; $q_X=0.6$ raises this to 1\%. If $q_X\leq e_A$, a more costly program has no resilience-only advantage. These sensitivities identify the required evidence, not a universal threat threshold.

If the program changes terrestrial prevention, resources, or induced hazards, Eq.~\eqref{eq:resilience} must use the revised event distribution and common dependencies. Its baseline product alone is then not a causal net benefit. Scientific and option value can precede refuge readiness; net strategic benefit requires the matched comparison in Eq.~\eqref{eq:strategy}.

\subsection{Threat classes and the terrestrial counterfactual}
Climate hazards rise with warming, while mitigation and adaptation retain substantial scope \cite{IPCC2023}. Regional habitability and food losses must be compared against protected terrestrial production and recovery, which retain Earth's pressure, gravity, and accessible water. An off-world refuge and evacuation of affected populations serve different endpoints.

The Double Asteroid Redirection Test (DART) changed its target binary's period by $33\pm1$~min \cite{DART2023}; applicability of deflection depends on detection lead time and object properties. War, volcanism, and other sunlight-reducing events instead motivate protected food, energy, inventories, and distribution \cite{Baum2015}. Comparisons must retain these mechanism-specific alternatives and distinguish food disruption from irreversible population loss.

Conflict can disrupt both Earth and the launches and institutions sustaining a refuge. Biological and technological threats require containment and functional diversity as well as separation. AI-assisted engineering can improve diagnosis while introducing shared-software and misuse risks \cite{AISafety2026}; independently enforced safe states and diversified control paths belong in the same assessment.

A resilience-driven program must demonstrate a threat-specific benefit after accounting for readiness, conditional survival, common dependencies, lifecycle cost, and feasible alternatives. Years of warning cannot support a decades-long construction response unless independent capacity already exists. Experiments and serviceable infrastructure can nevertheless create an option with scientific or economic value before relocation is selected.

\subsection{Science, resources, and commercially supported residence}
Comparative planetary science, extraterrestrial materials, and indigenous-life investigations have value without a reproductive colony \cite{NASEMThriving2023,PlanetaryProtection}. Compare crewed and robotic programs at matched return, contamination control, delay, and cost. Commercial residence requires external demand for actual services; transfers among dependent projects are not additional customers.

Resource use first becomes attractive where it displaces difficult imports. For an incremental plant with total delivered mass $M_{\rm plant}$ and annual avoided and support-import flows, a screening payback is
\begin{equation}
 T_{\rm pay}=\frac{M_{\rm plant}}{\dot M_{\rm displaced}-\dot M_{\rm support}},\qquad
 \dot M_{\rm displaced}>\dot M_{\rm support}.
 \label{eq:payback}
\end{equation}
Sections~\ref{sec:fabrication} and \ref{sec:servicecost} extend this screen to accepted yield, retired-material return, intervention success, demand, and money. Cheaper Earth supply can improve the viability of residence while weakening its short-term incentive for industrial independence. Local cover, repair, and reusable structural stock can nevertheless have high avoided-mass and recovery value. Export to Earth requires a separate destination-cost and market comparison; abundance is not a business case.

Development retaining Earth support can be justified by specific research and service objectives. Relocation for civilization-scale resilience additionally requires timely healthy population regeneration and essential-product renewal, with an advantage over terrestrial alternatives. Building that option is not a substitute for protecting Earth.

\section{Feasibility, development horizons, and decisive tests}
\label{sec:roadmap}
Technology readiness level (TRL) describes a specified implementation; biological qualification concerns the supported health and functional outcomes.
Successive adult crews sustain orbital occupation with Earth deliveries and ground support \cite{NASAISS25,OIGCrew2026}. Tables~\ref{tab:stages} and~\ref{tab:architectures} distinguish the next commitments and their destination-specific gates; Table~\ref{tab:trl} identifies component evidence. Development horizons presume sustained funding, flight access, and successful precursors. They describe campaigns for missing engineering capabilities, while population regeneration additionally awaits biological evidence.

\subsection{Readiness, confidence, and the meaning of a date}
\label{sec:readinessmethod}
NASA's nine-level scale applies to a specified implementation \cite{NASAReadiness2023}. Table~\ref{tab:trl} interprets the cited component evidence; biological, clinical, and societal adequacy are assessed separately. In this paper, ``10--20 years'' denotes a funded integration objective and ``several decades'' successive development and scale-up. Both depend on facilities, flight access, and successful qualification.

A date should be attached to a dependency graph. For scenario $s$, let $\mathcal P_s$ contain its prerequisite paths, and let $\tau_a$ be the duration of activity $a$. Then
\begin{equation}
 t_{{\rm ready},s}\geq t_0+
 \max_{p\in\mathcal P_s}\sum_{a\in p}\tau_a.
 \label{eq:readinesspath}
\end{equation}
Independent activities can run in parallel; shared facilities, lineage observations, commissioning, and launch windows impose actual dependencies. Section~\ref{sec:precursortiming} calculates construction and window contributions. Unknown durations remain explicit. NASA's illustrative 2039 Mars round trip is likewise a trajectory example rather than certification of the supporting systems \cite{NASAHumansMars2026}.

Demonstration also precedes scale-up. If commissioned supported capacity $C$ grows at net rate $g_C$ \emph{after} replacement and losses, an assumed exponential trajectory gives
\begin{equation}
 \Delta t=\frac{\ln(C_1/C_0)}{g_C}.
 \label{eq:capacitygrowthtime}
\end{equation}
A tenfold increase takes 46 years at 5\%/year or 23 years at 10\%/year after replacement. Imported plants and different class-specific growth rates can change those conditional intervals. The operating trajectory must still satisfy biological, clinical, and service constraints.

\begin{table}[!htbp]
\caption{Technology-level interpretation of cited evidence using NASA's technology readiness level (TRL) definitions \cite{NASAReadiness2023}. Values describe the demonstrated scope, not agency-certified ratings of the proposed architectures. A new operating environment or scale requires a new assessment; no composite settlement or biological TRL is assigned.}
\label{tab:trl}
\begin{tabular}{p{.29\textwidth}p{.34\textwidth}p{.32\textwidth}}
\toprule
Function and evidence & Evidence-consistent readiness description & What is not established\\\midrule
Repeated orbital crew/cargo service, prepared-interface exchange, collected-water recovery \cite{OIGCrew2026,CSADextre2024,Boyce2023}. & TRL~9-type operational evidence for the particular flown implementation and supported mission envelope. & Mars transfer, altered-gravity operation, lifelong health, or unsupported manufacture of its replacements.\\
MOXIE and orbital metal deposition \cite{Hoffman2022,ESAmetal2026}. & TRL~7-type space-demonstration evidence for bounded prototypes and processes; this is not a formal project rating. & Continuous kg/hour commodity output, qualified tanks, complete part-family acceptance, or factory self-renewal.\\
Robotic assembly, excavation, and regolith forming \cite{Gregg2024,Schuler2019,Gines2023}. & Laboratory proof/breadboard evidence, approximately TRL~3--4 for tested functions; relevant-environment evidence must be evaluated separately. & Complete autonomous field-duty fleets, prepared landing surfaces, or a pressure-tight inhabited structure.\\
Kilowatt nuclear test and proposed surface reactor \cite{Poston2020,NASAFSP2025}. & Ground nuclear-system test and subsequent development target; no precise project rating inferred here. A representative complete system demonstration is a TRL~6-type gate. & A qualified 100-kW lunar plant, the 19.5-MW reference grid, fuel renewal, or the specified failure reserve.\\
Continuous occupied modular rotation and surface centrifuges. & Analytical system concepts with higher-maturity constituent technologies; approximately TRL~2--3 evidence for the integrated concepts assessed here. & A flight-qualified rotating habitat, uninterrupted clinical adequacy, emergency despin, and protected developmental residence.\\
Industrial independence and population regeneration. & No integrated hardware rating or biological TRL assigned. Required product-network and health evidence is incomplete. & A date of self-sufficiency inferred from a printer, a high local mass fraction, or an adult flight duration.\\
\bottomrule
\end{tabular}
\end{table}

\subsection{Which bottleneck governs the next commitment?}
Initial adult occupation requires destination services that survive a missed delivery and the crew's arrival limitations. At a 26-month cadence, one missed opportunity gives 52 months between arrivals; food, clinical validity, fuel, and machinery must share that horizon. Passage of a landing test or successful initial reproduction does not replace these service requirements. Expansion is bounded by the first indispensable function whose accepted additions and restoration fail to cover demand. Population regeneration adds serial developmental observations and an age structure capable of supporting dependents and replacing expertise; industrial throughput alone cannot supply them.

\subsection{Experiments that would change the assessment}
Three campaigns should proceed in parallel. \emph{Integrated terrestrial trials} should combine a complete diet, crew and crop water, gas control, dispatch, maintenance, and waste retention. Test condensate-return and cooling failures separately as in Sec.~\ref{sec:food}, along with crop loss and contamination; measure purity, reserve restoration, and accepted replacement. Include at least one complete clinical pathway with expiring inputs, interrupted updates, and a qualified substitute (Sec.~\ref{sec:medicine}). Age-structured scenarios should test temporary fertility reductions, migration interruption, dependent care, and loss of essential skills without assuming compensatory reproductive decisions.

\emph{Orbital physiology and arrival trials} should combine the graded-gravity and lineage design of Sec.~\ref{sec:biodesign} with continuous-rotation and representative post-transit task tests. Assess mobility, vision, circulatory function, protected transfer, and emergency actions under the proposed assistance scheme before withdrawing it. Replicated skeletal, ocular, cardiovascular, vestibular, and reproductive adequacy at $0.38\gzero$ against flight $1\gzero$ controls would weaken the case for additional continuous loading for those endpoints. Deficits corrected by the flight controls favor rotation; persistent deficits identify other requirements. Longitudinal human evidence and qualified obstetric, pediatric, and chronic care remain necessary for species translation.

\emph{Robotic resource and renewal trials} should demonstrate the excavation-to-placement or feed-to-accepted-product routes of Sec.~\ref{sec:isrucommission} before replacing bulk imports. Record commissioning, interventions, rejected work, and replacement of the production system's own wearing parts, controls, and metrology. Increase occupancy within measured service capacity, retain imports needed by incomplete processes, and lengthen interruption trials only when inventories are restored rather than depleted. Funding continuity, protected agency, and the threat-specific comparison in Sec.~\ref{sec:purpose} remain conditions for larger commitments. These campaigns can progress within the planning horizons above; successful component trials cannot set a date for fully autonomous population continuity.

\subsection{Stage gates that determine whether a scenario advances}
\label{sec:stagegates}
Table~\ref{tab:stages} summarizes the decisions. Each commitment retains the preceding requirements: commissioned capacity before departure, health and reserve restoration before occupancy grows, and essential-product renewal before Earth support is withdrawn. A failed gate specifies which capability, retained supply, or architecture must change.

\section{Conclusions}
\label{sec:conclusions}

Sustained habitation becomes feasible in stages, with a different limiting requirement at each commitment. Earth-supported adult occupation is established in low Earth orbit; protected destinations and progressively localized supply have identifiable engineering routes. Independent population continuity additionally requires healthy development and reproduction, recoverable essential services, and renewal of the equipment that sustains them. The calculations support the following design choices and advancement criteria.

\emph{Local construction and qualified production.} Bulk cover and civil works should use destination material where surveyed resources, excavation, and structural acceptance permit. The specified 100-t/year product line reduces ten-year Earth delivery from 1,000 to 419~t after a three-year ramp, while retaining 16.1~t/year of inputs and support at rated output. This is a useful intermediate capability: localization displaces freight by installing maintained productive capacity. Accepted output must exceed net retirement before capacity grows. For the reference stock, 3,000~t/year reaches the target in about 13 rather than 9 years when retirement is included; 1,000~t/year never reaches it at constant throughput. The transferable limit is replacement-adjusted, product-compatible supply, not a universal population ceiling.

\emph{Power and heat as coupled constraints.} Manufacturing the full equipment-renewal stream locally raises the specified 1,000-resident mean demand from 11.8 to 19.5~MW; retention-time sensitivity gives 17--26~MW at fixed processing intensity. The 30-MW concurrent-power allocation accepts the reference continuous-production schedule but excludes half-time manufacture during crop illumination. Scheduling and annual energy are therefore insufficient without matching machine, generator, and cooling capacity. Under the stated fission conversion model, 19.5~MW electrical requires 95.5~MW thermal input. Assigning electrical dissipation to 300~K and conversion heat to 650~K gives 92,400~m$^2$ of total emitting area in the Mars-B case; routing the specified fraction of manufacturing heat to 450~K reduces this to 70,500~m$^2$. Heat quality and utility availability materially change the installation.

\emph{Solar baseline, fission option, and hybrid supply.} Solar generation is the reference option to evaluate at well-illuminated sites with manageable storage and array renewal; its advantage depends on end-of-life output and the longest credible generation interruption. At the stated storage performance, sustaining 10~MW for 336~h requires 13,440~t of cells alone. Long darkness, severe attenuation, or weak outer-system sunlight therefore strengthen the case for fission, subject to qualified conversion, shielding, heat rejection, fuel, and replacement. A hybrid can assign firm essential loads to fission and flexible production to solar only when curtailed work and subsequent catch-up fit the same inventories and utilities. Neither source is universally preferred: compare delivered service, survival of the specified outage, and lifecycle burden before selecting the mix (Sec.~\ref{sec:dispatch}).

\emph{Protected occupation and controllable gravity.} The published Martian transport profile requires about 63,000~t/ha of planar cover for the selected 10-mSv/year galactic-cosmic-ray allocation; occupied time outside protection and acute solar events impose separate requirements. That mass favors local emplacement and robotic exterior work, not a universal shield thickness or biological safety limit. Earth-like acceleration at 2~rpm requires a 224-m rotation radius, but does not require pressurizing a vessel of that radius. Modular rotating accommodation preserves a controllable-loading option while native-gravity adequacy remains unresolved. Pressure boundaries, docking, emergency refuge, and the complete exposure history still require qualification.

\emph{Recovery and clinical continuity.} Ten-day processor outages every 35 days require $k_w\geq1.4$ to prevent accumulating wastewater; strict excess restores inherited margin. Larger stores cannot correct persistent insufficient throughput. Crop-water return, atmospheric moisture removal, and personal-water processing require different surviving equipment and buffers. A diagnostic instrument likewise provides no continuing clinical service after its necessary inputs expire unless a qualified replacement or alternative exists. Compartmentation, independent utility paths, valid supplies, and tested recovery must be specified by failed function, with restoration completed before its usable reserve is depleted.

\emph{Modular refresh and robotic servicing.} Retain qualified pressure, structural, and thermal assets while replacing accessible life-limiting functions. In the 100-unit example, modular exchange reduces recurrent imports from 4.0 to 1.74~t/year including servicing support; the additional 10-t installation repays its mass in about 4.4 years. Whole-unit replacement remains appropriate when interfaces, damage extent, restored reliability, or demand invalidate that advantage. An unattended servicing fleet needs verified recovery of failed servicers, independent tools, and qualified substitutes. Reducing repair time eightfold gives only a 1.36-fold loss-frequency improvement in the example with the specified common-cause contribution: duplicate robots sharing the failed service do not supply independent recovery.

\emph{Radiation-tolerant compute and requalification.} Protection of residents does not qualify control electronics. Parts, boards, software recovery, and independent safe states must be verified against device dose, displacement damage, and single-event effects for the intended mission interval. Planned processor replacement can limit cumulative exposure, but cannot substitute for tolerance of a destructive event before exchange. Each changed hardware generation requires renewed evidence; common interfaces preserve serviceability, not automatic radiation qualification (Sec.~\ref{sec:acceptance}). No universal compute-refresh cadence follows from the present model.

\emph{Fleet scale and successive hardware generations.} Shared workshops and service hubs gain value from reachable, compatible demand rather than satellite or habitat count alone. The stated financial comparison requires about 2.3 interventions/year when avoiding an \$8-million replacement, versus 18.7 when avoiding a \$2-million replacement; these are price sensitivities, not market forecasts. Constellation geometry, transfer cost, retained interfaces, and continued demand determine whether support can be pooled. Comparisons spanning generations must charge renewed active hardware, mixed-generation spares, testing, software migration, and eventual retirement, while crediting shared structures only for the service they actually retain. Faster technology turnover can otherwise consume the apparent gain from cheaper launch or longer structural life.

\emph{Healthy population renewal.} Radiation transport and rotation establish environmental control options, not complete-life-cycle adequacy. The critical evidence concerns gestation, development, skeletal and circulatory function, vision, subsequent fertility, infection control, and aging under the same protected exposure. Results at $0.38\gzero$ against matched flight $1\gzero$ controls can discriminate native loading from additional continuous gravity; longitudinal human observations remain necessary for translation. Contemporary healthspan and a complete diet are the baseline. Age structure, genetic diversity, dependent care, and succession of expertise must survive cohort disturbances. Neither a founder headcount nor a reproductive biobank independently establishes that continuity.

\emph{Stage gates and development commitments.} Advance from relevant-environment trials to verified destination services, then increase occupancy within demonstrated nutrition, clinical, power, and recovery margins. Orbital facilities, robotic resource pilots, bounded production, and controlled-gravity investigations are credible 10--20-year objectives with sustained investment; maintained industrial grids and broad renewal networks require further multidecade integration. The three-year ramp and 13-year stock calculation start with installed capability, not with research funding. Reproductive and industrial autonomy remains undated. Run engineering and life-cycle campaigns in parallel, retain essential imports until their functions are renewable, and test multiple equipment replacements rather than endurance purchased solely with initial stocks (Table~\ref{tab:stages}).

\emph{Institutional and economic continuity.} Sleep, privacy, resident agency, voluntary reproductive choices, independent clinical and safety authority, and finance through maintenance and operator failure sustain the service system as directly as its hardware. Strategic resilience additionally requires readiness before a specified threat, survival of shared dependencies, and comparison with terrestrial prevention and recovery at matched cost and outcome. The supported decision is progressive, recoverable habitation: increase occupancy only within demonstrated health and net-capacity margins, and reduce Earth support only as qualified renewal replaces it.

\section*{Acknowledgments}
The work described here was carried out at the Jet Propulsion Laboratory, California Institute of Technology, Pasadena, California, under a contract with the National Aeronautics and Space Administration. \copyright\ 2026. California Institute of Technology. Government sponsorship acknowledged.

\appendix

\section{Input provenance and interpretation of uncertainty}
\label{app:repro}

The common scenario, ideal trajectory and deployment estimates, thermochemical bounds, reproductive and independence screens, surface-centrifuge kinematics, gas buffers, and periodic-backlog recurrence are calculations, not additional experimental observations. Constants are $g_0=9.80665$~m/s$^2$, $R=8.314462618$~J/(mol\,K), and $\sigma=5.670374419\times10^{-8}$~W/(m$^2$K$^4$); throughput accounting uses 365 days/year. Hardware quantities are tonnes per resident unless otherwise stated; conversion to kilograms is explicit in Eq.~\eqref{eq:capacityscreen}. Unrounded reference electricity is $p_h=11.784722$~kW/resident, manufacturing is 7.748288~MW per 1,000 residents, and the total is 19.533010~MW. Text and tables round these computed values according to their scenario interpretation.

Table~\ref{tab:inputs} separates measured or published inputs from design selections. Two endpoint calculations make the main service and capital intervals reproducible. The commissioning pilot in Eq.~\eqref{eq:productionramp} retains unrounded $M_F=122.83105$~t and $s_F=6.14155$~t/year from Eq.~\eqref{eq:isrumass}; its 419.2466-t ten-year delivery and 3.0746-year mass crossing are calculations under a selected three-year ramp. Local equipment-production fractions change site processing, not the already specified farm and gas/water loads. Predeployment time, cargo batches, additional reserves, and changing process coefficients require their own histories. With crop area $a_c$ in m$^2$/resident, lighting $p_L$ in kW/m$^2$, 16 h/day illumination, 4.5~kg/(m$^2$\,day) evapotranspiration, 2.44~MJ/kg latent heat, and coefficient of performance 3,
\begin{equation*}
 p_h=3.0+a_c\Big(\frac{16p_L}{24}+\frac{4.5\times2.44}{86.4\times3}\Big)\quad\mathrm{kW/resident}.
\end{equation*}
The paired endpoints $(a_c,p_L)=(30,0.15)$ and $(100,0.30)$ give 7.2708 and 27.2361~kW/resident. The constant 86.4 converts MJ/day to kW; transpiration scales with area only in this declared sensitivity.

For Table~\ref{tab:capital}, let $a_{\rm th}=92.4247$~m$^2$/resident be the total Mars-B emitting area at $p_\star=19.53301$~kW/resident. At fixed conversion efficiency and radiative boundaries, the thermal allocation is
\begin{equation*}
 k_{\rm th}=\frac{\mu_A a_{\rm th}(p/p_\star)+25p}{10^3}+k_s
 \quad\mathrm{t/resident},
\end{equation*}
where $\mu_A$ is kg per m$^2$ of emitting surface, $p$ is kW/resident, and $k_s$ is support tonnes/resident. The lower selection $(p,\mu_A,k_s)=(p_\star,5,0.1)$ gives 1.050~t/resident; $(50,10,0.3)$ gives 3.916~t/resident. Rounding yields the tabulated 1.1--3.9 interval. These allocations do not establish contingency independence.

The ISRU calculations use matched useful products, with $\mu$ distinguishing a heavier local alternative, $\iota$ recurring imported process input per accepted output, and the power/cooling installation explicitly added to the seed mass. The 100-t pilot, its rate envelope, availability, and component-retirement coefficients are design selections. Prototype excavation duty, laboratory material strength, and modeled oxygen energy are separate evidence, not calibration of that complete plant. Grade and recovery sensitivities do not represent surveyed reserves.

The retained readiness calculations use the original coefficients: five- and eight-year ramps give 509.2466 and 644.2466~t by year ten; 26-month opportunities and 259-day legs give 2.8763- and 5.0429-year precursor-to-crew intervals for the stated validation periods. Biological interpretation is triangulated against the systematic review and developmental roadmap \cite{Gimunova2024,Jain2023}.

The integrated crop-fault example assumes ten equal sectors, each with $J=22.5$~t/day at the reference flux. Return-pump and condensation failures have different surviving services; no new water loss is inferred from temporary relocation to storage. The humidity estimate uses $V_f=25,000$~m$^3$, $T=298.15$~K, $\varphi_0=0.60$, $\varphi_{\max}=0.80$, and $p_{\rm sat}=10^{5.40221-1838.675/(T-31.737)}$ bar \cite{NISTWaterVapor}. Ideal-gas conversion gives $\rho_{v,\rm sat}=0.023014$~kg/m$^3$ (rounded); constant initial flux and absence of other sinks define the short-time estimate. Full plant dynamics, food yield, and shared-failure response require testing.

The demographic perturbation uses 75 annual classes, initially $n_x=1,000/75$, unit survival through age 74 and removal at age 75, and $n_{0,t+1}=0.05\sum_{x=20}^{39}n_{x,t}$; only the first five birth transitions are multiplied by 0.75. Afterward the initial age-specific coefficients resume, not a forced birth count. All reported counts are deterministic expectations; the age ratio is not a care-hour conversion. The diagnostic validity calculation assumes a fresh six-month batch at interruption onset and valid storage; transport age or a storage excursion can shorten that interval. The calculation archive contains independent cohort, farm-flow, humidity, and expiry checks.

\begin{samepage}
The joint power domain holds class masses, yield, and services fixed and combines retention and gross-energy intervals. Each $q_K$ within the endpoint sums is attainable by interpolating the six retirement rates. The dispatch screen keeps installed capacity fixed. The industrial benchmark reproduces the published initial ledger and nominal throughput; the acceptance factor and fixed-active-power rule are added sensitivities. Exposure-duration cases, inputs, and independent numerical checks are archived.

Measured inputs carry their reported uncertainty; published models retain their stated geometry and domain; engineering selections define conditional comparisons; and acceptance criteria state the required performance. Scenario ranges support deterministic sensitivity and threshold tests, not confidence intervals. Probabilistic propagation requires justified input distributions and dependencies. The present numerical examples neither calibrate a complete settlement nor optimize its development program. Dates refer to conditional funded campaigns, while experimental durations and commissioning times refer to their own stated origins. Radiation dose metrics, biological species and endpoints, product specifications, and simultaneously available utilities must remain explicit wherever they change an inference.
\end{samepage}

\begin{table}[!htbp]
\caption{Provenance and scope of the main inputs. Design intervals are not confidence intervals.}
\label{tab:inputs}
\begin{tabular}{p{.17\textwidth}p{.30\textwidth}p{.48\textwidth}}
\toprule
Domain & Observation or published model & Engineering selection or unresolved quantity\\ \midrule
Radiation & RAD dose equivalents; HZETRN subsurface profile; historical-event organ-dose tables. & 10-mSv/year allocation, residence schedule, actual geometry and site transport, lifelong developmental risk.\\
Pressure and rotation & Fluid/circulatory, sensorimotor, ocular, and musculoskeletal responses; established mechanics. & 100-kPa pressure, steel-like density and 200-MPa allowable stress; population tolerance and complete structural design.\\
Chemistry and water & Reference metabolic rates; demonstrated recovery of collected ISS water. & Complete collection, throughput, reserves, repair duration, purity budget, and coupled performance at settlement scale.\\
Food and thermal transfer & Controlled crop studies and the cited wheat transpiration flux. & Joint diet, area, illumination, dehumidification, and complete-farm hardware; crop-specific margins after failure.\\
Electricity and renewal & Engineering accounting methods and component precedents. & 11.78-kW/resident service reference; six class inventories/retention times; 1\% growth, yield, process intensity, and fraction actually made at the site.\\
Heat and storage & Radiative and thermodynamic relations; surface-energy observations. & Half-Carnot conversion, two radiator temperatures, Mars A/B boundaries, installed mass and contingency performance.\\
Population and reliability & Genetic and statistical model structure; lineage-specific experiments. & Effective population, age-specific fertility/survival, care, clinical margins, rare-event targets, and recovery distributions.\\
Servicing and fabrication & Mission reports, specific ground/flight process demonstrations, and aerospace qualification practice. & Compatible interventions, restored reliability, complete plant mass, material return, rejection, intervention hours, and throughput.\\
Access and strategy & Published launch quotation, operational audits, and historical mission-cost analysis. & Fixed-architecture elasticity, hypothetical destination prices, relocation cadence, hazard timing and conditional survival; not market or threat forecasts.\\
Lifecycle economics & Financial and orbital-transfer accounting. & Hypothetical common-date costs, demand, discount rate, collateral loss, and material payback; not market forecasts.\\
\bottomrule
\end{tabular}
\end{table}

% Flush the final provenance float before the bibliography.
%\clearpage
% APS/Physical Review numerical citation style is used throughout; avoid author-date citation prose.
\bibliographystyle{apsrev4-2}
%\bibliography{human_settlement}

\begin{thebibliography}{143}%
\makeatletter
\providecommand \@ifxundefined [1]{%
 \@ifx{#1\undefined}
}%
\providecommand \@ifnum [1]{%
 \ifnum #1\expandafter \@firstoftwo
 \else \expandafter \@secondoftwo
 \fi
}%
\providecommand \@ifx [1]{%
 \ifx #1\expandafter \@firstoftwo
 \else \expandafter \@secondoftwo
 \fi
}%
\providecommand \natexlab [1]{#1}%
\providecommand \enquote  [1]{``#1''}%
\providecommand \bibnamefont  [1]{#1}%
\providecommand \bibfnamefont [1]{#1}%
\providecommand \citenamefont [1]{#1}%
\providecommand \href@noop [0]{\@secondoftwo}%
\providecommand \href [0]{\begingroup \@sanitize@url \@href}%
\providecommand \@href[1]{\@@startlink{#1}\@@href}%
\providecommand \@@href[1]{\endgroup#1\@@endlink}%
\providecommand \@sanitize@url [0]{\catcode `\\12\catcode `\$12\catcode
  `\&12\catcode `\#12\catcode `\^12\catcode `\_12\catcode `\%12\relax}%
\providecommand \@@startlink[1]{}%
\providecommand \@@endlink[0]{}%
\providecommand \url  [0]{\begingroup\@sanitize@url \@url }%
\providecommand \@url [1]{\endgroup\@href {#1}{\urlprefix }}%
\providecommand \urlprefix  [0]{URL }%
\providecommand \Eprint [0]{\href }%
\providecommand \doibase [0]{https://doi.org/}%
\providecommand \selectlanguage [0]{\@gobble}%
\providecommand \bibinfo  [0]{\@secondoftwo}%
\providecommand \bibfield  [0]{\@secondoftwo}%
\providecommand \translation [1]{[#1]}%
\providecommand \BibitemOpen [0]{}%
\providecommand \bibitemStop [0]{}%
\providecommand \bibitemNoStop [0]{.\EOS\space}%
\providecommand \EOS [0]{\spacefactor3000\relax}%
\providecommand \BibitemShut  [1]{\csname bibitem#1\endcsname}%
\let\auto@bib@innerbib\@empty
%</preamble>
\bibitem [{\citenamefont {{NASA}}(2025{\natexlab{a}})}]{NASAISS25}%
  \BibitemOpen
  \bibfield  {author} {\bibinfo {author} {\bibnamefont {{NASA}}},\ }\href
  {https://www.nasa.gov/international-space-station/iss25/} {\bibinfo {title}
  {International space station: 25 years}} (\bibinfo {year}
  {2025}{\natexlab{a}}),\ \bibinfo {note} {mission history of continuous
  occupation by successive crews}\BibitemShut {NoStop}%
\bibitem [{\citenamefont {{NASA Office of Inspector
  General}}(2026{\natexlab{a}})}]{OIGCrew2026}%
  \BibitemOpen
  \bibfield  {author} {\bibinfo {author} {\bibnamefont {{NASA Office of
  Inspector General}}},\ }\href
  {https://oig.nasa.gov/audits/nasas-management-of-its-commercial-crew-program/}
  {\emph {\bibinfo {title} {{NASA's Management of Its Commercial Crew
  Program}}}},\ \bibinfo {type} {Tech. Rep.}\ \bibinfo {number} {IG-26-011}\
  (\bibinfo  {institution} {NASA},\ \bibinfo {year} {2026})\BibitemShut
  {NoStop}%
\bibitem [{\citenamefont {Jones}(2018)}]{Jones2018}%
  \BibitemOpen
  \bibfield  {author} {\bibinfo {author} {\bibfnamefont {H.~W.}\ \bibnamefont
  {Jones}},\ }\bibfield  {title} {\bibinfo {title} {The recent large reduction
  in space launch cost},\ }in\ \href
  {https://ntrs.nasa.gov/citations/20200001093} {\emph {\bibinfo {booktitle}
  {48th International Conference on Environmental Systems}}},\ \bibinfo {series
  and number} {\bibinfo {number} {ICES-2018-81}}\ (\bibinfo {year} {2018})\
  \bibinfo {note} {{NASA} NTRS 20200001093}\BibitemShut {NoStop}%
\bibitem [{\citenamefont {{Canadian Space Agency}}(2024)}]{CSADextre2024}%
  \BibitemOpen
  \bibfield  {author} {\bibinfo {author} {\bibnamefont {{Canadian Space
  Agency}}},\ }\href {https://www.asc-csa.gc.ca/eng/iss/dextre/about.asp}
  {\bibinfo {title} {About {Dextre}}} (\bibinfo {year} {2024}),\ \bibinfo
  {note} {official operational description; modified 4 June 2024}\BibitemShut
  {NoStop}%
\bibitem [{\citenamefont {{Northrop Grumman}}(2026)}]{SpaceLogistics2026}%
  \BibitemOpen
  \bibfield  {author} {\bibinfo {author} {\bibnamefont {{Northrop Grumman}}},\
  }\href
  {https://www.northropgrumman.com/what-we-do/space/space-logistics-services}
  {\bibinfo {title} {{SpaceLogistics}: mission status and services}} (\bibinfo
  {year} {2026}),\ \bibinfo {note} {primary operator report; explicit mission
  updates distinguished from future-service descriptions}\BibitemShut {NoStop}%
\bibitem [{\citenamefont {Boyce}\ \emph {et~al.}(2023)\citenamefont {Boyce},
  \citenamefont {Joyce}, \citenamefont {Pasadilla} \emph {et~al.}}]{Boyce2023}%
  \BibitemOpen
  \bibfield  {author} {\bibinfo {author} {\bibfnamefont {S.~P.}\ \bibnamefont
  {Boyce}}, \bibinfo {author} {\bibfnamefont {C.~J.}\ \bibnamefont {Joyce}},
  \bibinfo {author} {\bibfnamefont {P.}~\bibnamefont {Pasadilla}}, \emph
  {et~al.},\ }\bibfield  {title} {\bibinfo {title} {Brine processor assembly: A
  year of successful operation on the {International Space Station}},\ }in\
  \href {https://doi.org/10.32865/2346/94707} {\emph {\bibinfo {booktitle}
  {52nd International Conference on Environmental Systems}}},\ \bibinfo {series
  and number} {\bibinfo {number} {ICES-2023-292}}\ (\bibinfo {year}
  {2023})\BibitemShut {NoStop}%
\bibitem [{\citenamefont {Gregg}\ \emph {et~al.}(2024)\citenamefont {Gregg},
  \citenamefont {Catanoso}, \citenamefont {Formoso} \emph
  {et~al.}}]{Gregg2024}%
  \BibitemOpen
  \bibfield  {author} {\bibinfo {author} {\bibfnamefont {C.~E.}\ \bibnamefont
  {Gregg}}, \bibinfo {author} {\bibfnamefont {D.}~\bibnamefont {Catanoso}},
  \bibinfo {author} {\bibfnamefont {O.~I.~B.}\ \bibnamefont {Formoso}}, \emph
  {et~al.},\ }\bibfield  {title} {\bibinfo {title} {Ultralight, strong, and
  self-reprogrammable mechanical metamaterials},\ }\href
  {https://doi.org/10.1126/scirobotics.adi2746} {\bibfield  {journal} {\bibinfo
   {journal} {Science Robotics}\ }\textbf {\bibinfo {volume} {9}},\ \bibinfo
  {pages} {eadi2746} (\bibinfo {year} {2024})}\BibitemShut {NoStop}%
\bibitem [{\citenamefont {{European Space Agency}}(2026)}]{ESAmetal2026}%
  \BibitemOpen
  \bibfield  {author} {\bibinfo {author} {\bibnamefont {{European Space
  Agency}}},\ }\href
  {https://www.esa.int/ESA_Multimedia/Videos/2026/07/3D-printed_metal}
  {\bibinfo {title} {{3D}-printed metal}} (\bibinfo {year} {2026}),\ \bibinfo
  {note} {2 July 2026; fourth ISS metal-printing sample retrieval}\BibitemShut
  {NoStop}%
\bibitem [{\citenamefont {Ewert}\ \emph {et~al.}(2022)\citenamefont {Ewert},
  \citenamefont {Chen},\ and\ \citenamefont {Powell}}]{BVAD2022}%
  \BibitemOpen
  \bibfield  {author} {\bibinfo {author} {\bibfnamefont {M.~K.}\ \bibnamefont
  {Ewert}}, \bibinfo {author} {\bibfnamefont {T.~T.}\ \bibnamefont {Chen}},\
  and\ \bibinfo {author} {\bibfnamefont {C.~D.}\ \bibnamefont {Powell}},\
  }\href {https://ntrs.nasa.gov/citations/20210024855} {\emph {\bibinfo {title}
  {Life Support Baseline Values and Assumptions Document}}},\ \bibinfo {type}
  {Tech. Rep.}\ \bibinfo {number} {NASA/TP-2015-218570/REV2}\ (\bibinfo
  {institution} {NASA},\ \bibinfo {year} {2022})\BibitemShut {NoStop}%
\bibitem [{\citenamefont {Wheeler}(2017)}]{Wheeler2017}%
  \BibitemOpen
  \bibfield  {author} {\bibinfo {author} {\bibfnamefont {R.~M.}\ \bibnamefont
  {Wheeler}},\ }\bibfield  {title} {\bibinfo {title} {Agriculture for space:
  People and places paving the way},\ }\href
  {https://doi.org/10.1515/opag-2017-0002} {\bibfield  {journal} {\bibinfo
  {journal} {Open Agriculture}\ }\textbf {\bibinfo {volume} {2}},\ \bibinfo
  {pages} {14--32} (\bibinfo {year} {2017})}\BibitemShut {NoStop}%
\bibitem [{\citenamefont {Do}\ \emph {et~al.}(2016)\citenamefont {Do},
  \citenamefont {Owens}, \citenamefont {Ho}, \citenamefont {Schreiner},\ and\
  \citenamefont {de~Weck}}]{Do2016}%
  \BibitemOpen
  \bibfield  {author} {\bibinfo {author} {\bibfnamefont {S.}~\bibnamefont
  {Do}}, \bibinfo {author} {\bibfnamefont {A.}~\bibnamefont {Owens}}, \bibinfo
  {author} {\bibfnamefont {K.}~\bibnamefont {Ho}}, \bibinfo {author}
  {\bibfnamefont {S.}~\bibnamefont {Schreiner}},\ and\ \bibinfo {author}
  {\bibfnamefont {O.~L.}\ \bibnamefont {de~Weck}},\ }\bibfield  {title}
  {\bibinfo {title} {An independent assessment of the technical feasibility of
  the {Mars One} mission plan---updated analysis},\ }\href
  {https://doi.org/10.1016/j.actaastro.2015.11.025} {\bibfield  {journal}
  {\bibinfo  {journal} {Acta Astronautica}\ }\textbf {\bibinfo {volume}
  {120}},\ \bibinfo {pages} {192--228} (\bibinfo {year} {2016})}\BibitemShut
  {NoStop}%
\bibitem [{\citenamefont {Ho}\ \emph {et~al.}(2022)\citenamefont {Ho},
  \citenamefont {Makrygiorgos}, \citenamefont {Hill},\ and\ \citenamefont
  {Berliner}}]{Ho2022}%
  \BibitemOpen
  \bibfield  {author} {\bibinfo {author} {\bibfnamefont {D.}~\bibnamefont
  {Ho}}, \bibinfo {author} {\bibfnamefont {G.}~\bibnamefont {Makrygiorgos}},
  \bibinfo {author} {\bibfnamefont {A.}~\bibnamefont {Hill}},\ and\ \bibinfo
  {author} {\bibfnamefont {A.~J.}\ \bibnamefont {Berliner}},\ }\bibfield
  {title} {\bibinfo {title} {Towards an extension of equivalent system mass for
  human exploration missions on {Mars}},\ }\href
  {https://doi.org/10.1038/s41526-022-00214-7} {\bibfield  {journal} {\bibinfo
  {journal} {npj Microgravity}\ }\textbf {\bibinfo {volume} {8}},\ \bibinfo
  {pages} {30} (\bibinfo {year} {2022})}\BibitemShut {NoStop}%
\bibitem [{\citenamefont {Owens}\ and\ \citenamefont
  {de~Weck}(2016)}]{Owens2016}%
  \BibitemOpen
  \bibfield  {author} {\bibinfo {author} {\bibfnamefont {A.~C.}\ \bibnamefont
  {Owens}}\ and\ \bibinfo {author} {\bibfnamefont {O.~L.}\ \bibnamefont
  {de~Weck}},\ }\bibfield  {title} {\bibinfo {title} {Limitations of
  reliability for long-endurance human spaceflight},\ }in\ \href
  {https://doi.org/10.2514/6.2016-5308} {\emph {\bibinfo {booktitle} {AIAA
  SPACE 2016}}},\ \bibinfo {series and number} {\bibinfo {number} {AIAA
  2016-5308}}\ (\bibinfo {year} {2016})\BibitemShut {NoStop}%
\bibitem [{\citenamefont {Owens}\ and\ \citenamefont
  {de~Weck}(2018)}]{Owens2018}%
  \BibitemOpen
  \bibfield  {author} {\bibinfo {author} {\bibfnamefont {A.~C.}\ \bibnamefont
  {Owens}}\ and\ \bibinfo {author} {\bibfnamefont {O.~L.}\ \bibnamefont
  {de~Weck}},\ }\bibfield  {title} {\bibinfo {title} {International space
  station operational experience and its impacts on future mission
  supportability},\ }in\ \href
  {https://ntrs.nasa.gov/api/citations/20200002558/downloads/20200002558.pdf}
  {\emph {\bibinfo {booktitle} {48th International Conference on Environmental
  Systems}}},\ \bibinfo {series and number} {\bibinfo {number}
  {ICES-2018-198}}\ (\bibinfo {year} {2018})\BibitemShut {NoStop}%
\bibitem [{\citenamefont {Metzger}\ \emph {et~al.}(2013)\citenamefont
  {Metzger}, \citenamefont {Muscatello}, \citenamefont {Mueller},\ and\
  \citenamefont {Mantovani}}]{Metzger2013}%
  \BibitemOpen
  \bibfield  {author} {\bibinfo {author} {\bibfnamefont {P.~T.}\ \bibnamefont
  {Metzger}}, \bibinfo {author} {\bibfnamefont {A.}~\bibnamefont {Muscatello}},
  \bibinfo {author} {\bibfnamefont {R.~P.}\ \bibnamefont {Mueller}},\ and\
  \bibinfo {author} {\bibfnamefont {J.}~\bibnamefont {Mantovani}},\ }\bibfield
  {title} {\bibinfo {title} {Affordable, rapid bootstrapping of the space
  industry and solar system civilization},\ }\href
  {https://doi.org/10.1061/(ASCE)AS.1943-5525.0000236} {\bibfield  {journal}
  {\bibinfo  {journal} {Journal of Aerospace Engineering}\ }\textbf {\bibinfo
  {volume} {26}},\ \bibinfo {pages} {18--29} (\bibinfo {year}
  {2013})}\BibitemShut {NoStop}%
\bibitem [{\citenamefont {Salotti}(2020)}]{Salotti2020}%
  \BibitemOpen
  \bibfield  {author} {\bibinfo {author} {\bibfnamefont {J.-M.}\ \bibnamefont
  {Salotti}},\ }\bibfield  {title} {\bibinfo {title} {Minimum number of
  settlers for survival on another planet},\ }\href
  {https://doi.org/10.1038/s41598-020-66740-0} {\bibfield  {journal} {\bibinfo
  {journal} {Scientific Reports}\ }\textbf {\bibinfo {volume} {10}},\ \bibinfo
  {pages} {9700} (\bibinfo {year} {2020})}\BibitemShut {NoStop}%
\bibitem [{\citenamefont {Tsuji}\ \emph {et~al.}(2026)\citenamefont {Tsuji},
  \citenamefont {Fujita}, \citenamefont {Hayashi} \emph {et~al.}}]{Tsuji2026}%
  \BibitemOpen
  \bibfield  {author} {\bibinfo {author} {\bibfnamefont {R.}~\bibnamefont
  {Tsuji}}, \bibinfo {author} {\bibfnamefont {R.}~\bibnamefont {Fujita}},
  \bibinfo {author} {\bibfnamefont {T.}~\bibnamefont {Hayashi}}, \emph
  {et~al.},\ }\bibfield  {title} {\bibinfo {title} {{0.33g} mitigates muscle
  atrophy while {0.67g} preserves muscle function and myofiber type composition
  in mice during spaceflight},\ }\href {https://doi.org/10.1126/sciadv.aed2258}
  {\bibfield  {journal} {\bibinfo  {journal} {Science Advances}\ }\textbf
  {\bibinfo {volume} {12}},\ \bibinfo {pages} {eaed2258} (\bibinfo {year}
  {2026})}\BibitemShut {NoStop}%
\bibitem [{\citenamefont {{National Academies of Sciences, Engineering, and
  Medicine}}(2023)}]{NASEMThriving2023}%
  \BibitemOpen
  \bibfield  {author} {\bibinfo {author} {\bibnamefont {{National Academies of
  Sciences, Engineering, and Medicine}}},\ }\href
  {https://doi.org/10.17226/26750} {\emph {\bibinfo {title} {Thriving in Space:
  Ensuring the Future of Biological and Physical Sciences Research: A Decadal
  Survey for 2023--2032}}}\ (\bibinfo  {publisher} {The National Academies
  Press},\ \bibinfo {address} {Washington, DC},\ \bibinfo {year}
  {2023})\BibitemShut {NoStop}%
\bibitem [{\citenamefont {{National Academies of Sciences, Engineering, and
  Medicine}}(2021{\natexlab{a}})}]{NASEMRadiation2021}%
  \BibitemOpen
  \bibfield  {author} {\bibinfo {author} {\bibnamefont {{National Academies of
  Sciences, Engineering, and Medicine}}},\ }\href
  {https://doi.org/10.17226/26155} {\emph {\bibinfo {title} {{Space Radiation
  and Astronaut Health: Managing and Communicating Cancer Risks}}}}\ (\bibinfo
  {publisher} {The National Academies Press},\ \bibinfo {address} {Washington,
  DC},\ \bibinfo {year} {2021})\BibitemShut {NoStop}%
\bibitem [{\citenamefont {{SpaceX}}(2026)}]{SpaceX2026Pricing}%
  \BibitemOpen
  \bibfield  {author} {\bibinfo {author} {\bibnamefont {{SpaceX}}},\ }\href
  {https://www.spacex.com/assets/media/Capabilities\%26Services.pdf} {\bibinfo
  {title} {Capabilities \& services}} (\bibinfo {year} {2026}),\ \bibinfo
  {note} {published standard payment plan through 2026; maximum performance
  separately labeled fully expendable}\BibitemShut {NoStop}%
\bibitem [{\citenamefont {{NASA Office of Inspector
  General}}(2026{\natexlab{b}})}]{OIGHLS2026}%
  \BibitemOpen
  \bibfield  {author} {\bibinfo {author} {\bibnamefont {{NASA Office of
  Inspector General}}},\ }\href
  {https://oig.nasa.gov/audits/nasas-management-of-the-human-landing-system-contracts/}
  {\emph {\bibinfo {title} {{NASA}'s Management of the Human Landing System
  Contracts}}},\ \bibinfo {type} {Tech. Rep.}\ \bibinfo {number} {IG-26-004}\
  (\bibinfo  {institution} {National Aeronautics and Space Administration},\
  \bibinfo {year} {2026})\BibitemShut {NoStop}%
\bibitem [{\citenamefont {{NASA Office of Inspector
  General}}(2026{\natexlab{c}})}]{OIGInfrastructure2026}%
  \BibitemOpen
  \bibfield  {author} {\bibinfo {author} {\bibnamefont {{NASA Office of
  Inspector General}}},\ }\href
  {https://oig.nasa.gov/audits/nasas-launch-infrastructure/} {\emph {\bibinfo
  {title} {{NASA}'s Launch Infrastructure}}},\ \bibinfo {type} {Tech. Rep.}\
  \bibinfo {number} {IG-26-010}\ (\bibinfo  {institution} {NASA},\ \bibinfo
  {year} {2026})\BibitemShut {NoStop}%
\bibitem [{\citenamefont {Polsgrove}\ and\ \citenamefont
  {Dwyer-Cianciolo}(2016)}]{Cianciolo2016}%
  \BibitemOpen
  \bibfield  {author} {\bibinfo {author} {\bibfnamefont {T.}~\bibnamefont
  {Polsgrove}}\ and\ \bibinfo {author} {\bibfnamefont {A.~M.}\ \bibnamefont
  {Dwyer-Cianciolo}},\ }\bibfield  {title} {\bibinfo {title} {Human {Mars}
  entry, descent and landing architecture study overview},\ }in\ \href
  {https://doi.org/10.2514/6.2016-5494} {\emph {\bibinfo {booktitle} {AIAA
  SPACE 2016}}},\ \bibinfo {series and number} {\bibinfo {number} {AIAA
  2016-5494}}\ (\bibinfo {year} {2016})\BibitemShut {NoStop}%
\bibitem [{\citenamefont {{NASA Jet Propulsion
  Laboratory}}(2026{\natexlab{a}})}]{JPLPlanets}%
  \BibitemOpen
  \bibfield  {author} {\bibinfo {author} {\bibnamefont {{NASA Jet Propulsion
  Laboratory}}},\ }\href {https://ssd.jpl.nasa.gov/planets/phys_par.html}
  {\bibinfo {title} {Planetary physical parameters}} (\bibinfo {year}
  {2026}{\natexlab{a}})\BibitemShut {NoStop}%
\bibitem [{\citenamefont {{National Academies of Sciences, Engineering, and
  Medicine}}(2021{\natexlab{b}})}]{NASEMNuclear2021}%
  \BibitemOpen
  \bibfield  {author} {\bibinfo {author} {\bibnamefont {{National Academies of
  Sciences, Engineering, and Medicine}}},\ }\href
  {https://doi.org/10.17226/25977} {\emph {\bibinfo {title} {{Space Nuclear
  Propulsion for Human Mars Exploration}}}}\ (\bibinfo  {publisher} {The
  National Academies Press},\ \bibinfo {address} {Washington, DC},\ \bibinfo
  {year} {2021})\BibitemShut {NoStop}%
\bibitem [{\citenamefont {{NASA}}(2026{\natexlab{a}})}]{NASA3001V2F}%
  \BibitemOpen
  \bibfield  {author} {\bibinfo {author} {\bibnamefont {{NASA}}},\ }\href
  {https://www.nasa.gov/wp-content/uploads/2025/05/nasa-std-3001-vol-2.pdf}
  {\emph {\bibinfo {title} {{NASA Spaceflight Human-System Standard, Volume 2:
  Human Factors, Habitability, and Environmental Health}}}},\ \bibinfo {type}
  {Tech. Rep.}\ \bibinfo {number} {NASA-STD-3001, Volume 2, Revision F}\
  (\bibinfo  {institution} {National Aeronautics and Space Administration},\
  \bibinfo {year} {2026})\BibitemShut {NoStop}%
\bibitem [{\citenamefont {Tays}\ \emph {et~al.}(2021)\citenamefont {Tays},
  \citenamefont {Hupfeld}, \citenamefont {McGregor}, \citenamefont {Salazar},
  \citenamefont {De~Dios}, \citenamefont {Beltran}, \citenamefont
  {Reuter-Lorenz}, \citenamefont {Kofman}, \citenamefont {Wood}, \citenamefont
  {Bloomberg}, \citenamefont {Mulavara},\ and\ \citenamefont
  {Seidler}}]{Tays2021}%
  \BibitemOpen
  \bibfield  {author} {\bibinfo {author} {\bibfnamefont {G.~D.}\ \bibnamefont
  {Tays}}, \bibinfo {author} {\bibfnamefont {K.~E.}\ \bibnamefont {Hupfeld}},
  \bibinfo {author} {\bibfnamefont {H.~R.}\ \bibnamefont {McGregor}}, \bibinfo
  {author} {\bibfnamefont {A.~P.}\ \bibnamefont {Salazar}}, \bibinfo {author}
  {\bibfnamefont {Y.~E.}\ \bibnamefont {De~Dios}}, \bibinfo {author}
  {\bibfnamefont {N.~E.}\ \bibnamefont {Beltran}}, \bibinfo {author}
  {\bibfnamefont {P.~A.}\ \bibnamefont {Reuter-Lorenz}}, \bibinfo {author}
  {\bibfnamefont {I.~S.}\ \bibnamefont {Kofman}}, \bibinfo {author}
  {\bibfnamefont {S.~J.}\ \bibnamefont {Wood}}, \bibinfo {author}
  {\bibfnamefont {J.~J.}\ \bibnamefont {Bloomberg}}, \bibinfo {author}
  {\bibfnamefont {A.~P.}\ \bibnamefont {Mulavara}},\ and\ \bibinfo {author}
  {\bibfnamefont {R.~D.}\ \bibnamefont {Seidler}},\ }\bibfield  {title}
  {\bibinfo {title} {The effects of long duration spaceflight on sensorimotor
  control and cognition},\ }\href {https://doi.org/10.3389/fncir.2021.723504}
  {\bibfield  {journal} {\bibinfo  {journal} {Frontiers in Neural Circuits}\
  }\textbf {\bibinfo {volume} {15}},\ \bibinfo {pages} {723504} (\bibinfo
  {year} {2021})}\BibitemShut {NoStop}%
\bibitem [{\citenamefont {Urban}\ \emph {et~al.}(2024)\citenamefont {Urban},
  \citenamefont {Ruff}, \citenamefont {Ferkul}, \citenamefont {Easton},
  \citenamefont {Johnston}, \citenamefont {Owens}, \citenamefont {Olson},
  \citenamefont {Fortenberry}, \citenamefont {Graf}, \citenamefont {George},
  \citenamefont {Toth}, \citenamefont {Meyer}, \citenamefont {Eigenbrod},
  \citenamefont {T'ien}, \citenamefont {Liao}, \citenamefont {Fernandez-Pello},
  \citenamefont {Legros}, \citenamefont {Guibaud}, \citenamefont {Smirnov},
  \citenamefont {Fujita}, \citenamefont {Rojas-Alva},\ and\ \citenamefont
  {Jomaas}}]{Urban2024}%
  \BibitemOpen
  \bibfield  {author} {\bibinfo {author} {\bibfnamefont {D.~L.}\ \bibnamefont
  {Urban}}, \bibinfo {author} {\bibfnamefont {G.~A.}\ \bibnamefont {Ruff}},
  \bibinfo {author} {\bibfnamefont {P.}~\bibnamefont {Ferkul}}, \bibinfo
  {author} {\bibfnamefont {J.}~\bibnamefont {Easton}}, \bibinfo {author}
  {\bibfnamefont {M.}~\bibnamefont {Johnston}}, \bibinfo {author}
  {\bibfnamefont {J.}~\bibnamefont {Owens}}, \bibinfo {author} {\bibfnamefont
  {S.}~\bibnamefont {Olson}}, \bibinfo {author} {\bibfnamefont
  {C.}~\bibnamefont {Fortenberry}}, \bibinfo {author} {\bibfnamefont
  {J.}~\bibnamefont {Graf}}, \bibinfo {author} {\bibfnamefont {O.}~\bibnamefont
  {George}}, \bibinfo {author} {\bibfnamefont {B.}~\bibnamefont {Toth}},
  \bibinfo {author} {\bibfnamefont {F.}~\bibnamefont {Meyer}}, \bibinfo
  {author} {\bibfnamefont {C.}~\bibnamefont {Eigenbrod}}, \bibinfo {author}
  {\bibfnamefont {J.~S.}\ \bibnamefont {T'ien}}, \bibinfo {author}
  {\bibfnamefont {Y.-T.~T.}\ \bibnamefont {Liao}}, \bibinfo {author}
  {\bibfnamefont {A.~C.}\ \bibnamefont {Fernandez-Pello}}, \bibinfo {author}
  {\bibfnamefont {G.}~\bibnamefont {Legros}}, \bibinfo {author} {\bibfnamefont
  {A.}~\bibnamefont {Guibaud}}, \bibinfo {author} {\bibfnamefont
  {N.}~\bibnamefont {Smirnov}}, \bibinfo {author} {\bibfnamefont
  {O.}~\bibnamefont {Fujita}}, \bibinfo {author} {\bibfnamefont
  {U.}~\bibnamefont {Rojas-Alva}},\ and\ \bibinfo {author} {\bibfnamefont
  {G.}~\bibnamefont {Jomaas}},\ }\bibfield  {title} {\bibinfo {title}
  {Preliminary results from the {Saffire VI} experiment},\ }in\ \href
  {https://doi.org/10.32865/2346/98994} {\emph {\bibinfo {booktitle} {53rd
  International Conference on Environmental Systems}}},\ \bibinfo {series and
  number} {\bibinfo {number} {ICES-2024-365}}\ (\bibinfo {address} {Louisville,
  Kentucky},\ \bibinfo {year} {2024})\BibitemShut {NoStop}%
\bibitem [{\citenamefont {Christiansen}(2003)}]{Christiansen2003}%
  \BibitemOpen
  \bibfield  {author} {\bibinfo {author} {\bibfnamefont {E.~L.}\ \bibnamefont
  {Christiansen}},\ }\href {https://ntrs.nasa.gov/citations/20030068423} {\emph
  {\bibinfo {title} {Meteoroid/debris shielding}}},\ \bibinfo {type} {Tech.
  Rep.}\ \bibinfo {number} {NASA/TP-2003-210788}\ (\bibinfo  {institution}
  {NASA Johnson Space Center},\ \bibinfo {year} {2003})\BibitemShut {NoStop}%
\bibitem [{\citenamefont {Zeitlin}\ \emph {et~al.}(2013)\citenamefont
  {Zeitlin}, \citenamefont {Hassler}, \citenamefont {Cucinotta} \emph
  {et~al.}}]{Zeitlin2013}%
  \BibitemOpen
  \bibfield  {author} {\bibinfo {author} {\bibfnamefont {C.}~\bibnamefont
  {Zeitlin}}, \bibinfo {author} {\bibfnamefont {D.~M.}\ \bibnamefont
  {Hassler}}, \bibinfo {author} {\bibfnamefont {F.~A.}\ \bibnamefont
  {Cucinotta}}, \emph {et~al.},\ }\bibfield  {title} {\bibinfo {title}
  {Measurements of energetic particle radiation in transit to {Mars} on the
  {Mars Science Laboratory}},\ }\href {https://doi.org/10.1126/science.1235989}
  {\bibfield  {journal} {\bibinfo  {journal} {Science}\ }\textbf {\bibinfo
  {volume} {340}},\ \bibinfo {pages} {1080--1084} (\bibinfo {year}
  {2013})}\BibitemShut {NoStop}%
\bibitem [{\citenamefont {Hassler}\ \emph {et~al.}(2014)\citenamefont
  {Hassler}, \citenamefont {Zeitlin}, \citenamefont {Wimmer-Schweingruber}
  \emph {et~al.}}]{Hassler2014}%
  \BibitemOpen
  \bibfield  {author} {\bibinfo {author} {\bibfnamefont {D.~M.}\ \bibnamefont
  {Hassler}}, \bibinfo {author} {\bibfnamefont {C.}~\bibnamefont {Zeitlin}},
  \bibinfo {author} {\bibfnamefont {R.~F.}\ \bibnamefont
  {Wimmer-Schweingruber}}, \emph {et~al.},\ }\bibfield  {title} {\bibinfo
  {title} {{Mars}' surface radiation environment measured with the {Mars
  Science Laboratory}'s {Curiosity} rover},\ }\href
  {https://doi.org/10.1126/science.1244797} {\bibfield  {journal} {\bibinfo
  {journal} {Science}\ }\textbf {\bibinfo {volume} {343}},\ \bibinfo {pages}
  {1244797} (\bibinfo {year} {2014})}\BibitemShut {NoStop}%
\bibitem [{\citenamefont {{NASA}}(2026{\natexlab{b}})}]{NASA3001}%
  \BibitemOpen
  \bibfield  {author} {\bibinfo {author} {\bibnamefont {{NASA}}},\ }\href
  {https://www.nasa.gov/reference/4-0-human-performance/} {\bibinfo {title}
  {{NASA Spaceflight Human-System Standard}, volume 1: Section 4.0, human
  performance}} (\bibinfo {year} {2026}{\natexlab{b}}),\ \bibinfo {note}
  {online text updated 18 March 2026; requirement V1 4030}\BibitemShut
  {NoStop}%
\bibitem [{\citenamefont {Hayne}\ \emph {et~al.}(2017)\citenamefont {Hayne},
  \citenamefont {Bandfield}, \citenamefont {Siegler}, \citenamefont {Vasavada},
  \citenamefont {Ghent}, \citenamefont {Williams}, \citenamefont {Greenhagen},
  \citenamefont {Aharonson}, \citenamefont {Elder}, \citenamefont {Lucey},\
  and\ \citenamefont {Paige}}]{Hayne2017}%
  \BibitemOpen
  \bibfield  {author} {\bibinfo {author} {\bibfnamefont {P.~O.}\ \bibnamefont
  {Hayne}}, \bibinfo {author} {\bibfnamefont {J.~L.}\ \bibnamefont
  {Bandfield}}, \bibinfo {author} {\bibfnamefont {M.~A.}\ \bibnamefont
  {Siegler}}, \bibinfo {author} {\bibfnamefont {A.~R.}\ \bibnamefont
  {Vasavada}}, \bibinfo {author} {\bibfnamefont {R.~R.}\ \bibnamefont {Ghent}},
  \bibinfo {author} {\bibfnamefont {J.-P.}\ \bibnamefont {Williams}}, \bibinfo
  {author} {\bibfnamefont {B.~T.}\ \bibnamefont {Greenhagen}}, \bibinfo
  {author} {\bibfnamefont {O.}~\bibnamefont {Aharonson}}, \bibinfo {author}
  {\bibfnamefont {C.~M.}\ \bibnamefont {Elder}}, \bibinfo {author}
  {\bibfnamefont {P.~G.}\ \bibnamefont {Lucey}},\ and\ \bibinfo {author}
  {\bibfnamefont {D.~A.}\ \bibnamefont {Paige}},\ }\bibfield  {title} {\bibinfo
  {title} {Global regolith thermophysical properties of the {Moon} from the
  {Diviner Lunar Radiometer Experiment}},\ }\href
  {https://doi.org/10.1002/2017JE005387} {\bibfield  {journal} {\bibinfo
  {journal} {JGR: Planets}\ }\textbf {\bibinfo {volume} {122}},\ \bibinfo
  {pages} {2371--2400} (\bibinfo {year} {2017})}\BibitemShut {NoStop}%
\bibitem [{\citenamefont {R{\"o}stel}\ \emph {et~al.}(2020)\citenamefont
  {R{\"o}stel}, \citenamefont {Guo}, \citenamefont {Banjac}, \citenamefont
  {Wimmer-Schweingruber},\ and\ \citenamefont {Heber}}]{Rostel2020}%
  \BibitemOpen
  \bibfield  {author} {\bibinfo {author} {\bibfnamefont {L.}~\bibnamefont
  {R{\"o}stel}}, \bibinfo {author} {\bibfnamefont {J.}~\bibnamefont {Guo}},
  \bibinfo {author} {\bibfnamefont {S.}~\bibnamefont {Banjac}}, \bibinfo
  {author} {\bibfnamefont {R.~F.}\ \bibnamefont {Wimmer-Schweingruber}},\ and\
  \bibinfo {author} {\bibfnamefont {B.}~\bibnamefont {Heber}},\ }\bibfield
  {title} {\bibinfo {title} {Subsurface radiation environment of {Mars} and its
  implication for shielding protection of future habitats},\ }\href
  {https://doi.org/10.1029/2019JE006246} {\bibfield  {journal} {\bibinfo
  {journal} {JGR: Planets}\ }\textbf {\bibinfo {volume} {125}},\ \bibinfo
  {pages} {e2019JE006246} (\bibinfo {year} {2020})}\BibitemShut {NoStop}%
\bibitem [{\citenamefont {Matthi{\"a}}\ and\ \citenamefont
  {Berger}(2024)}]{Matthia2024}%
  \BibitemOpen
  \bibfield  {author} {\bibinfo {author} {\bibfnamefont {D.}~\bibnamefont
  {Matthi{\"a}}}\ and\ \bibinfo {author} {\bibfnamefont {T.}~\bibnamefont
  {Berger}},\ }\bibfield  {title} {\bibinfo {title} {Radiation exposure and
  shielding effects on the lunar surface},\ }\href
  {https://doi.org/10.1029/2024SW004095} {\bibfield  {journal} {\bibinfo
  {journal} {Space Weather}\ }\textbf {\bibinfo {volume} {22}},\ \bibinfo
  {pages} {e2024SW004095} (\bibinfo {year} {2024})}\BibitemShut {NoStop}%
\bibitem [{\citenamefont {Wilson}\ \emph {et~al.}(1997)\citenamefont {Wilson},
  \citenamefont {Shinn}, \citenamefont {Simonsen}, \citenamefont {Cucinotta},
  \citenamefont {Dubey}, \citenamefont {Jordan}, \citenamefont {Jones},
  \citenamefont {Chang},\ and\ \citenamefont {Kim}}]{Wilson1997b}%
  \BibitemOpen
  \bibfield  {author} {\bibinfo {author} {\bibfnamefont {J.~W.}\ \bibnamefont
  {Wilson}}, \bibinfo {author} {\bibfnamefont {J.~L.}\ \bibnamefont {Shinn}},
  \bibinfo {author} {\bibfnamefont {L.~C.}\ \bibnamefont {Simonsen}}, \bibinfo
  {author} {\bibfnamefont {F.~A.}\ \bibnamefont {Cucinotta}}, \bibinfo {author}
  {\bibfnamefont {R.~R.}\ \bibnamefont {Dubey}}, \bibinfo {author}
  {\bibfnamefont {W.~R.}\ \bibnamefont {Jordan}}, \bibinfo {author}
  {\bibfnamefont {T.~D.}\ \bibnamefont {Jones}}, \bibinfo {author}
  {\bibfnamefont {C.~K.}\ \bibnamefont {Chang}},\ and\ \bibinfo {author}
  {\bibfnamefont {M.~Y.}\ \bibnamefont {Kim}},\ }\href
  {https://www.cs.odu.edu/~mln/ltrs-pdfs/NASA-97-tp3668.pdf} {\emph {\bibinfo
  {title} {Exposures to Solar Particle Events in Deep Space Missions}}},\
  \bibinfo {type} {Tech. Rep.}\ \bibinfo {number} {NASA TP-3668}\ (\bibinfo
  {institution} {NASA Langley Research Center},\ \bibinfo {year}
  {1997})\BibitemShut {NoStop}%
\bibitem [{\citenamefont {Simonsen}\ \emph {et~al.}(2020)\citenamefont
  {Simonsen}, \citenamefont {Slaba}, \citenamefont {Guida},\ and\ \citenamefont
  {Rusek}}]{Simonsen2020}%
  \BibitemOpen
  \bibfield  {author} {\bibinfo {author} {\bibfnamefont {L.~C.}\ \bibnamefont
  {Simonsen}}, \bibinfo {author} {\bibfnamefont {T.~C.}\ \bibnamefont {Slaba}},
  \bibinfo {author} {\bibfnamefont {P.}~\bibnamefont {Guida}},\ and\ \bibinfo
  {author} {\bibfnamefont {A.}~\bibnamefont {Rusek}},\ }\bibfield  {title}
  {\bibinfo {title} {{NASA}'s first ground-based galactic cosmic ray simulator:
  Enabling a new era in space radiobiology research},\ }\href
  {https://doi.org/10.1371/journal.pbio.3000669} {\bibfield  {journal}
  {\bibinfo  {journal} {PLOS Biology}\ }\textbf {\bibinfo {volume} {18}},\
  \bibinfo {pages} {e3000669} (\bibinfo {year} {2020})}\BibitemShut {NoStop}%
\bibitem [{\citenamefont {Marshall-Goebel}\ \emph {et~al.}(2019)\citenamefont
  {Marshall-Goebel}, \citenamefont {Laurie}, \citenamefont {Alferova} \emph
  {et~al.}}]{MarshallGoebel2019}%
  \BibitemOpen
  \bibfield  {author} {\bibinfo {author} {\bibfnamefont {K.}~\bibnamefont
  {Marshall-Goebel}}, \bibinfo {author} {\bibfnamefont {S.~S.}\ \bibnamefont
  {Laurie}}, \bibinfo {author} {\bibfnamefont {I.~V.}\ \bibnamefont
  {Alferova}}, \emph {et~al.},\ }\bibfield  {title} {\bibinfo {title}
  {Assessment of jugular venous blood flow stasis and thrombosis during
  spaceflight},\ }\href {https://doi.org/10.1001/jamanetworkopen.2019.15011}
  {\bibfield  {journal} {\bibinfo  {journal} {JAMA Network Open}\ }\textbf
  {\bibinfo {volume} {2}},\ \bibinfo {pages} {e1915011} (\bibinfo {year}
  {2019})}\BibitemShut {NoStop}%
\bibitem [{\citenamefont {Macias}\ \emph {et~al.}(2020)\citenamefont {Macias},
  \citenamefont {Patel}, \citenamefont {Gibson}, \citenamefont {Samuels},
  \citenamefont {Laurie}, \citenamefont {Otto}, \citenamefont {Ferguson},
  \citenamefont {Lee}, \citenamefont {Ploutz-Snyder}, \citenamefont {Kramer},
  \citenamefont {Mader}, \citenamefont {Brunstetter},\ and\ \citenamefont
  {Stenger}}]{Macias2020}%
  \BibitemOpen
  \bibfield  {author} {\bibinfo {author} {\bibfnamefont {B.~R.}\ \bibnamefont
  {Macias}}, \bibinfo {author} {\bibfnamefont {N.~B.}\ \bibnamefont {Patel}},
  \bibinfo {author} {\bibfnamefont {C.~R.}\ \bibnamefont {Gibson}}, \bibinfo
  {author} {\bibfnamefont {B.~C.}\ \bibnamefont {Samuels}}, \bibinfo {author}
  {\bibfnamefont {S.~S.}\ \bibnamefont {Laurie}}, \bibinfo {author}
  {\bibfnamefont {C.}~\bibnamefont {Otto}}, \bibinfo {author} {\bibfnamefont
  {C.~R.}\ \bibnamefont {Ferguson}}, \bibinfo {author} {\bibfnamefont
  {S.~M.~C.}\ \bibnamefont {Lee}}, \bibinfo {author} {\bibfnamefont
  {R.}~\bibnamefont {Ploutz-Snyder}}, \bibinfo {author} {\bibfnamefont {L.~A.}\
  \bibnamefont {Kramer}}, \bibinfo {author} {\bibfnamefont {T.~H.}\
  \bibnamefont {Mader}}, \bibinfo {author} {\bibfnamefont {T.}~\bibnamefont
  {Brunstetter}},\ and\ \bibinfo {author} {\bibfnamefont {M.~B.}\ \bibnamefont
  {Stenger}},\ }\bibfield  {title} {\bibinfo {title} {Association of
  long-duration spaceflight with anterior and posterior ocular structure
  changes in astronauts and their recovery},\ }\href
  {https://doi.org/10.1001/jamaophthalmol.2020.0673} {\bibfield  {journal}
  {\bibinfo  {journal} {JAMA Ophthalmology}\ }\textbf {\bibinfo {volume}
  {138}},\ \bibinfo {pages} {553--559} (\bibinfo {year} {2020})}\BibitemShut
  {NoStop}%
\bibitem [{\citenamefont {Gabel}\ \emph {et~al.}(2022)\citenamefont {Gabel},
  \citenamefont {Liphardt}, \citenamefont {Hulme}, \citenamefont {Heer},
  \citenamefont {Zwart}, \citenamefont {Sibonga}, \citenamefont {Smith},\ and\
  \citenamefont {Boyd}}]{Gabel2022}%
  \BibitemOpen
  \bibfield  {author} {\bibinfo {author} {\bibfnamefont {L.}~\bibnamefont
  {Gabel}}, \bibinfo {author} {\bibfnamefont {A.-M.}\ \bibnamefont {Liphardt}},
  \bibinfo {author} {\bibfnamefont {P.~A.}\ \bibnamefont {Hulme}}, \bibinfo
  {author} {\bibfnamefont {M.}~\bibnamefont {Heer}}, \bibinfo {author}
  {\bibfnamefont {S.~R.}\ \bibnamefont {Zwart}}, \bibinfo {author}
  {\bibfnamefont {J.~D.}\ \bibnamefont {Sibonga}}, \bibinfo {author}
  {\bibfnamefont {S.~M.}\ \bibnamefont {Smith}},\ and\ \bibinfo {author}
  {\bibfnamefont {S.~K.}\ \bibnamefont {Boyd}},\ }\bibfield  {title} {\bibinfo
  {title} {Incomplete recovery of bone strength and trabecular
  microarchitecture at the distal tibia 1 year after return from long duration
  spaceflight},\ }\href {https://doi.org/10.1038/s41598-022-13461-1} {\bibfield
   {journal} {\bibinfo  {journal} {Scientific Reports}\ }\textbf {\bibinfo
  {volume} {12}},\ \bibinfo {pages} {9446} (\bibinfo {year}
  {2022})}\BibitemShut {NoStop}%
\bibitem [{\citenamefont {Kramer}\ \emph {et~al.}(2021)\citenamefont {Kramer}
  \emph {et~al.}}]{Kramer2021}%
  \BibitemOpen
  \bibfield  {author} {\bibinfo {author} {\bibfnamefont {A.}~\bibnamefont
  {Kramer}} \emph {et~al.},\ }\bibfield  {title} {\bibinfo {title} {Daily
  30-min exposure to artificial gravity during 60 days of bed rest does not
  maintain aerobic exercise capacity but mitigates some deteriorations of
  muscle function: results from the {AGBRESA RCT}},\ }\href
  {https://doi.org/10.1007/s00421-021-04673-w} {\bibfield  {journal} {\bibinfo
  {journal} {European Journal of Applied Physiology}\ }\textbf {\bibinfo
  {volume} {121}},\ \bibinfo {pages} {2015--2026} (\bibinfo {year}
  {2021})}\BibitemShut {NoStop}%
\bibitem [{\citenamefont {Hayashi}\ \emph {et~al.}(2023)\citenamefont
  {Hayashi}, \citenamefont {Fujita}, \citenamefont {Okada} \emph
  {et~al.}}]{Hayashi2023}%
  \BibitemOpen
  \bibfield  {author} {\bibinfo {author} {\bibfnamefont {T.}~\bibnamefont
  {Hayashi}}, \bibinfo {author} {\bibfnamefont {R.}~\bibnamefont {Fujita}},
  \bibinfo {author} {\bibfnamefont {R.}~\bibnamefont {Okada}}, \emph {et~al.},\
  }\bibfield  {title} {\bibinfo {title} {Lunar gravity prevents skeletal muscle
  atrophy but not myofiber type shift in mice},\ }\href
  {https://doi.org/10.1038/s42003-023-04769-3} {\bibfield  {journal} {\bibinfo
  {journal} {Communications Biology}\ }\textbf {\bibinfo {volume} {6}},\
  \bibinfo {pages} {424} (\bibinfo {year} {2023})}\BibitemShut {NoStop}%
\bibitem [{\citenamefont {Okamura}\ \emph {et~al.}(2024)\citenamefont
  {Okamura}, \citenamefont {Gochi}, \citenamefont {Ishikawa}, \citenamefont
  {Hayashi}, \citenamefont {Fuseya}, \citenamefont {Suzuki}, \citenamefont
  {Kanai}, \citenamefont {Inoue}, \citenamefont {Murakami}, \citenamefont
  {Sadaki}, \citenamefont {Jeon}, \citenamefont {Hayama}, \citenamefont
  {Ishii}, \citenamefont {Tsunakawa}, \citenamefont {Ochi}, \citenamefont
  {Sato}, \citenamefont {Hamada}, \citenamefont {Abe}, \citenamefont {Morita},
  \citenamefont {Okada}, \citenamefont {Shiba}, \citenamefont {Muratani},
  \citenamefont {Shinohara}, \citenamefont {Akiyama}, \citenamefont {Kudo},\
  and\ \citenamefont {Takahashi}}]{Okamura2024}%
  \BibitemOpen
  \bibfield  {author} {\bibinfo {author} {\bibfnamefont {Y.}~\bibnamefont
  {Okamura}}, \bibinfo {author} {\bibfnamefont {K.}~\bibnamefont {Gochi}},
  \bibinfo {author} {\bibfnamefont {T.}~\bibnamefont {Ishikawa}}, \bibinfo
  {author} {\bibfnamefont {T.}~\bibnamefont {Hayashi}}, \bibinfo {author}
  {\bibfnamefont {S.}~\bibnamefont {Fuseya}}, \bibinfo {author} {\bibfnamefont
  {R.}~\bibnamefont {Suzuki}}, \bibinfo {author} {\bibfnamefont
  {M.}~\bibnamefont {Kanai}}, \bibinfo {author} {\bibfnamefont
  {Y.}~\bibnamefont {Inoue}}, \bibinfo {author} {\bibfnamefont
  {Y.}~\bibnamefont {Murakami}}, \bibinfo {author} {\bibfnamefont
  {S.}~\bibnamefont {Sadaki}}, \bibinfo {author} {\bibfnamefont
  {H.}~\bibnamefont {Jeon}}, \bibinfo {author} {\bibfnamefont {M.}~\bibnamefont
  {Hayama}}, \bibinfo {author} {\bibfnamefont {H.}~\bibnamefont {Ishii}},
  \bibinfo {author} {\bibfnamefont {Y.}~\bibnamefont {Tsunakawa}}, \bibinfo
  {author} {\bibfnamefont {H.}~\bibnamefont {Ochi}}, \bibinfo {author}
  {\bibfnamefont {S.}~\bibnamefont {Sato}}, \bibinfo {author} {\bibfnamefont
  {M.}~\bibnamefont {Hamada}}, \bibinfo {author} {\bibfnamefont
  {C.}~\bibnamefont {Abe}}, \bibinfo {author} {\bibfnamefont {H.}~\bibnamefont
  {Morita}}, \bibinfo {author} {\bibfnamefont {R.}~\bibnamefont {Okada}},
  \bibinfo {author} {\bibfnamefont {D.}~\bibnamefont {Shiba}}, \bibinfo
  {author} {\bibfnamefont {M.}~\bibnamefont {Muratani}}, \bibinfo {author}
  {\bibfnamefont {M.}~\bibnamefont {Shinohara}}, \bibinfo {author}
  {\bibfnamefont {T.}~\bibnamefont {Akiyama}}, \bibinfo {author} {\bibfnamefont
  {T.}~\bibnamefont {Kudo}},\ and\ \bibinfo {author} {\bibfnamefont
  {S.}~\bibnamefont {Takahashi}},\ }\bibfield  {title} {\bibinfo {title}
  {Impact of microgravity and lunar gravity on murine skeletal and immune
  systems during space travel},\ }\href
  {https://doi.org/10.1038/s41598-024-79315-0} {\bibfield  {journal} {\bibinfo
  {journal} {Scientific Reports}\ }\textbf {\bibinfo {volume} {14}},\ \bibinfo
  {pages} {28774} (\bibinfo {year} {2024})}\BibitemShut {NoStop}%
\bibitem [{\citenamefont {Gimunov{\'a}}\ \emph {et~al.}(2024)\citenamefont
  {Gimunov{\'a}}, \citenamefont {Paludo}, \citenamefont {Bernacikov{\'a}} \emph
  {et~al.}}]{Gimunova2024}%
  \BibitemOpen
  \bibfield  {author} {\bibinfo {author} {\bibfnamefont {M.}~\bibnamefont
  {Gimunov{\'a}}}, \bibinfo {author} {\bibfnamefont {A.~C.}\ \bibnamefont
  {Paludo}}, \bibinfo {author} {\bibfnamefont {M.}~\bibnamefont
  {Bernacikov{\'a}}}, \emph {et~al.},\ }\bibfield  {title} {\bibinfo {title}
  {The effect of space travel on human reproductive health: a systematic
  review},\ }\href {https://doi.org/10.1038/s41526-024-00351-1} {\bibfield
  {journal} {\bibinfo  {journal} {npj Microgravity}\ }\textbf {\bibinfo
  {volume} {10}},\ \bibinfo {pages} {10} (\bibinfo {year} {2024})}\BibitemShut
  {NoStop}%
\bibitem [{\citenamefont {Jain}\ \emph {et~al.}(2023)\citenamefont {Jain},
  \citenamefont {Chuva~de Sousa~Lopes}, \citenamefont {Benotmane},
  \citenamefont {Verratti}, \citenamefont {Mitchell},\ and\ \citenamefont
  {Stukenborg}}]{Jain2023}%
  \BibitemOpen
  \bibfield  {author} {\bibinfo {author} {\bibfnamefont {V.}~\bibnamefont
  {Jain}}, \bibinfo {author} {\bibfnamefont {S.~M.}\ \bibnamefont {Chuva~de
  Sousa~Lopes}}, \bibinfo {author} {\bibfnamefont {M.~A.}\ \bibnamefont
  {Benotmane}}, \bibinfo {author} {\bibfnamefont {V.}~\bibnamefont {Verratti}},
  \bibinfo {author} {\bibfnamefont {R.~T.}\ \bibnamefont {Mitchell}},\ and\
  \bibinfo {author} {\bibfnamefont {J.-B.}\ \bibnamefont {Stukenborg}},\
  }\bibfield  {title} {\bibinfo {title} {Human development and reproduction in
  space---a european perspective},\ }\href
  {https://doi.org/10.1038/s41526-023-00272-5} {\bibfield  {journal} {\bibinfo
  {journal} {npj Microgravity}\ }\textbf {\bibinfo {volume} {9}},\ \bibinfo
  {pages} {24} (\bibinfo {year} {2023})}\BibitemShut {NoStop}%
\bibitem [{\citenamefont {Wakayama}\ \emph {et~al.}(2021)\citenamefont
  {Wakayama}, \citenamefont {Ito}, \citenamefont {Kamada} \emph
  {et~al.}}]{Wakayama2021}%
  \BibitemOpen
  \bibfield  {author} {\bibinfo {author} {\bibfnamefont {S.}~\bibnamefont
  {Wakayama}}, \bibinfo {author} {\bibfnamefont {D.}~\bibnamefont {Ito}},
  \bibinfo {author} {\bibfnamefont {Y.}~\bibnamefont {Kamada}}, \emph
  {et~al.},\ }\bibfield  {title} {\bibinfo {title} {Evaluating the long-term
  effect of space radiation on the reproductive normality of mammalian sperm
  preserved on the {International Space Station}},\ }\href
  {https://doi.org/10.1126/sciadv.abg5554} {\bibfield  {journal} {\bibinfo
  {journal} {Science Advances}\ }\textbf {\bibinfo {volume} {7}},\ \bibinfo
  {pages} {eabg5554} (\bibinfo {year} {2021})}\BibitemShut {NoStop}%
\bibitem [{\citenamefont {Wakayama}\ \emph {et~al.}(2023)\citenamefont
  {Wakayama}, \citenamefont {Kikuchi}, \citenamefont {Soejima} \emph
  {et~al.}}]{Wakayama2023}%
  \BibitemOpen
  \bibfield  {author} {\bibinfo {author} {\bibfnamefont {S.}~\bibnamefont
  {Wakayama}}, \bibinfo {author} {\bibfnamefont {Y.}~\bibnamefont {Kikuchi}},
  \bibinfo {author} {\bibfnamefont {M.}~\bibnamefont {Soejima}}, \emph
  {et~al.},\ }\bibfield  {title} {\bibinfo {title} {Effect of microgravity on
  mammalian embryo development evaluated at the {International Space
  Station}},\ }\href {https://doi.org/10.1016/j.isci.2023.108177} {\bibfield
  {journal} {\bibinfo  {journal} {iScience}\ }\textbf {\bibinfo {volume}
  {26}},\ \bibinfo {pages} {108177} (\bibinfo {year} {2023})}\BibitemShut
  {NoStop}%
\bibitem [{\citenamefont {Wong}\ and\ \citenamefont
  {DeSantis}(1997)}]{Wong1997}%
  \BibitemOpen
  \bibfield  {author} {\bibinfo {author} {\bibfnamefont {A.~M.}\ \bibnamefont
  {Wong}}\ and\ \bibinfo {author} {\bibfnamefont {M.}~\bibnamefont
  {DeSantis}},\ }\bibfield  {title} {\bibinfo {title} {Rat gestation during
  space flight: Outcomes for dams and their offspring born after return to
  {Earth}},\ }\href {https://doi.org/10.1007/BF02688630} {\bibfield  {journal}
  {\bibinfo  {journal} {Integrative Physiological and Behavioral Science}\
  }\textbf {\bibinfo {volume} {32}},\ \bibinfo {pages} {322--342} (\bibinfo
  {year} {1997})}\BibitemShut {NoStop}%
\bibitem [{\citenamefont {Lyons}\ \emph {et~al.}(2026)\citenamefont {Lyons},
  \citenamefont {Nikitaras}, \citenamefont {Arman}, \citenamefont
  {McIlfatrick}, \citenamefont {Nottle}, \citenamefont {Gonzalez},\ and\
  \citenamefont {McPherson}}]{Lyons2026}%
  \BibitemOpen
  \bibfield  {author} {\bibinfo {author} {\bibfnamefont {H.~E.}\ \bibnamefont
  {Lyons}}, \bibinfo {author} {\bibfnamefont {V.}~\bibnamefont {Nikitaras}},
  \bibinfo {author} {\bibfnamefont {B.~M.}\ \bibnamefont {Arman}}, \bibinfo
  {author} {\bibfnamefont {S.~M.}\ \bibnamefont {McIlfatrick}}, \bibinfo
  {author} {\bibfnamefont {M.~B.}\ \bibnamefont {Nottle}}, \bibinfo {author}
  {\bibfnamefont {M.~B.}\ \bibnamefont {Gonzalez}},\ and\ \bibinfo {author}
  {\bibfnamefont {N.~O.}\ \bibnamefont {McPherson}},\ }\bibfield  {title}
  {\bibinfo {title} {Simulated microgravity alters sperm navigation,
  fertilization and embryo development in mammals},\ }\href
  {https://doi.org/10.1038/s42003-026-09734-4} {\bibfield  {journal} {\bibinfo
  {journal} {Communications Biology}\ }\textbf {\bibinfo {volume} {9}},\
  \bibinfo {pages} {401} (\bibinfo {year} {2026})}\BibitemShut {NoStop}%
\bibitem [{\citenamefont {Hong}\ \emph {et~al.}(2026)\citenamefont {Hong},
  \citenamefont {Dalouchi}, \citenamefont {Prom}, \citenamefont {Ponte},
  \citenamefont {Hockey}, \citenamefont {Braun}, \citenamefont {Nies},
  \citenamefont {Stoltz}, \citenamefont {Chakravarthi}, \citenamefont {Sinha},
  \citenamefont {Choi}, \citenamefont {Ronca}, \citenamefont {Alwood},
  \citenamefont {Fajardo}, \citenamefont {Morris}, \citenamefont {Puukila},\
  and\ \citenamefont {Christenson}}]{Hong2026}%
  \BibitemOpen
  \bibfield  {author} {\bibinfo {author} {\bibfnamefont {X.}~\bibnamefont
  {Hong}}, \bibinfo {author} {\bibfnamefont {F.}~\bibnamefont {Dalouchi}},
  \bibinfo {author} {\bibfnamefont {J.~C.}\ \bibnamefont {Prom}}, \bibinfo
  {author} {\bibfnamefont {M.~E.}\ \bibnamefont {Ponte}}, \bibinfo {author}
  {\bibfnamefont {B.~L.}\ \bibnamefont {Hockey}}, \bibinfo {author}
  {\bibfnamefont {J.~L.}\ \bibnamefont {Braun}}, \bibinfo {author}
  {\bibfnamefont {P.}~\bibnamefont {Nies}}, \bibinfo {author} {\bibfnamefont
  {A.}~\bibnamefont {Stoltz}}, \bibinfo {author} {\bibfnamefont {V.~P.}\
  \bibnamefont {Chakravarthi}}, \bibinfo {author} {\bibfnamefont
  {M.}~\bibnamefont {Sinha}}, \bibinfo {author} {\bibfnamefont {S.~Y.}\
  \bibnamefont {Choi}}, \bibinfo {author} {\bibfnamefont {A.~E.}\ \bibnamefont
  {Ronca}}, \bibinfo {author} {\bibfnamefont {J.~S.}\ \bibnamefont {Alwood}},
  \bibinfo {author} {\bibfnamefont {V.~A.}\ \bibnamefont {Fajardo}}, \bibinfo
  {author} {\bibfnamefont {E.~M.}\ \bibnamefont {Morris}}, \bibinfo {author}
  {\bibfnamefont {S.}~\bibnamefont {Puukila}},\ and\ \bibinfo {author}
  {\bibfnamefont {L.~K.}\ \bibnamefont {Christenson}},\ }\bibfield  {title}
  {\bibinfo {title} {Female reproductive dysfunction and transgenerational
  consequences following prolonged spaceflight exposure},\ }\href
  {https://doi.org/10.1073/pnas.2606092123} {\bibfield  {journal} {\bibinfo
  {journal} {Proceedings of the National Academy of Sciences}\ }\textbf
  {\bibinfo {volume} {123}},\ \bibinfo {pages} {e2606092123} (\bibinfo {year}
  {2026})}\BibitemShut {NoStop}%
\bibitem [{\citenamefont {Gifani}\ and\ \citenamefont
  {Burns}(2026)}]{Gifani2026}%
  \BibitemOpen
  \bibfield  {author} {\bibinfo {author} {\bibfnamefont {M.}~\bibnamefont
  {Gifani}}\ and\ \bibinfo {author} {\bibfnamefont {G.~W.}\ \bibnamefont
  {Burns}},\ }\bibfield  {title} {\bibinfo {title} {The transgenerational toll
  of maternal spaceflight},\ }\href {https://doi.org/10.1073/pnas.2621879123}
  {\bibfield  {journal} {\bibinfo  {journal} {Proceedings of the National
  Academy of Sciences}\ }\textbf {\bibinfo {volume} {123}},\ \bibinfo {pages}
  {e2621879123} (\bibinfo {year} {2026})},\ \bibinfo {note}
  {commentary}\BibitemShut {NoStop}%
\bibitem [{\citenamefont {Graybiel}\ \emph {et~al.}(1965)\citenamefont
  {Graybiel}, \citenamefont {Kennedy}, \citenamefont {Knoblock}, \citenamefont
  {Guedry}, \citenamefont {Mertz}, \citenamefont {McLeod}, \citenamefont
  {Colehour}, \citenamefont {Miller},\ and\ \citenamefont
  {Fregly}}]{Graybiel1965}%
  \BibitemOpen
  \bibfield  {author} {\bibinfo {author} {\bibfnamefont {A.}~\bibnamefont
  {Graybiel}}, \bibinfo {author} {\bibfnamefont {R.~S.}\ \bibnamefont
  {Kennedy}}, \bibinfo {author} {\bibfnamefont {E.~C.}\ \bibnamefont
  {Knoblock}}, \bibinfo {author} {\bibfnamefont {F.~E.}\ \bibnamefont
  {Guedry}}, \bibinfo {author} {\bibfnamefont {W.}~\bibnamefont {Mertz}},
  \bibinfo {author} {\bibfnamefont {M.~E.}\ \bibnamefont {McLeod}}, \bibinfo
  {author} {\bibfnamefont {J.~K.}\ \bibnamefont {Colehour}}, \bibinfo {author}
  {\bibfnamefont {E.~F.}\ \bibnamefont {Miller}},\ and\ \bibinfo {author}
  {\bibfnamefont {A.~R.}\ \bibnamefont {Fregly}},\ }\href
  {https://pubmed.ncbi.nlm.nih.gov/14313061/} {\emph {\bibinfo {title} {The
  Effects of Exposure to a Rotating Environment (10 RPM) on Four Aviators for a
  Period of Twelve Days}}},\ \bibinfo {type} {Tech. Rep.}\ \bibinfo {number}
  {NSAM-923}\ (\bibinfo  {institution} {U.S. Naval School of Aviation
  Medicine},\ \bibinfo {year} {1965})\BibitemShut {NoStop}%
\bibitem [{\citenamefont {Bretl}\ and\ \citenamefont
  {Clark}(2020)}]{Bretl2020}%
  \BibitemOpen
  \bibfield  {author} {\bibinfo {author} {\bibfnamefont {K.~N.}\ \bibnamefont
  {Bretl}}\ and\ \bibinfo {author} {\bibfnamefont {T.~K.}\ \bibnamefont
  {Clark}},\ }\bibfield  {title} {\bibinfo {title} {Improved feasibility of
  astronaut short-radius artificial gravity through a 50-day incremental,
  personalized, vestibular acclimation protocol},\ }\href
  {https://doi.org/10.1038/s41526-020-00112-w} {\bibfield  {journal} {\bibinfo
  {journal} {npj Microgravity}\ }\textbf {\bibinfo {volume} {6}},\ \bibinfo
  {pages} {22} (\bibinfo {year} {2020})}\BibitemShut {NoStop}%
\bibitem [{\citenamefont {Schuirmann}(1987)}]{Schuirmann1987}%
  \BibitemOpen
  \bibfield  {author} {\bibinfo {author} {\bibfnamefont {D.~J.}\ \bibnamefont
  {Schuirmann}},\ }\bibfield  {title} {\bibinfo {title} {A comparison of the
  two one-sided tests procedure and the power approach for assessing the
  equivalence of average bioavailability},\ }\href
  {https://doi.org/10.1007/BF01068419} {\bibfield  {journal} {\bibinfo
  {journal} {Journal of Pharmacokinetics and Biopharmaceutics}\ }\textbf
  {\bibinfo {volume} {15}},\ \bibinfo {pages} {657--680} (\bibinfo {year}
  {1987})}\BibitemShut {NoStop}%
\bibitem [{\citenamefont {Phillips}(1990)}]{Phillips1990}%
  \BibitemOpen
  \bibfield  {author} {\bibinfo {author} {\bibfnamefont {K.~F.}\ \bibnamefont
  {Phillips}},\ }\bibfield  {title} {\bibinfo {title} {Power of the two
  one-sided tests procedure in bioequivalence},\ }\href
  {https://doi.org/10.1007/BF01063556} {\bibfield  {journal} {\bibinfo
  {journal} {Journal of Pharmacokinetics and Biopharmaceutics}\ }\textbf
  {\bibinfo {volume} {18}},\ \bibinfo {pages} {137--144} (\bibinfo {year}
  {1990})}\BibitemShut {NoStop}%
\bibitem [{\citenamefont {Leslie}(1945)}]{Leslie1945}%
  \BibitemOpen
  \bibfield  {author} {\bibinfo {author} {\bibfnamefont {P.~H.}\ \bibnamefont
  {Leslie}},\ }\bibfield  {title} {\bibinfo {title} {On the use of matrices in
  certain population mathematics},\ }\href
  {https://doi.org/10.1093/biomet/33.3.183} {\bibfield  {journal} {\bibinfo
  {journal} {Biometrika}\ }\textbf {\bibinfo {volume} {33}},\ \bibinfo {pages}
  {183--212} (\bibinfo {year} {1945})}\BibitemShut {NoStop}%
\bibitem [{\citenamefont {Wright}(1931)}]{Wright1931}%
  \BibitemOpen
  \bibfield  {author} {\bibinfo {author} {\bibfnamefont {S.}~\bibnamefont
  {Wright}},\ }\bibfield  {title} {\bibinfo {title} {Evolution in {Mendelian}
  populations},\ }\href {https://doi.org/10.1093/genetics/16.2.97} {\bibfield
  {journal} {\bibinfo  {journal} {Genetics}\ }\textbf {\bibinfo {volume}
  {16}},\ \bibinfo {pages} {97--159} (\bibinfo {year} {1931})}\BibitemShut
  {NoStop}%
\bibitem [{\citenamefont {Morton}\ \emph {et~al.}(1956)\citenamefont {Morton},
  \citenamefont {Crow},\ and\ \citenamefont {Muller}}]{Morton1956}%
  \BibitemOpen
  \bibfield  {author} {\bibinfo {author} {\bibfnamefont {N.~E.}\ \bibnamefont
  {Morton}}, \bibinfo {author} {\bibfnamefont {J.~F.}\ \bibnamefont {Crow}},\
  and\ \bibinfo {author} {\bibfnamefont {H.~J.}\ \bibnamefont {Muller}},\
  }\bibfield  {title} {\bibinfo {title} {An estimate of the mutational damage
  in man from data on consanguineous marriages},\ }\href
  {https://doi.org/10.1073/pnas.42.11.855} {\bibfield  {journal} {\bibinfo
  {journal} {Proceedings of the National Academy of Sciences}\ }\textbf
  {\bibinfo {volume} {42}},\ \bibinfo {pages} {855--863} (\bibinfo {year}
  {1956})}\BibitemShut {NoStop}%
\bibitem [{\citenamefont {Kalinowski}\ and\ \citenamefont
  {Hedrick}(1998)}]{Kalinowski1998}%
  \BibitemOpen
  \bibfield  {author} {\bibinfo {author} {\bibfnamefont {S.~T.}\ \bibnamefont
  {Kalinowski}}\ and\ \bibinfo {author} {\bibfnamefont {P.~W.}\ \bibnamefont
  {Hedrick}},\ }\bibfield  {title} {\bibinfo {title} {An improved method for
  estimating inbreeding depression in pedigrees},\ }\href
  {https://doi.org/10.1002/(SICI)1098-2361(1998)17:6<481::AID-ZOO2>3.0.CO;2-G}
  {\bibfield  {journal} {\bibinfo  {journal} {Zoo Biology}\ }\textbf {\bibinfo
  {volume} {17}},\ \bibinfo {pages} {481--497} (\bibinfo {year}
  {1998})}\BibitemShut {NoStop}%
\bibitem [{\citenamefont {Smith}(2014)}]{Smith2014}%
  \BibitemOpen
  \bibfield  {author} {\bibinfo {author} {\bibfnamefont {C.~M.}\ \bibnamefont
  {Smith}},\ }\bibfield  {title} {\bibinfo {title} {Estimation of a genetically
  viable population for multigenerational interstellar voyaging: Review and
  data for project {Hyperion}},\ }\href
  {https://doi.org/10.1016/j.actaastro.2013.12.013} {\bibfield  {journal}
  {\bibinfo  {journal} {Acta Astronautica}\ }\textbf {\bibinfo {volume} {97}},\
  \bibinfo {pages} {16--29} (\bibinfo {year} {2014})}\BibitemShut {NoStop}%
\bibitem [{\citenamefont {Marin}\ and\ \citenamefont
  {Beluffi}(2018)}]{Marin2018}%
  \BibitemOpen
  \bibfield  {author} {\bibinfo {author} {\bibfnamefont {F.}~\bibnamefont
  {Marin}}\ and\ \bibinfo {author} {\bibfnamefont {C.}~\bibnamefont
  {Beluffi}},\ }\bibfield  {title} {\bibinfo {title} {Computing the minimal
  crew for a multi-generational space journey towards {Proxima Centauri b}},\
  }\href {https://arxiv.org/abs/1806.03856} {\bibfield  {journal} {\bibinfo
  {journal} {Journal of the British Interplanetary Society}\ }\textbf {\bibinfo
  {volume} {71}},\ \bibinfo {pages} {431--438} (\bibinfo {year} {2018})},\
  \Eprint {https://arxiv.org/abs/1806.03856} {arXiv:1806.03856
  [physics.pop-ph]} \BibitemShut {NoStop}%
\bibitem [{\citenamefont {Lewandowski}\ \emph {et~al.}(2024)\citenamefont
  {Lewandowski}, \citenamefont {Schkurko}, \citenamefont {Miller},
  \citenamefont {Valentine}, \citenamefont {Calaway}, \citenamefont {Yang},
  \citenamefont {Ebert}, \citenamefont {Sargsyan}, \citenamefont {Byrne},
  \citenamefont {Walton}, \citenamefont {Lemery}, \citenamefont {Suresh},
  \citenamefont {Thompson}, \citenamefont {Easter},\ and\ \citenamefont
  {Lehnhardt}}]{Lewandowski2024}%
  \BibitemOpen
  \bibfield  {author} {\bibinfo {author} {\bibfnamefont {B.~E.}\ \bibnamefont
  {Lewandowski}}, \bibinfo {author} {\bibfnamefont {C.~M.}\ \bibnamefont
  {Schkurko}}, \bibinfo {author} {\bibfnamefont {R.~S.}\ \bibnamefont
  {Miller}}, \bibinfo {author} {\bibfnamefont {R.~W.}\ \bibnamefont
  {Valentine}}, \bibinfo {author} {\bibfnamefont {K.~M.}\ \bibnamefont
  {Calaway}}, \bibinfo {author} {\bibfnamefont {J.~D.}\ \bibnamefont {Yang}},
  \bibinfo {author} {\bibfnamefont {D.~J.}\ \bibnamefont {Ebert}}, \bibinfo
  {author} {\bibfnamefont {A.}~\bibnamefont {Sargsyan}}, \bibinfo {author}
  {\bibfnamefont {V.}~\bibnamefont {Byrne}}, \bibinfo {author} {\bibfnamefont
  {M.}~\bibnamefont {Walton}}, \bibinfo {author} {\bibfnamefont
  {J.}~\bibnamefont {Lemery}}, \bibinfo {author} {\bibfnamefont
  {R.}~\bibnamefont {Suresh}}, \bibinfo {author} {\bibfnamefont {M.~S.}\
  \bibnamefont {Thompson}}, \bibinfo {author} {\bibfnamefont {B.~D.}\
  \bibnamefont {Easter}},\ and\ \bibinfo {author} {\bibfnamefont {K.~R.}\
  \bibnamefont {Lehnhardt}},\ }\bibfield  {title} {\bibinfo {title} {Technology
  modification, development, and demonstrations for future spaceflight medical
  systems at {NASA}},\ }\href {https://doi.org/10.3389/frspt.2024.1384457}
  {\bibfield  {journal} {\bibinfo  {journal} {Frontiers in Space Technologies}\
  }\textbf {\bibinfo {volume} {5}},\ \bibinfo {pages} {1384457} (\bibinfo
  {year} {2024})}\BibitemShut {NoStop}%
\bibitem [{\citenamefont {Diaz}\ \emph {et~al.}(2024)\citenamefont {Diaz},
  \citenamefont {Ives}, \citenamefont {Lazare},\ and\ \citenamefont
  {Buckland}}]{Diaz2024}%
  \BibitemOpen
  \bibfield  {author} {\bibinfo {author} {\bibfnamefont {T.~E.}\ \bibnamefont
  {Diaz}}, \bibinfo {author} {\bibfnamefont {E.~C.}\ \bibnamefont {Ives}},
  \bibinfo {author} {\bibfnamefont {D.~I.}\ \bibnamefont {Lazare}},\ and\
  \bibinfo {author} {\bibfnamefont {D.~M.}\ \bibnamefont {Buckland}},\
  }\bibfield  {title} {\bibinfo {title} {Expiration analysis of the
  {International Space Station} formulary for exploration mission planning},\
  }\href {https://doi.org/10.1038/s41526-024-00414-3} {\bibfield  {journal}
  {\bibinfo  {journal} {npj Microgravity}\ }\textbf {\bibinfo {volume} {10}},\
  \bibinfo {pages} {76} (\bibinfo {year} {2024})}\BibitemShut {NoStop}%
\bibitem [{\citenamefont {Plante}\ \emph {et~al.}(2025)\citenamefont {Plante},
  \citenamefont {Daniels}, \citenamefont {Young}, \citenamefont {Gaza},
  \citenamefont {Wu},\ and\ \citenamefont {Reichard}}]{Plante2025}%
  \BibitemOpen
  \bibfield  {author} {\bibinfo {author} {\bibfnamefont {I.}~\bibnamefont
  {Plante}}, \bibinfo {author} {\bibfnamefont {V.}~\bibnamefont {Daniels}},
  \bibinfo {author} {\bibfnamefont {M.}~\bibnamefont {Young}}, \bibinfo
  {author} {\bibfnamefont {R.}~\bibnamefont {Gaza}}, \bibinfo {author}
  {\bibfnamefont {H.}~\bibnamefont {Wu}},\ and\ \bibinfo {author}
  {\bibfnamefont {J.~F.}\ \bibnamefont {Reichard}},\ }\bibfield  {title}
  {\bibinfo {title} {The long-term stability of solid-state oral
  pharmaceuticals exposed to simulated intravehicular space radiation},\ }\href
  {https://doi.org/10.1038/s41526-025-00469-w} {\bibfield  {journal} {\bibinfo
  {journal} {npj Microgravity}\ }\textbf {\bibinfo {volume} {11}},\ \bibinfo
  {pages} {17} (\bibinfo {year} {2025})}\BibitemShut {NoStop}%
\bibitem [{\citenamefont {Kim}\ \emph {et~al.}(2024)\citenamefont {Kim},
  \citenamefont {Tierney}, \citenamefont {Overbey} \emph {et~al.}}]{Kim2024}%
  \BibitemOpen
  \bibfield  {author} {\bibinfo {author} {\bibfnamefont {J.}~\bibnamefont
  {Kim}}, \bibinfo {author} {\bibfnamefont {B.~T.}\ \bibnamefont {Tierney}},
  \bibinfo {author} {\bibfnamefont {E.~G.}\ \bibnamefont {Overbey}}, \emph
  {et~al.},\ }\bibfield  {title} {\bibinfo {title} {Single-cell multi-ome and
  immune profiles of the {Inspiration4} crew reveal conserved, cell-type, and
  sex-specific responses to spaceflight},\ }\href
  {https://doi.org/10.1038/s41467-024-49211-2} {\bibfield  {journal} {\bibinfo
  {journal} {Nature Communications}\ }\textbf {\bibinfo {volume} {15}},\
  \bibinfo {pages} {4954} (\bibinfo {year} {2024})}\BibitemShut {NoStop}%
\bibitem [{\citenamefont {Overbey}\ \emph {et~al.}(2024)\citenamefont
  {Overbey}, \citenamefont {Kim}, \citenamefont {Tierney} \emph
  {et~al.}}]{Overbey2024}%
  \BibitemOpen
  \bibfield  {author} {\bibinfo {author} {\bibfnamefont {E.~G.}\ \bibnamefont
  {Overbey}}, \bibinfo {author} {\bibfnamefont {J.}~\bibnamefont {Kim}},
  \bibinfo {author} {\bibfnamefont {B.~T.}\ \bibnamefont {Tierney}}, \emph
  {et~al.},\ }\bibfield  {title} {\bibinfo {title} {The space omics and medical
  atlas ({SOMA}) and international astronaut biobank},\ }\href
  {https://doi.org/10.1038/s41586-024-07639-y} {\bibfield  {journal} {\bibinfo
  {journal} {Nature}\ }\textbf {\bibinfo {volume} {632}},\ \bibinfo {pages}
  {1145--1154} (\bibinfo {year} {2024})}\BibitemShut {NoStop}%
\bibitem [{\citenamefont {Salido}\ \emph {et~al.}(2025)\citenamefont {Salido},
  \citenamefont {Zhao}, \citenamefont {McDonald} \emph {et~al.}}]{Salido2025}%
  \BibitemOpen
  \bibfield  {author} {\bibinfo {author} {\bibfnamefont {R.~A.}\ \bibnamefont
  {Salido}}, \bibinfo {author} {\bibfnamefont {H.~N.}\ \bibnamefont {Zhao}},
  \bibinfo {author} {\bibfnamefont {D.}~\bibnamefont {McDonald}}, \emph
  {et~al.},\ }\bibfield  {title} {\bibinfo {title} {The {International Space
  Station} has a unique and extreme microbial and chemical environment driven
  by use patterns},\ }\href {https://doi.org/10.1016/j.cell.2025.01.039}
  {\bibfield  {journal} {\bibinfo  {journal} {Cell}\ }\textbf {\bibinfo
  {volume} {188}},\ \bibinfo {pages} {2022--2041.e23} (\bibinfo {year}
  {2025})}\BibitemShut {NoStop}%
\bibitem [{\citenamefont {Waziry}\ \emph {et~al.}(2023)\citenamefont {Waziry}
  \emph {et~al.}}]{Waziry2023}%
  \BibitemOpen
  \bibfield  {author} {\bibinfo {author} {\bibfnamefont {R.}~\bibnamefont
  {Waziry}} \emph {et~al.},\ }\bibfield  {title} {\bibinfo {title} {Effect of
  long-term caloric restriction on {DNA} methylation measures of biological
  aging in healthy adults from the {CALERIE} trial},\ }\href
  {https://doi.org/10.1038/s43587-022-00357-y} {\bibfield  {journal} {\bibinfo
  {journal} {Nature Aging}\ }\textbf {\bibinfo {volume} {3}},\ \bibinfo {pages}
  {248--257} (\bibinfo {year} {2023})}\BibitemShut {NoStop}%
\bibitem [{\citenamefont {Yang}\ \emph {et~al.}(2023)\citenamefont {Yang} \emph
  {et~al.}}]{Yang2023}%
  \BibitemOpen
  \bibfield  {author} {\bibinfo {author} {\bibfnamefont {Y.}~\bibnamefont
  {Yang}} \emph {et~al.},\ }\bibfield  {title} {\bibinfo {title} {Induction of
  a torpor-like hypothermic and hypometabolic state in rodents by ultrasound},\
  }\href {https://doi.org/10.1038/s42255-023-00804-z} {\bibfield  {journal}
  {\bibinfo  {journal} {Nature Metabolism}\ }\textbf {\bibinfo {volume} {5}},\
  \bibinfo {pages} {789--803} (\bibinfo {year} {2023})}\BibitemShut {NoStop}%
\bibitem [{\citenamefont {Basner}\ \emph {et~al.}(2013)\citenamefont {Basner},
  \citenamefont {Dinges}, \citenamefont {Mollicone} \emph
  {et~al.}}]{Basner2013}%
  \BibitemOpen
  \bibfield  {author} {\bibinfo {author} {\bibfnamefont {M.}~\bibnamefont
  {Basner}}, \bibinfo {author} {\bibfnamefont {D.~F.}\ \bibnamefont {Dinges}},
  \bibinfo {author} {\bibfnamefont {D.~J.}\ \bibnamefont {Mollicone}}, \emph
  {et~al.},\ }\bibfield  {title} {\bibinfo {title} {{Mars} 520-d mission
  simulation reveals protracted crew hypokinesis and alterations of sleep
  duration and timing},\ }\href {https://doi.org/10.1073/pnas.1212646110}
  {\bibfield  {journal} {\bibinfo  {journal} {Proceedings of the National
  Academy of Sciences}\ }\textbf {\bibinfo {volume} {110}},\ \bibinfo {pages}
  {2635--2640} (\bibinfo {year} {2013})}\BibitemShut {NoStop}%
\bibitem [{\citenamefont {Basner}\ \emph {et~al.}(2014)\citenamefont {Basner},
  \citenamefont {Dinges}, \citenamefont {Mollicone} \emph
  {et~al.}}]{Basner2014}%
  \BibitemOpen
  \bibfield  {author} {\bibinfo {author} {\bibfnamefont {M.}~\bibnamefont
  {Basner}}, \bibinfo {author} {\bibfnamefont {D.~F.}\ \bibnamefont {Dinges}},
  \bibinfo {author} {\bibfnamefont {D.~J.}\ \bibnamefont {Mollicone}}, \emph
  {et~al.},\ }\bibfield  {title} {\bibinfo {title} {Psychological and
  behavioral changes during confinement in a 520-day simulated interplanetary
  mission to {Mars}},\ }\href {https://doi.org/10.1371/journal.pone.0093298}
  {\bibfield  {journal} {\bibinfo  {journal} {PLOS ONE}\ }\textbf {\bibinfo
  {volume} {9}},\ \bibinfo {pages} {e93298} (\bibinfo {year}
  {2014})}\BibitemShut {NoStop}%
\bibitem [{\citenamefont {Van~Dongen}\ \emph {et~al.}(2003)\citenamefont
  {Van~Dongen}, \citenamefont {Maislin}, \citenamefont {Mullington},\ and\
  \citenamefont {Dinges}}]{VanDongen2003}%
  \BibitemOpen
  \bibfield  {author} {\bibinfo {author} {\bibfnamefont {H.~P.~A.}\
  \bibnamefont {Van~Dongen}}, \bibinfo {author} {\bibfnamefont
  {G.}~\bibnamefont {Maislin}}, \bibinfo {author} {\bibfnamefont {J.~M.}\
  \bibnamefont {Mullington}},\ and\ \bibinfo {author} {\bibfnamefont {D.~F.}\
  \bibnamefont {Dinges}},\ }\bibfield  {title} {\bibinfo {title} {The
  cumulative cost of additional wakefulness: Dose-response effects on
  neurobehavioral functions and sleep physiology from chronic sleep restriction
  and total sleep deprivation},\ }\href
  {https://doi.org/10.1093/sleep/26.2.117} {\bibfield  {journal} {\bibinfo
  {journal} {Sleep}\ }\textbf {\bibinfo {volume} {26}},\ \bibinfo {pages}
  {117--126} (\bibinfo {year} {2003})}\BibitemShut {NoStop}%
\bibitem [{\citenamefont {{NASA}}(2026{\natexlab{c}})}]{MarsRelay}%
  \BibitemOpen
  \bibfield  {author} {\bibinfo {author} {\bibnamefont {{NASA}}},\ }\href
  {https://science.nasa.gov/mars/mars-relay-network/} {\bibinfo {title} {{Mars
  Relay Network}}} (\bibinfo {year} {2026}{\natexlab{c}})\BibitemShut {NoStop}%
\bibitem [{\citenamefont {Ostrom}(2009)}]{Ostrom2009}%
  \BibitemOpen
  \bibfield  {author} {\bibinfo {author} {\bibfnamefont {E.}~\bibnamefont
  {Ostrom}},\ }\bibfield  {title} {\bibinfo {title} {A general framework for
  analyzing sustainability of social-ecological systems},\ }\href
  {https://doi.org/10.1126/science.1172133} {\bibfield  {journal} {\bibinfo
  {journal} {Science}\ }\textbf {\bibinfo {volume} {325}},\ \bibinfo {pages}
  {419--422} (\bibinfo {year} {2009})}\BibitemShut {NoStop}%
\bibitem [{\citenamefont {Jones}(2017)}]{Jones2017}%
  \BibitemOpen
  \bibfield  {author} {\bibinfo {author} {\bibfnamefont {H.~W.}\ \bibnamefont
  {Jones}},\ }\href {https://ntrs.nasa.gov/citations/20180007067} {\emph
  {\bibinfo {title} {The Impact of Lower Launch Cost on Space Life Support}}},\
  \bibinfo {type} {Tech. Rep.}\ \bibinfo {number} {NTRS 20180007067}\ (\bibinfo
   {institution} {NASA},\ \bibinfo {year} {2017})\BibitemShut {NoStop}%
\bibitem [{\citenamefont {{NASA}}(2023{\natexlab{a}})}]{NASAWater2023}%
  \BibitemOpen
  \bibfield  {author} {\bibinfo {author} {\bibnamefont {{NASA}}},\ }\href
  {https://www.nasa.gov/missions/station/iss-research/nasa-achieves-water-recovery-milestone-on-international-space-station/}
  {\bibinfo {title} {{NASA Achieves Water Recovery Milestone on International
  Space Station}}} (\bibinfo {year} {2023}{\natexlab{a}})\BibitemShut {NoStop}%
\bibitem [{\citenamefont {Greenwood}\ \emph {et~al.}(2015)\citenamefont
  {Greenwood} \emph {et~al.}}]{Greenwood2015}%
  \BibitemOpen
  \bibfield  {author} {\bibinfo {author} {\bibfnamefont {Z.~W.}\ \bibnamefont
  {Greenwood}} \emph {et~al.},\ }\href
  {https://ntrs.nasa.gov/citations/20150016511} {\emph {\bibinfo {title}
  {Increased Oxygen Recovery from {Sabatier} Systems Using Plasma Pyrolysis
  Technology and Metal Hydride Separation}}},\ \bibinfo {type} {Tech. Rep.}\
  \bibinfo {number} {NTRS 20150016511}\ (\bibinfo  {institution} {NASA},\
  \bibinfo {year} {2015})\BibitemShut {NoStop}%
\bibitem [{\citenamefont {Chase}(1998)}]{Chase1998}%
  \BibitemOpen
  \bibfield  {author} {\bibinfo {author} {\bibfnamefont {M.~W.}\ \bibnamefont
  {Chase}, \bibfnamefont {Jr.}},\ }\href {https://webbook.nist.gov/chemistry/}
  {\emph {\bibinfo {title} {{NIST-JANAF} Thermochemical Tables}}},\ \bibinfo
  {edition} {4th}\ ed.,\ Journal of Physical and Chemical Reference Data,
  Monograph 9\ (\bibinfo  {publisher} {American Institute of Physics},\
  \bibinfo {year} {1998})\ \bibinfo {note} {standard-state entries for methane,
  carbon dioxide, liquid water, and oxygen; numerical values also available
  through NIST Chemistry WebBook, SRD 69}\BibitemShut {NoStop}%
\bibitem [{\citenamefont {Hoffman}\ \emph {et~al.}(2022)\citenamefont
  {Hoffman}, \citenamefont {Hecht}, \citenamefont {Rapp} \emph
  {et~al.}}]{Hoffman2022}%
  \BibitemOpen
  \bibfield  {author} {\bibinfo {author} {\bibfnamefont {J.~A.}\ \bibnamefont
  {Hoffman}}, \bibinfo {author} {\bibfnamefont {M.~H.}\ \bibnamefont {Hecht}},
  \bibinfo {author} {\bibfnamefont {D.}~\bibnamefont {Rapp}}, \emph {et~al.},\
  }\bibfield  {title} {\bibinfo {title} {{Mars Oxygen ISRU Experiment
  (MOXIE)}---preparing for human {Mars} exploration},\ }\href
  {https://doi.org/10.1126/sciadv.abp8636} {\bibfield  {journal} {\bibinfo
  {journal} {Science Advances}\ }\textbf {\bibinfo {volume} {8}},\ \bibinfo
  {pages} {eabp8636} (\bibinfo {year} {2022})}\BibitemShut {NoStop}%
\bibitem [{\citenamefont {Schaezler}\ \emph {et~al.}(2011)\citenamefont
  {Schaezler}, \citenamefont {Ghariani}, \citenamefont {Cook},\ and\
  \citenamefont {Leonard}}]{Schaezler2011}%
  \BibitemOpen
  \bibfield  {author} {\bibinfo {author} {\bibfnamefont {R.~N.}\ \bibnamefont
  {Schaezler}}, \bibinfo {author} {\bibfnamefont {A.}~\bibnamefont {Ghariani}},
  \bibinfo {author} {\bibfnamefont {A.~J.}\ \bibnamefont {Cook}},\ and\
  \bibinfo {author} {\bibfnamefont {D.~J.}\ \bibnamefont {Leonard}},\
  }\bibfield  {title} {\bibinfo {title} {Trending of overboard leakage of {ISS}
  cabin atmosphere},\ }in\ \href {https://doi.org/10.2514/6.2011-5149} {\emph
  {\bibinfo {booktitle} {41st International Conference on Environmental
  Systems}}},\ \bibinfo {series and number} {\bibinfo {number} {AIAA
  2011-5149}}\ (\bibinfo {year} {2011})\BibitemShut {NoStop}%
\bibitem [{\citenamefont {Trainer}\ \emph {et~al.}(2019)\citenamefont
  {Trainer}, \citenamefont {Wong}, \citenamefont {McConnochie}, \citenamefont
  {Franz}, \citenamefont {Atreya}, \citenamefont {Conrad}, \citenamefont
  {Lef{\`e}vre}, \citenamefont {Mahaffy}, \citenamefont {Malespin},
  \citenamefont {Manning}, \citenamefont {Mart{\'i}n-Torres}, \citenamefont
  {Mart{\'i}nez}, \citenamefont {McKay}, \citenamefont {Navarro-Gonz{\'a}lez},
  \citenamefont {Vicente-Retortillo}, \citenamefont {Webster},\ and\
  \citenamefont {Zorzano}}]{Trainer2019}%
  \BibitemOpen
  \bibfield  {author} {\bibinfo {author} {\bibfnamefont {M.~G.}\ \bibnamefont
  {Trainer}}, \bibinfo {author} {\bibfnamefont {M.~H.}\ \bibnamefont {Wong}},
  \bibinfo {author} {\bibfnamefont {T.~H.}\ \bibnamefont {McConnochie}},
  \bibinfo {author} {\bibfnamefont {H.~B.}\ \bibnamefont {Franz}}, \bibinfo
  {author} {\bibfnamefont {S.~K.}\ \bibnamefont {Atreya}}, \bibinfo {author}
  {\bibfnamefont {P.~G.}\ \bibnamefont {Conrad}}, \bibinfo {author}
  {\bibfnamefont {F.}~\bibnamefont {Lef{\`e}vre}}, \bibinfo {author}
  {\bibfnamefont {P.~R.}\ \bibnamefont {Mahaffy}}, \bibinfo {author}
  {\bibfnamefont {C.~A.}\ \bibnamefont {Malespin}}, \bibinfo {author}
  {\bibfnamefont {H.~L.~K.}\ \bibnamefont {Manning}}, \bibinfo {author}
  {\bibfnamefont {J.}~\bibnamefont {Mart{\'i}n-Torres}}, \bibinfo {author}
  {\bibfnamefont {G.~M.}\ \bibnamefont {Mart{\'i}nez}}, \bibinfo {author}
  {\bibfnamefont {C.~P.}\ \bibnamefont {McKay}}, \bibinfo {author}
  {\bibfnamefont {R.}~\bibnamefont {Navarro-Gonz{\'a}lez}}, \bibinfo {author}
  {\bibfnamefont {{\'A}.}~\bibnamefont {Vicente-Retortillo}}, \bibinfo {author}
  {\bibfnamefont {C.~R.}\ \bibnamefont {Webster}},\ and\ \bibinfo {author}
  {\bibfnamefont {M.-P.}\ \bibnamefont {Zorzano}},\ }\bibfield  {title}
  {\bibinfo {title} {Seasonal variations in atmospheric composition as measured
  in {Gale} crater, {Mars}},\ }\href {https://doi.org/10.1029/2019JE006175}
  {\bibfield  {journal} {\bibinfo  {journal} {JGR: Planets}\ }\textbf {\bibinfo
  {volume} {124}},\ \bibinfo {pages} {3000--3024} (\bibinfo {year}
  {2019})}\BibitemShut {NoStop}%
\bibitem [{\citenamefont {Colaprete}\ \emph {et~al.}(2010)\citenamefont
  {Colaprete}, \citenamefont {Schultz}, \citenamefont {Heldmann} \emph
  {et~al.}}]{Colaprete2010}%
  \BibitemOpen
  \bibfield  {author} {\bibinfo {author} {\bibfnamefont {A.}~\bibnamefont
  {Colaprete}}, \bibinfo {author} {\bibfnamefont {P.}~\bibnamefont {Schultz}},
  \bibinfo {author} {\bibfnamefont {J.}~\bibnamefont {Heldmann}}, \emph
  {et~al.},\ }\bibfield  {title} {\bibinfo {title} {Detection of water in the
  {LCROSS} ejecta plume},\ }\href {https://doi.org/10.1126/science.1186986}
  {\bibfield  {journal} {\bibinfo  {journal} {Science}\ }\textbf {\bibinfo
  {volume} {330}},\ \bibinfo {pages} {463--468} (\bibinfo {year}
  {2010})}\BibitemShut {NoStop}%
\bibitem [{\citenamefont {Poulet}\ \emph {et~al.}(2022)\citenamefont {Poulet}
  \emph {et~al.}}]{Poulet2022}%
  \BibitemOpen
  \bibfield  {author} {\bibinfo {author} {\bibfnamefont {L.}~\bibnamefont
  {Poulet}} \emph {et~al.},\ }\bibfield  {title} {\bibinfo {title} {Large-scale
  crop production for the moon and mars: Current gaps and future
  perspectives},\ }\href {https://doi.org/10.3389/fspas.2021.733944} {\bibfield
   {journal} {\bibinfo  {journal} {Frontiers in Astronomy and Space Sciences}\
  }\textbf {\bibinfo {volume} {8}},\ \bibinfo {pages} {733944} (\bibinfo {year}
  {2022})}\BibitemShut {NoStop}%
\bibitem [{\citenamefont {Wheeler}\ and\ \citenamefont
  {Sager}(1990)}]{WheelerSager1990}%
  \BibitemOpen
  \bibfield  {author} {\bibinfo {author} {\bibfnamefont {R.~M.}\ \bibnamefont
  {Wheeler}}\ and\ \bibinfo {author} {\bibfnamefont {J.~C.}\ \bibnamefont
  {Sager}},\ }\href {https://ntrs.nasa.gov/citations/19900016137} {\emph
  {\bibinfo {title} {Carbon Dioxide and Water Exchange Rates by a Wheat Crop in
  {NASA}'s Biomass Production Chamber: Results from an 86-Day Study (January to
  April 1989)}}},\ \bibinfo {type} {Tech. Rep.}\ \bibinfo {number}
  {NASA-TM-102788}\ (\bibinfo  {institution} {NASA, Kennedy Space Center},\
  \bibinfo {year} {1990})\BibitemShut {NoStop}%
\bibitem [{\citenamefont {{National Institute of Standards and
  Technology}}(2026)}]{NISTWaterVapor}%
  \BibitemOpen
  \bibfield  {author} {\bibinfo {author} {\bibnamefont {{National Institute of
  Standards and Technology}}},\ }\href
  {https://webbook.nist.gov/cgi/cbook.cgi?ID=C7732185\&Mask=4} {\bibinfo
  {title} {Water: Antoine equation parameters}},\ \bibinfo {howpublished}
  {{NIST} Chemistry WebBook, Standard Reference Database 69} (\bibinfo {year}
  {2026}),\ \bibinfo {note} {273--303 K correlation from the Bridgeman and
  Aldrich data; accessed 14 September 2026}\BibitemShut {NoStop}%
\bibitem [{\citenamefont {Fu}\ \emph {et~al.}(2025)\citenamefont {Fu},
  \citenamefont {Liu}, \citenamefont {Liu}, \citenamefont {Hu}, \citenamefont
  {Xie}, \citenamefont {Liu}, \citenamefont {Yi},\ and\ \citenamefont
  {Liu}}]{Fu2025}%
  \BibitemOpen
  \bibfield  {author} {\bibinfo {author} {\bibfnamefont {Y.}~\bibnamefont
  {Fu}}, \bibinfo {author} {\bibfnamefont {H.}~\bibnamefont {Liu}}, \bibinfo
  {author} {\bibfnamefont {D.}~\bibnamefont {Liu}}, \bibinfo {author}
  {\bibfnamefont {D.}~\bibnamefont {Hu}}, \bibinfo {author} {\bibfnamefont
  {B.}~\bibnamefont {Xie}}, \bibinfo {author} {\bibfnamefont {G.}~\bibnamefont
  {Liu}}, \bibinfo {author} {\bibfnamefont {Z.}~\bibnamefont {Yi}},\ and\
  \bibinfo {author} {\bibfnamefont {H.}~\bibnamefont {Liu}},\ }\bibfield
  {title} {\bibinfo {title} {A case for supporting human long-term survival on
  the moon: `{Lunar Palace 365}' mission},\ }\href
  {https://doi.org/10.1016/j.actaastro.2024.12.002} {\bibfield  {journal}
  {\bibinfo  {journal} {Acta Astronautica}\ }\textbf {\bibinfo {volume}
  {228}},\ \bibinfo {pages} {131--140} (\bibinfo {year} {2025})}\BibitemShut
  {NoStop}%
\bibitem [{\citenamefont {Severinghaus}\ \emph {et~al.}(1994)\citenamefont
  {Severinghaus}, \citenamefont {Broecker}, \citenamefont {Dempster},
  \citenamefont {MacCallum},\ and\ \citenamefont {Wahlen}}]{Severinghaus1994}%
  \BibitemOpen
  \bibfield  {author} {\bibinfo {author} {\bibfnamefont {J.~P.}\ \bibnamefont
  {Severinghaus}}, \bibinfo {author} {\bibfnamefont {W.~S.}\ \bibnamefont
  {Broecker}}, \bibinfo {author} {\bibfnamefont {W.~F.}\ \bibnamefont
  {Dempster}}, \bibinfo {author} {\bibfnamefont {T.}~\bibnamefont
  {MacCallum}},\ and\ \bibinfo {author} {\bibfnamefont {M.}~\bibnamefont
  {Wahlen}},\ }\bibfield  {title} {\bibinfo {title} {Oxygen loss in {Biosphere
  2}},\ }\href {https://doi.org/10.1029/94EO00285} {\bibfield  {journal}
  {\bibinfo  {journal} {Eos, Transactions American Geophysical Union}\ }\textbf
  {\bibinfo {volume} {75}},\ \bibinfo {pages} {33--37} (\bibinfo {year}
  {1994})}\BibitemShut {NoStop}%
\bibitem [{\citenamefont {Leger}\ \emph {et~al.}(2021)\citenamefont {Leger},
  \citenamefont {Matassa}, \citenamefont {Noor}, \citenamefont {Shepon},
  \citenamefont {Milo},\ and\ \citenamefont {Bar-Even}}]{Leger2021}%
  \BibitemOpen
  \bibfield  {author} {\bibinfo {author} {\bibfnamefont {D.}~\bibnamefont
  {Leger}}, \bibinfo {author} {\bibfnamefont {S.}~\bibnamefont {Matassa}},
  \bibinfo {author} {\bibfnamefont {E.}~\bibnamefont {Noor}}, \bibinfo {author}
  {\bibfnamefont {A.}~\bibnamefont {Shepon}}, \bibinfo {author} {\bibfnamefont
  {R.}~\bibnamefont {Milo}},\ and\ \bibinfo {author} {\bibfnamefont
  {A.}~\bibnamefont {Bar-Even}},\ }\bibfield  {title} {\bibinfo {title}
  {Photovoltaic-driven microbial protein production can use land and sunlight
  more efficiently than conventional crops},\ }\href
  {https://doi.org/10.1073/pnas.2015025118} {\bibfield  {journal} {\bibinfo
  {journal} {Proc. Nat. Acad. Sci.}\ }\textbf {\bibinfo {volume} {118}},\
  \bibinfo {pages} {e2015025118} (\bibinfo {year} {2021})}\BibitemShut
  {NoStop}%
\bibitem [{\citenamefont {Hecht}\ \emph {et~al.}(2009)\citenamefont {Hecht},
  \citenamefont {Kounaves}, \citenamefont {Quinn} \emph {et~al.}}]{Hecht2009}%
  \BibitemOpen
  \bibfield  {author} {\bibinfo {author} {\bibfnamefont {M.~H.}\ \bibnamefont
  {Hecht}}, \bibinfo {author} {\bibfnamefont {S.~P.}\ \bibnamefont {Kounaves}},
  \bibinfo {author} {\bibfnamefont {R.~C.}\ \bibnamefont {Quinn}}, \emph
  {et~al.},\ }\bibfield  {title} {\bibinfo {title} {Detection of perchlorate
  and the soluble chemistry of {Martian} soil at the {Phoenix} lander site},\
  }\href {https://doi.org/10.1126/science.1172466} {\bibfield  {journal}
  {\bibinfo  {journal} {Science}\ }\textbf {\bibinfo {volume} {325}},\ \bibinfo
  {pages} {64--67} (\bibinfo {year} {2009})}\BibitemShut {NoStop}%
\bibitem [{\citenamefont {Oze}\ \emph {et~al.}(2021)\citenamefont {Oze},
  \citenamefont {Beisel}, \citenamefont {Dabsys}, \citenamefont {Dall},
  \citenamefont {North}, \citenamefont {Scott}, \citenamefont {Lopez},
  \citenamefont {Holmes},\ and\ \citenamefont {Fendorf}}]{Oze2021}%
  \BibitemOpen
  \bibfield  {author} {\bibinfo {author} {\bibfnamefont {C.}~\bibnamefont
  {Oze}}, \bibinfo {author} {\bibfnamefont {J.}~\bibnamefont {Beisel}},
  \bibinfo {author} {\bibfnamefont {E.}~\bibnamefont {Dabsys}}, \bibinfo
  {author} {\bibfnamefont {J.}~\bibnamefont {Dall}}, \bibinfo {author}
  {\bibfnamefont {G.}~\bibnamefont {North}}, \bibinfo {author} {\bibfnamefont
  {A.}~\bibnamefont {Scott}}, \bibinfo {author} {\bibfnamefont {A.~M.}\
  \bibnamefont {Lopez}}, \bibinfo {author} {\bibfnamefont {R.}~\bibnamefont
  {Holmes}},\ and\ \bibinfo {author} {\bibfnamefont {S.}~\bibnamefont
  {Fendorf}},\ }\bibfield  {title} {\bibinfo {title} {Perchlorate and
  agriculture on {Mars}},\ }\href {https://doi.org/10.3390/soilsystems5030037}
  {\bibfield  {journal} {\bibinfo  {journal} {Soil Systems}\ }\textbf {\bibinfo
  {volume} {5}},\ \bibinfo {pages} {37} (\bibinfo {year} {2021})}\BibitemShut
  {NoStop}%
\bibitem [{\citenamefont {Coker}\ \emph {et~al.}(2026)\citenamefont {Coker},
  \citenamefont {Denvir},\ and\ \citenamefont {Howe}}]{Coker2026}%
  \BibitemOpen
  \bibfield  {author} {\bibinfo {author} {\bibfnamefont {H.~R.}\ \bibnamefont
  {Coker}}, \bibinfo {author} {\bibfnamefont {A.~C.}\ \bibnamefont {Denvir}},\
  and\ \bibinfo {author} {\bibfnamefont {J.~A.}\ \bibnamefont {Howe}},\
  }\bibfield  {title} {\bibinfo {title} {Simple and effective remediation
  strategies of {Martian} perchlorates},\ }\href
  {https://doi.org/10.1002/saj2.70201} {\bibfield  {journal} {\bibinfo
  {journal} {Soil Science Society of America Journal}\ }\textbf {\bibinfo
  {volume} {90}},\ \bibinfo {pages} {e70201} (\bibinfo {year}
  {2026})}\BibitemShut {NoStop}%
\bibitem [{\citenamefont {Turyshev}(2026{\natexlab{a}})}]{Turyshev2026}%
  \BibitemOpen
  \bibfield  {author} {\bibinfo {author} {\bibfnamefont {S.~G.}\ \bibnamefont
  {Turyshev}},\ }\bibfield  {title} {\bibinfo {title} {Lunar dust: Formation,
  microphysics, and transport},\ }\href {https://doi.org/10.1103/hn1p-nx25}
  {\bibfield  {journal} {\bibinfo  {journal} {Physical Review Research}\
  }\textbf {\bibinfo {volume} {8}},\ \bibinfo {pages} {013307} (\bibinfo {year}
  {2026}{\natexlab{a}})},\ \Eprint {https://arxiv.org/abs/2511.08503}
  {arXiv:2511.08503 [astro-ph.EP]} \BibitemShut {NoStop}%
\bibitem [{\citenamefont {Johnson}\ \emph {et~al.}(1971)\citenamefont
  {Johnson}, \citenamefont {Kendall},\ and\ \citenamefont
  {Roberts}}]{Johnson1971}%
  \BibitemOpen
  \bibfield  {author} {\bibinfo {author} {\bibfnamefont {K.~L.}\ \bibnamefont
  {Johnson}}, \bibinfo {author} {\bibfnamefont {K.}~\bibnamefont {Kendall}},\
  and\ \bibinfo {author} {\bibfnamefont {A.~D.}\ \bibnamefont {Roberts}},\
  }\bibfield  {title} {\bibinfo {title} {Surface energy and the contact of
  elastic solids},\ }\href {https://doi.org/10.1098/rspa.1971.0141} {\bibfield
  {journal} {\bibinfo  {journal} {Proceedings of the Royal Society of London.
  Series A}\ }\textbf {\bibinfo {volume} {324}},\ \bibinfo {pages} {301--313}
  (\bibinfo {year} {1971})}\BibitemShut {NoStop}%
\bibitem [{\citenamefont {Derjaguin}\ \emph {et~al.}(1975)\citenamefont
  {Derjaguin}, \citenamefont {Muller},\ and\ \citenamefont
  {Toporov}}]{Derjaguin1975}%
  \BibitemOpen
  \bibfield  {author} {\bibinfo {author} {\bibfnamefont {B.~V.}\ \bibnamefont
  {Derjaguin}}, \bibinfo {author} {\bibfnamefont {V.~M.}\ \bibnamefont
  {Muller}},\ and\ \bibinfo {author} {\bibfnamefont {Y.~P.}\ \bibnamefont
  {Toporov}},\ }\bibfield  {title} {\bibinfo {title} {Effect of contact
  deformations on the adhesion of particles},\ }\href
  {https://doi.org/10.1016/0021-9797(75)90018-1} {\bibfield  {journal}
  {\bibinfo  {journal} {Journal of Colloid and Interface Science}\ }\textbf
  {\bibinfo {volume} {53}},\ \bibinfo {pages} {314--326} (\bibinfo {year}
  {1975})}\BibitemShut {NoStop}%
\bibitem [{\citenamefont {{NASA}}(2025{\natexlab{b}})}]{NASAEDS2025}%
  \BibitemOpen
  \bibfield  {author} {\bibinfo {author} {\bibnamefont {{NASA}}},\ }\href
  {https://www.nasa.gov/image-article/nasas-dust-shield-successfully-repels-lunar-regolith-on-moon/}
  {\bibinfo {title} {{NASA}'s dust shield successfully repels lunar regolith on
  {Moon}}} (\bibinfo {year} {2025}{\natexlab{b}}),\ \bibinfo {note} {primary
  mission report, March 27, 2025}\BibitemShut {NoStop}%
\bibitem [{\citenamefont {Ebert}\ \emph {et~al.}(2026)\citenamefont {Ebert},
  \citenamefont {Lowe},\ and\ \citenamefont {Francisco}}]{MarsDust2026}%
  \BibitemOpen
  \bibfield  {author} {\bibinfo {author} {\bibfnamefont {D.}~\bibnamefont
  {Ebert}}, \bibinfo {author} {\bibfnamefont {K.}~\bibnamefont {Lowe}},\ and\
  \bibinfo {author} {\bibfnamefont {D.}~\bibnamefont {Francisco}},\ }\href
  {https://www.nasa.gov/wp-content/uploads/2026/07/martian-dust-exposure-limits-sp-20260005907.pdf}
  {\emph {\bibinfo {title} {{Martian} Dust Exposure Limits: Permissible
  Exposure Limits ({PELs}), Toxicological Risks, and Health Standards for
  {Martian} Dust as Discussed by the {Mars} Dust Working Group}}},\ \bibinfo
  {type} {Tech. Rep.}\ \bibinfo {number} {NASA/SP-20260005907}\ (\bibinfo
  {institution} {NASA},\ \bibinfo {year} {2026})\BibitemShut {NoStop}%
\bibitem [{\citenamefont {Hanford}(2006)}]{Hanford2006}%
  \BibitemOpen
  \bibfield  {author} {\bibinfo {author} {\bibfnamefont {A.~J.}\ \bibnamefont
  {Hanford}},\ }\href {https://spacearchitect.org/pubs/NASA-CR-2006-213693.pdf}
  {\emph {\bibinfo {title} {Exploration Life Support Baseline Values and
  Assumptions Document}}},\ \bibinfo {type} {Tech. Rep.}\ \bibinfo {number}
  {NASA CR-2006-213693}\ (\bibinfo  {institution} {NASA Johnson Space Center},\
  \bibinfo {year} {2006})\ \bibinfo {note} {edited by A. J. Hanford;
  thermal-control accounting in Table 3.2.14 and crew-time accounting in Sec.
  3.3}\BibitemShut {NoStop}%
\bibitem [{\citenamefont {Mohammad}\ \emph {et~al.}(2021)\citenamefont
  {Mohammad}, \citenamefont {Dupuis}, \citenamefont {Funkenbusch},\ and\
  \citenamefont {Kelley}}]{Mohammad2021}%
  \BibitemOpen
  \bibfield  {author} {\bibinfo {author} {\bibfnamefont {I.}~\bibnamefont
  {Mohammad}}, \bibinfo {author} {\bibfnamefont {M.}~\bibnamefont {Dupuis}},
  \bibinfo {author} {\bibfnamefont {P.~D.}\ \bibnamefont {Funkenbusch}},\ and\
  \bibinfo {author} {\bibfnamefont {D.~H.}\ \bibnamefont {Kelley}},\ }\href
  {https://doi.org/10.48550/arXiv.2106.12606} {\bibinfo {title} {Oscillating
  currents stabilize aluminium cells for efficient, low carbon production}}
  (\bibinfo {year} {2021}),\ \bibinfo {note} {preprint; the reported 2019
  production/electricity totals are used only as a terrestrial process-scale
  comparison},\ \Eprint {https://arxiv.org/abs/2106.12606} {arXiv:2106.12606
  [physics.flu-dyn]} \BibitemShut {NoStop}%
\bibitem [{\citenamefont {{International Aluminium
  Institute}}(2024)}]{IAISmelting2024}%
  \BibitemOpen
  \bibfield  {author} {\bibinfo {author} {\bibnamefont {{International
  Aluminium Institute}}},\ }\href
  {https://international-aluminium.org/statistics/primary-aluminium-smelting-energy-intensity/}
  {\bibinfo {title} {Primary aluminium smelting energy intensity}} (\bibinfo
  {year} {2024}),\ \bibinfo {note} {statistical boundary definitions; excludes
  casting and carbon-plant electricity}\BibitemShut {NoStop}%
\bibitem [{\citenamefont {Giandrea}\ \emph {et~al.}(2022)\citenamefont
  {Giandrea}, \citenamefont {Kornfeld}, \citenamefont {Meyer},\ and\
  \citenamefont {Powers}}]{Giandrea2022AssetLives}%
  \BibitemOpen
  \bibfield  {author} {\bibinfo {author} {\bibfnamefont {M.~D.}\ \bibnamefont
  {Giandrea}}, \bibinfo {author} {\bibfnamefont {R.~J.}\ \bibnamefont
  {Kornfeld}}, \bibinfo {author} {\bibfnamefont {P.~B.}\ \bibnamefont
  {Meyer}},\ and\ \bibinfo {author} {\bibfnamefont {S.~G.}\ \bibnamefont
  {Powers}},\ }\bibfield  {title} {\bibinfo {title} {Alternative capital asset
  depreciation rates for {U.S.} capital and total factor productivity
  measures},\ }\bibfield  {journal} {\bibinfo  {journal} {Monthly Labor
  Review}\ }\href {https://doi.org/10.21916/mlr.2022.24} {10.21916/mlr.2022.24}
  (\bibinfo {year} {2022})\BibitemShut {NoStop}%
\bibitem [{\citenamefont {Kopp}\ and\ \citenamefont
  {Lean}(2011)}]{KoppLean2011}%
  \BibitemOpen
  \bibfield  {author} {\bibinfo {author} {\bibfnamefont {G.}~\bibnamefont
  {Kopp}}\ and\ \bibinfo {author} {\bibfnamefont {J.~L.}\ \bibnamefont
  {Lean}},\ }\bibfield  {title} {\bibinfo {title} {A new, lower value of total
  solar irradiance: Evidence and climate significance},\ }\href
  {https://doi.org/10.1029/2010GL045777} {\bibfield  {journal} {\bibinfo
  {journal} {Geophys. Res. Lett.}\ }\textbf {\bibinfo {volume} {38}},\ \bibinfo
  {pages} {L01706} (\bibinfo {year} {2011})}\BibitemShut {NoStop}%
\bibitem [{\citenamefont {Poston}\ \emph {et~al.}(2020)\citenamefont {Poston},
  \citenamefont {Gibson}, \citenamefont {Sanchez},\ and\ \citenamefont
  {McClure}}]{Poston2020}%
  \BibitemOpen
  \bibfield  {author} {\bibinfo {author} {\bibfnamefont {D.~I.}\ \bibnamefont
  {Poston}}, \bibinfo {author} {\bibfnamefont {M.~A.}\ \bibnamefont {Gibson}},
  \bibinfo {author} {\bibfnamefont {R.~G.}\ \bibnamefont {Sanchez}},\ and\
  \bibinfo {author} {\bibfnamefont {P.~R.}\ \bibnamefont {McClure}},\
  }\bibfield  {title} {\bibinfo {title} {Results of the {KRUSTY} nuclear system
  test},\ }\href {https://doi.org/10.1080/00295450.2020.1730673} {\bibfield
  {journal} {\bibinfo  {journal} {Nuclear Technology}\ }\textbf {\bibinfo
  {volume} {206}},\ \bibinfo {pages} {S89--S117} (\bibinfo {year}
  {2020})}\BibitemShut {NoStop}%
\bibitem [{\citenamefont {Barrett}\ and\ \citenamefont
  {Johnson}(2005)}]{Barrett2005}%
  \BibitemOpen
  \bibfield  {author} {\bibinfo {author} {\bibfnamefont {M.~J.}\ \bibnamefont
  {Barrett}}\ and\ \bibinfo {author} {\bibfnamefont {P.~K.}\ \bibnamefont
  {Johnson}},\ }\bibfield  {title} {\bibinfo {title} {Performance and mass
  modeling subtleties in closed-{Brayton}-cycle space power systems},\ }in\
  \href {https://doi.org/10.2514/6.2005-5700} {\emph {\bibinfo {booktitle} {3rd
  International Energy Conversion Engineering Conference}}},\ \bibinfo {series
  and number} {\bibinfo {number} {AIAA 2005-5700}}\ (\bibinfo {year}
  {2005})\BibitemShut {NoStop}%
\bibitem [{\citenamefont {Mart{\'i}nez}\ \emph {et~al.}(2021)\citenamefont
  {Mart{\'i}nez}, \citenamefont {Vicente-Retortillo}, \citenamefont {Vasavada},
  \citenamefont {Newman}, \citenamefont {Fischer}, \citenamefont {Renn{\'o}}
  \emph {et~al.}}]{Martinez2021}%
  \BibitemOpen
  \bibfield  {author} {\bibinfo {author} {\bibfnamefont {G.~M.}\ \bibnamefont
  {Mart{\'i}nez}}, \bibinfo {author} {\bibfnamefont {A.}~\bibnamefont
  {Vicente-Retortillo}}, \bibinfo {author} {\bibfnamefont {A.~R.}\ \bibnamefont
  {Vasavada}}, \bibinfo {author} {\bibfnamefont {C.~E.}\ \bibnamefont
  {Newman}}, \bibinfo {author} {\bibfnamefont {E.}~\bibnamefont {Fischer}},
  \bibinfo {author} {\bibfnamefont {N.~O.}\ \bibnamefont {Renn{\'o}}}, \emph
  {et~al.},\ }\bibfield  {title} {\bibinfo {title} {The surface energy budget
  at {Gale Crater} during the first 2500 sols of the {Mars Science Laboratory}
  mission},\ }\href {https://doi.org/10.1029/2020JE006804} {\bibfield
  {journal} {\bibinfo  {journal} {JGR: Planets}\ }\textbf {\bibinfo {volume}
  {126}},\ \bibinfo {pages} {e2020JE006804} (\bibinfo {year}
  {2021})}\BibitemShut {NoStop}%
\bibitem [{\citenamefont {{NASA}}(2024{\natexlab{a}})}]{NASAFSP2024}%
  \BibitemOpen
  \bibfield  {author} {\bibinfo {author} {\bibnamefont {{NASA}}},\ }\href
  {https://www.nasa.gov/centers-and-facilities/glenn/nasas-fission-surface-power-project-energizes-lunar-exploration/}
  {\bibinfo {title} {{NASA}'s fission surface power project energizes lunar
  exploration}} (\bibinfo {year} {2024}{\natexlab{a}}),\ \bibinfo {note}
  {earlier Phase-I design targets: 40 kW electrical, less than 6 t, decade
  operation; not treated as achieved plant performance}\BibitemShut {NoStop}%
\bibitem [{\citenamefont {{NASA}}(2025{\natexlab{c}})}]{NASAFSP2025}%
  \BibitemOpen
  \bibfield  {author} {\bibinfo {author} {\bibnamefont {{NASA}}},\ }\href
  {https://www.nasa.gov/centers-and-facilities/glenn/nasa-seeks-industry-feedback-on-fission-surface-power/}
  {\bibinfo {title} {{NASA} seeks industry feedback on fission surface power}}
  (\bibinfo {year} {2025}{\natexlab{c}}),\ \bibinfo {note} {14 August 2025,
  updated 5 December 2025: at least 100 kW electrical and closed-Brayton
  conversion in the development solicitation}\BibitemShut {NoStop}%
\bibitem [{\citenamefont {{NASA}}(2026{\natexlab{d}})}]{NASAFSP2030}%
  \BibitemOpen
  \bibfield  {author} {\bibinfo {author} {\bibnamefont {{NASA}}},\ }\href
  {https://www.nasa.gov/news-release/nasa-department-of-energy-to-develop-lunar-surface-reactor-by-2030/}
  {\bibinfo {title} {{NASA}, department of energy to develop lunar surface
  reactor by 2030}} (\bibinfo {year} {2026}{\natexlab{d}}),\ \bibinfo {note}
  {release 26-004, 13 January 2026, updated 2 February 2026; program objective,
  not a demonstrated readiness date.}\BibitemShut {Stop}%
\bibitem [{\citenamefont {Hanley}\ and\ \citenamefont
  {Lippman-Hand}(1983)}]{Hanley1983}%
  \BibitemOpen
  \bibfield  {author} {\bibinfo {author} {\bibfnamefont {J.~A.}\ \bibnamefont
  {Hanley}}\ and\ \bibinfo {author} {\bibfnamefont {A.}~\bibnamefont
  {Lippman-Hand}},\ }\bibfield  {title} {\bibinfo {title} {If nothing goes
  wrong, is everything all right? interpreting zero numerators},\ }\href
  {https://doi.org/10.1001/jama.1983.03330370053031} {\bibfield  {journal}
  {\bibinfo  {journal} {JAMA}\ }\textbf {\bibinfo {volume} {249}},\ \bibinfo
  {pages} {1743--1745} (\bibinfo {year} {1983})}\BibitemShut {NoStop}%
\bibitem [{\citenamefont {{U.S. Nuclear Regulatory
  Commission}}(1990)}]{NUREG1150}%
  \BibitemOpen
  \bibfield  {author} {\bibinfo {author} {\bibnamefont {{U.S. Nuclear
  Regulatory Commission}}},\ }\href
  {https://www.nrc.gov/reading-rm/doc-collections/nuregs/staff/sr1150/v1/index}
  {\emph {\bibinfo {title} {Severe Accident Risks: An Assessment for Five
  {U.S.} Nuclear Power Plants---Final Summary Report}}},\ \bibinfo {type}
  {Tech. Rep.}\ \bibinfo {number} {NUREG-1150, Volume 1}\ (\bibinfo
  {institution} {U.S. Nuclear Regulatory Commission},\ \bibinfo {year}
  {1990})\BibitemShut {NoStop}%
\bibitem [{\citenamefont {{U.S. Nuclear Regulatory
  Commission}}(2018)}]{NRC174}%
  \BibitemOpen
  \bibfield  {author} {\bibinfo {author} {\bibnamefont {{U.S. Nuclear
  Regulatory Commission}}},\ }\href
  {https://www.nrc.gov/docs/ML1731/ML17317A256.pdf} {\emph {\bibinfo {title}
  {An Approach for Using Probabilistic Risk Assessment in Risk-Informed
  Decisions on Plant-Specific Changes to the Licensing Basis}}},\ \bibinfo
  {type} {Tech. Rep.}\ \bibinfo {number} {Regulatory Guide 1.174, Revision 3}\
  (\bibinfo  {institution} {U.S. Nuclear Regulatory Commission},\ \bibinfo
  {year} {2018})\BibitemShut {NoStop}%
\bibitem [{\citenamefont {{Federal Aviation Administration}}(2024)}]{FAA1309}%
  \BibitemOpen
  \bibfield  {author} {\bibinfo {author} {\bibnamefont {{Federal Aviation
  Administration}}},\ }\href
  {https://www.faa.gov/documentLibrary/media/Advisory_Circular/AC_25.1309-1B.pdf}
  {\emph {\bibinfo {title} {System Design and Analysis}}},\ \bibinfo {type}
  {Tech. Rep.}\ \bibinfo {number} {Advisory Circular 25.1309-1B}\ (\bibinfo
  {institution} {U.S. Department of Transportation},\ \bibinfo {year}
  {2024})\BibitemShut {NoStop}%
\bibitem [{\citenamefont {Schuler}\ \emph {et~al.}(2019)\citenamefont
  {Schuler}, \citenamefont {Smith}, \citenamefont {Mueller},\ and\
  \citenamefont {Nick}}]{Schuler2019}%
  \BibitemOpen
  \bibfield  {author} {\bibinfo {author} {\bibfnamefont {J.~M.}\ \bibnamefont
  {Schuler}}, \bibinfo {author} {\bibfnamefont {J.~D.}\ \bibnamefont {Smith}},
  \bibinfo {author} {\bibfnamefont {R.~P.}\ \bibnamefont {Mueller}},\ and\
  \bibinfo {author} {\bibfnamefont {A.~J.}\ \bibnamefont {Nick}},\ }\bibfield
  {title} {\bibinfo {title} {{RASSOR}, the reduced gravity excavator},\ }in\
  \href {https://www.hou.usra.edu/meetings/lunarisru2019/pdf/5061.pdf} {\emph
  {\bibinfo {booktitle} {Developing a New Space Economy: Lunar Resources and
  Their Utilization}}}\ (\bibinfo {year} {2019})\ \bibinfo {note} {abstract
  5061; prototype parameters and qualification status as reported}\BibitemShut
  {NoStop}%
\bibitem [{\citenamefont {Giel}\ \emph {et~al.}(2025)\citenamefont {Giel},
  \citenamefont {Hurrell}, \citenamefont {Santra}, \citenamefont {Mishra},
  \citenamefont {Uno},\ and\ \citenamefont {Yoshida}}]{Giel2025}%
  \BibitemOpen
  \bibfield  {author} {\bibinfo {author} {\bibfnamefont {S.}~\bibnamefont
  {Giel}}, \bibinfo {author} {\bibfnamefont {J.}~\bibnamefont {Hurrell}},
  \bibinfo {author} {\bibfnamefont {S.}~\bibnamefont {Santra}}, \bibinfo
  {author} {\bibfnamefont {A.}~\bibnamefont {Mishra}}, \bibinfo {author}
  {\bibfnamefont {K.}~\bibnamefont {Uno}},\ and\ \bibinfo {author}
  {\bibfnamefont {K.}~\bibnamefont {Yoshida}},\ }\href
  {https://arxiv.org/abs/2511.00492} {\bibinfo {title} {Design and development
  of a modular bucket drum excavator for lunar {ISRU}}} (\bibinfo {year}
  {2025}),\ \bibinfo {note} {preprint; measured terrestrial filling tests and
  separately modeled batch duty},\ \Eprint {https://arxiv.org/abs/2511.00492}
  {arXiv:2511.00492 [cs.RO]} \BibitemShut {NoStop}%
\bibitem [{\citenamefont {Gin{\'e}s-Palomares}\ \emph
  {et~al.}(2023)\citenamefont {Gin{\'e}s-Palomares}, \citenamefont {Fateri},
  \citenamefont {Kalh{\"o}fer}, \citenamefont {Schubert}, \citenamefont
  {Meyer}, \citenamefont {Kolsch}, \citenamefont {Brandi{\'c}~Lipi{\'n}ska},
  \citenamefont {Davenport}, \citenamefont {Imhof}, \citenamefont {Waclavicek},
  \citenamefont {Sperl}, \citenamefont {Makaya},\ and\ \citenamefont
  {G{\"u}nster}}]{Gines2023}%
  \BibitemOpen
  \bibfield  {author} {\bibinfo {author} {\bibfnamefont {J.-C.}\ \bibnamefont
  {Gin{\'e}s-Palomares}}, \bibinfo {author} {\bibfnamefont {M.}~\bibnamefont
  {Fateri}}, \bibinfo {author} {\bibfnamefont {E.}~\bibnamefont
  {Kalh{\"o}fer}}, \bibinfo {author} {\bibfnamefont {T.}~\bibnamefont
  {Schubert}}, \bibinfo {author} {\bibfnamefont {L.}~\bibnamefont {Meyer}},
  \bibinfo {author} {\bibfnamefont {N.}~\bibnamefont {Kolsch}}, \bibinfo
  {author} {\bibfnamefont {M.}~\bibnamefont {Brandi{\'c}~Lipi{\'n}ska}},
  \bibinfo {author} {\bibfnamefont {R.}~\bibnamefont {Davenport}}, \bibinfo
  {author} {\bibfnamefont {B.}~\bibnamefont {Imhof}}, \bibinfo {author}
  {\bibfnamefont {R.}~\bibnamefont {Waclavicek}}, \bibinfo {author}
  {\bibfnamefont {M.}~\bibnamefont {Sperl}}, \bibinfo {author} {\bibfnamefont
  {A.}~\bibnamefont {Makaya}},\ and\ \bibinfo {author} {\bibfnamefont
  {J.}~\bibnamefont {G{\"u}nster}},\ }\bibfield  {title} {\bibinfo {title}
  {Laser melting manufacturing of large elements of lunar regolith simulant for
  paving on the {Moon}},\ }\href {https://doi.org/10.1038/s41598-023-42008-1}
  {\bibfield  {journal} {\bibinfo  {journal} {Scientific Reports}\ }\textbf
  {\bibinfo {volume} {13}},\ \bibinfo {pages} {15593} (\bibinfo {year}
  {2023})}\BibitemShut {NoStop}%
\bibitem [{\citenamefont {{European Space Agency}}(2020)}]{ESAoxygen2020}%
  \BibitemOpen
  \bibfield  {author} {\bibinfo {author} {\bibnamefont {{European Space
  Agency}}},\ }\href
  {https://www.esa.int/Enabling_Support/Space_Engineering_Technology/ESA_opens_oxygen_plant_making_air_out_of_moondust}
  {\bibinfo {title} {{ESA} opens oxygen plant---making air out of moondust}}
  (\bibinfo {year} {2020}),\ \bibinfo {note} {primary laboratory report, 17
  January 2020; molten-salt process and product boundaries}\BibitemShut
  {NoStop}%
\bibitem [{\citenamefont {Prater}\ \emph {et~al.}(2019)\citenamefont {Prater},
  \citenamefont {Werkheiser}, \citenamefont {Ledbetter}, \citenamefont
  {Timucin}, \citenamefont {Wheeler},\ and\ \citenamefont
  {Snyder}}]{Prater2019}%
  \BibitemOpen
  \bibfield  {author} {\bibinfo {author} {\bibfnamefont {T.}~\bibnamefont
  {Prater}}, \bibinfo {author} {\bibfnamefont {N.}~\bibnamefont {Werkheiser}},
  \bibinfo {author} {\bibfnamefont {F.}~\bibnamefont {Ledbetter}}, \bibinfo
  {author} {\bibfnamefont {D.}~\bibnamefont {Timucin}}, \bibinfo {author}
  {\bibfnamefont {K.}~\bibnamefont {Wheeler}},\ and\ \bibinfo {author}
  {\bibfnamefont {M.}~\bibnamefont {Snyder}},\ }\bibfield  {title} {\bibinfo
  {title} {{3D} printing in zero g technology demonstration mission: complete
  experimental results and summary of related material modeling efforts},\
  }\href {https://doi.org/10.1007/s00170-018-2827-7} {\bibfield  {journal}
  {\bibinfo  {journal} {The International Journal of Advanced Manufacturing
  Technology}\ }\textbf {\bibinfo {volume} {101}},\ \bibinfo {pages} {391--417}
  (\bibinfo {year} {2019})}\BibitemShut {NoStop}%
\bibitem [{\citenamefont {{NASA}}(2021)}]{NASA6030}%
  \BibitemOpen
  \bibfield  {author} {\bibinfo {author} {\bibnamefont {{NASA}}},\ }\href
  {https://standards.nasa.gov/standard/NASA/NASA-STD-6030} {\emph {\bibinfo
  {title} {Additive Manufacturing Requirements for Spaceflight Systems}}},\
  \bibinfo {type} {Technical Standard}\ \bibinfo {number} {NASA-STD-6030}\
  (\bibinfo  {institution} {National Aeronautics and Space Administration},\
  \bibinfo {year} {2021})\ \bibinfo {note} {baseline approved 21 April 2021;
  especially Sections 4.8, 4.9, and 4.18}\BibitemShut {NoStop}%
\bibitem [{\citenamefont {Leger}\ \emph {et~al.}(2025)\citenamefont {Leger},
  \citenamefont {Ghaffari-Tabrizi}, \citenamefont {Shaw}, \citenamefont
  {Rasera}, \citenamefont {Dickson}, \citenamefont {Valentin}, \citenamefont
  {Morlock}, \citenamefont {Thoresen},\ and\ \citenamefont
  {Cowley}}]{Leger2025}%
  \BibitemOpen
  \bibfield  {author} {\bibinfo {author} {\bibfnamefont {D.}~\bibnamefont
  {Leger}}, \bibinfo {author} {\bibfnamefont {F.}~\bibnamefont
  {Ghaffari-Tabrizi}}, \bibinfo {author} {\bibfnamefont {M.}~\bibnamefont
  {Shaw}}, \bibinfo {author} {\bibfnamefont {J.}~\bibnamefont {Rasera}},
  \bibinfo {author} {\bibfnamefont {D.}~\bibnamefont {Dickson}}, \bibinfo
  {author} {\bibfnamefont {B.}~\bibnamefont {Valentin}}, \bibinfo {author}
  {\bibfnamefont {A.}~\bibnamefont {Morlock}}, \bibinfo {author} {\bibfnamefont
  {F.}~\bibnamefont {Thoresen}},\ and\ \bibinfo {author} {\bibfnamefont
  {A.}~\bibnamefont {Cowley}},\ }\bibfield  {title} {\bibinfo {title} {Modeling
  energy requirements for oxygen production on the {Moon}},\ }\href
  {https://doi.org/10.1073/pnas.2306146122} {\bibfield  {journal} {\bibinfo
  {journal} {Proc. Nat. Acad. Sci.}\ }\textbf {\bibinfo {volume} {122}},\
  \bibinfo {pages} {e2306146122} (\bibinfo {year} {2025})}\BibitemShut
  {NoStop}%
\bibitem [{\citenamefont {{NASA}}(2024{\natexlab{b}})}]{NASAOSAM2024}%
  \BibitemOpen
  \bibfield  {author} {\bibinfo {author} {\bibnamefont {{NASA}}},\ }\href
  {https://www.nasa.gov/mission/on-orbit-servicing-assembly-and-manufacturing-1/}
  {\bibinfo {title} {On-orbit servicing, assembly, and manufacturing 1
  ({OSAM-1}): mission update}} (\bibinfo {year} {2024}{\natexlab{b}}),\
  \bibinfo {note} {cancellation and orderly-shutdown announcement}\BibitemShut
  {NoStop}%
\bibitem [{\citenamefont {Figliozzi}(2024)}]{NASAARMADAS2024}%
  \BibitemOpen
  \bibfield  {author} {\bibinfo {author} {\bibfnamefont {G.}~\bibnamefont
  {Figliozzi}},\ }\href
  {https://www.nasa.gov/general/robot-team-builds-high-performance-digital-structure-for-nasa/}
  {\bibinfo {title} {Robot team builds high-performance digital structure for
  {NASA}}},\ \bibinfo {howpublished} {NASA Ames Research Center} (\bibinfo
  {year} {2024})\BibitemShut {NoStop}%
\bibitem [{\citenamefont {Stewart}\ \emph {et~al.}(2025)\citenamefont
  {Stewart}, \citenamefont {Chapin}, \citenamefont {Leontie},\ and\
  \citenamefont {Henshaw}}]{Stewart2025}%
  \BibitemOpen
  \bibfield  {author} {\bibinfo {author} {\bibfnamefont {K.}~\bibnamefont
  {Stewart}}, \bibinfo {author} {\bibfnamefont {S.}~\bibnamefont {Chapin}},
  \bibinfo {author} {\bibfnamefont {R.}~\bibnamefont {Leontie}},\ and\ \bibinfo
  {author} {\bibfnamefont {C.~G.}\ \bibnamefont {Henshaw}},\ }\bibfield
  {title} {\bibinfo {title} {Crossing the {Sim2Real} gap between simulation and
  ground testing to space deployment of autonomous free-flyer control},\ }in\
  \href {https://arxiv.org/abs/2512.03736} {\emph {\bibinfo {booktitle} {2025
  International Conference on Space Robotics (iSpaRo)}}}\ (\bibinfo {year}
  {2025})\ \bibinfo {note} {author manuscript consulted; conference publication
  identified by the arXiv record},\ \Eprint {https://arxiv.org/abs/2512.03736}
  {arXiv:2512.03736 [cs.RO]} \BibitemShut {NoStop}%
\bibitem [{\citenamefont {Gaskill}(2025)}]{NASAAM2025}%
  \BibitemOpen
  \bibfield  {author} {\bibinfo {author} {\bibfnamefont {M.~L.}\ \bibnamefont
  {Gaskill}},\ }\href
  {https://www.nasa.gov/missions/station/3d-printing-saving-weight-and-space-at-launch/}
  {\bibinfo {title} {{{3D} Printing: Saving Weight and Space at Launch}}},\
  \bibinfo {howpublished} {NASA ISS Research Communications} (\bibinfo {year}
  {2025})\BibitemShut {NoStop}%
\bibitem [{\citenamefont {{NASA Marshall Space Flight
  Center}}(2024)}]{NASAWelding2024}%
  \BibitemOpen
  \bibfield  {author} {\bibinfo {author} {\bibnamefont {{NASA Marshall Space
  Flight Center}}},\ }\href
  {https://www.nasa.gov/centers-and-facilities/marshall/nasa-to-transform-in-space-manufacturing-with-laser-beam-welding-collaboration/}
  {\bibinfo {title} {{NASA} to transform in-space manufacturing with laser beam
  welding collaboration}} (\bibinfo {year} {2024}),\ \bibinfo {note} {7
  November 2024; reduced-gravity test results}\BibitemShut {NoStop}%
\bibitem [{\citenamefont {Bengio}\ \emph {et~al.}(2026)\citenamefont {Bengio},
  \citenamefont {Clare}, \citenamefont {Prunkl}, \citenamefont {Murray},
  \citenamefont {Andriushchenko} \emph {et~al.}}]{AISafety2026}%
  \BibitemOpen
  \bibfield  {author} {\bibinfo {author} {\bibfnamefont {Y.}~\bibnamefont
  {Bengio}}, \bibinfo {author} {\bibfnamefont {S.}~\bibnamefont {Clare}},
  \bibinfo {author} {\bibfnamefont {C.}~\bibnamefont {Prunkl}}, \bibinfo
  {author} {\bibfnamefont {M.}~\bibnamefont {Murray}}, \bibinfo {author}
  {\bibfnamefont {M.}~\bibnamefont {Andriushchenko}}, \emph {et~al.},\ }\href
  {https://internationalaisafetyreport.org/publication/international-ai-safety-report-2026}
  {\emph {\bibinfo {title} {International {AI} Safety Report 2026}}},\ \bibinfo
  {type} {Tech. Rep.}\ \bibinfo {number} {DSIT 2026/001}\ (\bibinfo
  {institution} {Department for Science, Innovation and Technology},\ \bibinfo
  {year} {2026})\BibitemShut {NoStop}%
\bibitem [{\citenamefont {{European Cooperation for Space
  Standardization}}(2025)}]{ECSSRHA2025}%
  \BibitemOpen
  \bibfield  {author} {\bibinfo {author} {\bibnamefont {{European Cooperation
  for Space Standardization}}},\ }\href
  {https://ecss.nl/standard/ecss-q-st-60-15c-rev-1-radiation-hardness-assurance-20-march-2025/}
  {\emph {\bibinfo {title} {Space Product Assurance: Radiation Hardness
  Assurance---{EEE} Components}}},\ \bibinfo {type} {Tech. Rep.}\ \bibinfo
  {number} {ECSS-Q-ST-60-15C Rev.1}\ (\bibinfo  {institution} {ESA Requirements
  and Standards Section},\ \bibinfo {address} {Noordwijk, The Netherlands},\
  \bibinfo {year} {2025})\ \bibinfo {note} {20 March 2025; Secs. 3.2.6, 4.2,
  and 5.1--5.4}\BibitemShut {NoStop}%
\bibitem [{\citenamefont {{NASA Jet Propulsion
  Laboratory}}(2026{\natexlab{b}})}]{JPLSatellites}%
  \BibitemOpen
  \bibfield  {author} {\bibinfo {author} {\bibnamefont {{NASA Jet Propulsion
  Laboratory}}},\ }\href {https://ssd.jpl.nasa.gov/sats/phys_par/} {\bibinfo
  {title} {Planetary satellite physical parameters}} (\bibinfo {year}
  {2026}{\natexlab{b}})\BibitemShut {NoStop}%
\bibitem [{\citenamefont {{NASA}}(2026{\natexlab{e}})}]{NASATitan}%
  \BibitemOpen
  \bibfield  {author} {\bibinfo {author} {\bibnamefont {{NASA}}},\ }\href
  {https://science.nasa.gov/saturn/moons/titan/facts/} {\bibinfo {title}
  {{Titan: Facts}}} (\bibinfo {year} {2026}{\natexlab{e}})\BibitemShut
  {NoStop}%
\bibitem [{\citenamefont {Arney}\ and\ \citenamefont {Jones}(2015)}]{HAVOC}%
  \BibitemOpen
  \bibfield  {author} {\bibinfo {author} {\bibfnamefont {D.}~\bibnamefont
  {Arney}}\ and\ \bibinfo {author} {\bibfnamefont {C.}~\bibnamefont {Jones}},\
  }\href {https://ntrs.nasa.gov/citations/20160006329} {\emph {\bibinfo {title}
  {{HAVOC}: High Altitude {Venus} Operational Concept}}},\ \bibinfo {type}
  {Tech. Rep.}\ \bibinfo {number} {NTRS 20160006329}\ (\bibinfo  {institution}
  {NASA},\ \bibinfo {year} {2015})\BibitemShut {NoStop}%
\bibitem [{\citenamefont {{NASA Jet Propulsion Laboratory}}(2023)}]{MOXIE2023}%
  \BibitemOpen
  \bibfield  {author} {\bibinfo {author} {\bibnamefont {{NASA Jet Propulsion
  Laboratory}}},\ }\href
  {https://www.jpl.nasa.gov/news/nasas-oxygen-generating-experiment-moxie-completes-mars-mission/}
  {\bibinfo {title} {{NASA}'s oxygen-generating experiment {MOXIE} completes
  {Mars} mission}} (\bibinfo {year} {2023})\BibitemShut {NoStop}%
\bibitem [{\citenamefont {Jakosky}\ and\ \citenamefont
  {Edwards}(2018)}]{Jakosky2018}%
  \BibitemOpen
  \bibfield  {author} {\bibinfo {author} {\bibfnamefont {B.~M.}\ \bibnamefont
  {Jakosky}}\ and\ \bibinfo {author} {\bibfnamefont {C.~S.}\ \bibnamefont
  {Edwards}},\ }\bibfield  {title} {\bibinfo {title} {Inventory of {CO2}
  available for terraforming {Mars}},\ }\href
  {https://doi.org/10.1038/s41550-018-0529-6} {\bibfield  {journal} {\bibinfo
  {journal} {Nature Astronomy}\ }\textbf {\bibinfo {volume} {2}},\ \bibinfo
  {pages} {634--639} (\bibinfo {year} {2018})}\BibitemShut {NoStop}%
\bibitem [{\citenamefont {Turyshev}(2026{\natexlab{b}})}]{Terraform2026}%
  \BibitemOpen
  \bibfield  {author} {\bibinfo {author} {\bibfnamefont {S.~G.}\ \bibnamefont
  {Turyshev}},\ }\bibfield  {title} {\bibinfo {title} {Terraforming {Mars}:
  Mass, forcing, and industrial throughput constraints},\ }\href
  {https://doi.org/10.1103/krb8-h3v3} {\bibfield  {journal} {\bibinfo
  {journal} {APS Open Science}\ }\textbf {\bibinfo {volume} {1}},\ \bibinfo
  {pages} {000029} (\bibinfo {year} {2026}{\natexlab{b}})},\ \Eprint
  {https://arxiv.org/abs/2603.00402} {arXiv:2603.00402 [astro-ph.EP]}
  \BibitemShut {NoStop}%
\bibitem [{\citenamefont {Ansari}\ \emph {et~al.}(2024)\citenamefont {Ansari},
  \citenamefont {Kite}, \citenamefont {Ramirez}, \citenamefont {Steele},\ and\
  \citenamefont {Mohseni}}]{Ansari2024}%
  \BibitemOpen
  \bibfield  {author} {\bibinfo {author} {\bibfnamefont {S.}~\bibnamefont
  {Ansari}}, \bibinfo {author} {\bibfnamefont {E.~S.}\ \bibnamefont {Kite}},
  \bibinfo {author} {\bibfnamefont {R.}~\bibnamefont {Ramirez}}, \bibinfo
  {author} {\bibfnamefont {L.~J.}\ \bibnamefont {Steele}},\ and\ \bibinfo
  {author} {\bibfnamefont {H.}~\bibnamefont {Mohseni}},\ }\bibfield  {title}
  {\bibinfo {title} {Feasibility of keeping {Mars} warm with nanoparticles},\
  }\href {https://doi.org/10.1126/sciadv.adn4650} {\bibfield  {journal}
  {\bibinfo  {journal} {Science Advances}\ }\textbf {\bibinfo {volume} {10}},\
  \bibinfo {pages} {eadn4650} (\bibinfo {year} {2024})}\BibitemShut {NoStop}%
\bibitem [{\citenamefont {Richardson}\ \emph {et~al.}(2026)\citenamefont
  {Richardson}, \citenamefont {Ansari}, \citenamefont {Fan}, \citenamefont
  {Ramirez}, \citenamefont {Mohseni}, \citenamefont {Mischna}, \citenamefont
  {Hecht}, \citenamefont {Steele}, \citenamefont {Sharipov},\ and\
  \citenamefont {Kite}}]{Richardson2026}%
  \BibitemOpen
  \bibfield  {author} {\bibinfo {author} {\bibfnamefont {M.~I.}\ \bibnamefont
  {Richardson}}, \bibinfo {author} {\bibfnamefont {S.}~\bibnamefont {Ansari}},
  \bibinfo {author} {\bibfnamefont {B.}~\bibnamefont {Fan}}, \bibinfo {author}
  {\bibfnamefont {R.}~\bibnamefont {Ramirez}}, \bibinfo {author} {\bibfnamefont
  {H.}~\bibnamefont {Mohseni}}, \bibinfo {author} {\bibfnamefont {M.~A.}\
  \bibnamefont {Mischna}}, \bibinfo {author} {\bibfnamefont {M.~H.}\
  \bibnamefont {Hecht}}, \bibinfo {author} {\bibfnamefont {L.~J.}\ \bibnamefont
  {Steele}}, \bibinfo {author} {\bibfnamefont {F.}~\bibnamefont {Sharipov}},\
  and\ \bibinfo {author} {\bibfnamefont {E.~S.}\ \bibnamefont {Kite}},\
  }\bibfield  {title} {\bibinfo {title} {Atmospheric dynamics of {IR}-active
  particles released from {Mars}' surface},\ }\href
  {https://doi.org/10.1029/2025GL121051} {\bibfield  {journal} {\bibinfo
  {journal} {GRL}\ }\textbf {\bibinfo {volume} {53}},\ \bibinfo {pages}
  {e2025GL121051} (\bibinfo {year} {2026})}\BibitemShut {NoStop}%
\bibitem [{\citenamefont {{NASA}}(2026{\natexlab{f}})}]{PlanetaryProtection}%
  \BibitemOpen
  \bibfield  {author} {\bibinfo {author} {\bibnamefont {{NASA}}},\ }\href
  {https://sma.nasa.gov/sma-disciplines/planetary-protection} {\bibinfo {title}
  {{Planetary Protection}}} (\bibinfo {year} {2026}{\natexlab{f}})\BibitemShut
  {NoStop}%
\bibitem [{\citenamefont {{NASA}}(2026{\natexlab{g}})}]{NASAArtemisIII2026}%
  \BibitemOpen
  \bibfield  {author} {\bibinfo {author} {\bibnamefont {{NASA}}},\ }\href
  {https://www.nasa.gov/mission/artemis-iii/} {\bibinfo {title} {{Artemis
  III}}} (\bibinfo {year} {2026}{\natexlab{g}}),\ \bibinfo {note} {mission
  page: 2027 crewed low-Earth-orbit demonstration target, not a lunar
  landing}\BibitemShut {NoStop}%
\bibitem [{\citenamefont {{NASA}}(2026{\natexlab{h}})}]{NASAArtemisIV2026}%
  \BibitemOpen
  \bibfield  {author} {\bibinfo {author} {\bibnamefont {{NASA}}},\ }\href
  {https://www.nasa.gov/mission/artemis-iv/} {\bibinfo {title} {{Artemis IV}}}
  (\bibinfo {year} {2026}{\natexlab{h}}),\ \bibinfo {note} {mission page:
  early-2028 landing target; two crew for about one week on the
  surface}\BibitemShut {NoStop}%
\bibitem [{\citenamefont
  {{NASA}}(2026{\natexlab{i}})}]{NASAMarsArchitecture2026}%
  \BibitemOpen
  \bibfield  {author} {\bibinfo {author} {\bibnamefont {{NASA}}},\ }\href
  {https://www.nasa.gov/moontomarsarchitecture/} {\bibinfo {title} {Moon to
  {Mars} architecture}} (\bibinfo {year} {2026}{\natexlab{i}}),\ \bibinfo
  {note} {agency description of required capabilities and elements, not a
  mission manifest}\BibitemShut {NoStop}%
\bibitem [{\citenamefont {{NASA}}(2026{\natexlab{j}})}]{NASADragonfly2026}%
  \BibitemOpen
  \bibfield  {author} {\bibinfo {author} {\bibnamefont {{NASA}}},\ }\href
  {https://science.nasa.gov/mission/dragonfly/} {\bibinfo {title} {Dragonfly}}
  (\bibinfo {year} {2026}{\natexlab{j}}),\ \bibinfo {note} {mission page:
  no-earlier-than July 2028 launch and late-2034 arrival targets; robotic
  mission}\BibitemShut {NoStop}%
\bibitem [{\citenamefont {{Intergovernmental Panel on Climate
  Change}}(2023)}]{IPCC2023}%
  \BibitemOpen
  \bibfield  {author} {\bibinfo {author} {\bibnamefont {{Intergovernmental
  Panel on Climate Change}}},\ }\href
  {https://doi.org/10.59327/IPCC/AR6-9789291691647} {\emph {\bibinfo {title}
  {Climate Change 2023: Synthesis Report. Contribution of Working Groups I, II
  and III to the Sixth Assessment Report of the Intergovernmental Panel on
  Climate Change}}},\ edited by\ \bibinfo {editor} {\bibfnamefont
  {H.}~\bibnamefont {Lee}}\ and\ \bibinfo {editor} {\bibfnamefont
  {J.}~\bibnamefont {Romero}}\ (\bibinfo  {publisher} {IPCC},\ \bibinfo
  {address} {Geneva, Switzerland},\ \bibinfo {year} {2023})\ \bibinfo {note}
  {{Core Writing Team}; 184 pp.}\BibitemShut {Stop}%
\bibitem [{\citenamefont {{NASA}}(2023{\natexlab{b}})}]{DART2023}%
  \BibitemOpen
  \bibfield  {author} {\bibinfo {author} {\bibnamefont {{NASA}}},\ }\href
  {https://www.nasa.gov/centers-and-facilities/marshall/nasas-dart-data-validates-kinetic-impact-as-planetary-defense-method/}
  {\bibinfo {title} {{NASA}'s {DART} data validates kinetic impact as planetary
  defense method}} (\bibinfo {year} {2023}{\natexlab{b}})\BibitemShut {NoStop}%
\bibitem [{\citenamefont {Baum}\ \emph {et~al.}(2015)\citenamefont {Baum},
  \citenamefont {Denkenberger}, \citenamefont {Pearce}, \citenamefont
  {Robock},\ and\ \citenamefont {Winkler}}]{Baum2015}%
  \BibitemOpen
  \bibfield  {author} {\bibinfo {author} {\bibfnamefont {S.~D.}\ \bibnamefont
  {Baum}}, \bibinfo {author} {\bibfnamefont {D.~C.}\ \bibnamefont
  {Denkenberger}}, \bibinfo {author} {\bibfnamefont {J.~M.}\ \bibnamefont
  {Pearce}}, \bibinfo {author} {\bibfnamefont {A.}~\bibnamefont {Robock}},\
  and\ \bibinfo {author} {\bibfnamefont {R.}~\bibnamefont {Winkler}},\
  }\bibfield  {title} {\bibinfo {title} {Resilience to global food supply
  catastrophes},\ }\href {https://doi.org/10.1007/s10669-015-9549-2} {\bibfield
   {journal} {\bibinfo  {journal} {Environment Systems and Decisions}\ }\textbf
  {\bibinfo {volume} {35}},\ \bibinfo {pages} {301--313} (\bibinfo {year}
  {2015})}\BibitemShut {NoStop}%
\bibitem [{\citenamefont {{NASA}}(2023{\natexlab{c}})}]{NASAReadiness2023}%
  \BibitemOpen
  \bibfield  {author} {\bibinfo {author} {\bibnamefont {{NASA}}},\ }\href
  {https://www.nasa.gov/directorates/somd/space-communications-navigation-program/technology-readiness-levels/}
  {\bibinfo {title} {{Technology Readiness Levels}}} (\bibinfo {year}
  {2023}{\natexlab{c}})\BibitemShut {NoStop}%
\bibitem [{\citenamefont {{NASA}}(2026{\natexlab{k}})}]{NASAHumansMars2026}%
  \BibitemOpen
  \bibfield  {author} {\bibinfo {author} {\bibnamefont {{NASA}}},\ }\href
  {https://www.nasa.gov/humans-in-space/humans-to-mars/} {\bibinfo {title}
  {Humans to {Mars}}} (\bibinfo {year} {2026}{\natexlab{k}}),\ \bibinfo {note}
  {the illustrated 2039 round trip is an example trajectory, not a qualified
  mission schedule}\BibitemShut {NoStop}%
\end{thebibliography}


\end{document}